\documentclass[aps,prd,nofootinbib,twocolumn,superscriptaddress,letterpaper,preprintnumbers, longbibliography]{revtex4-1}

\usepackage{amssymb}
\usepackage{amsmath}
\usepackage{epsfig}
\usepackage[colorlinks=true
,urlcolor=blue
,anchorcolor=blue
,citecolor=blue
,filecolor=blue
,linkcolor=red
,menucolor=blue
,hyperfootnotes=false,
,linktocpage=true
,pdfproducer=medialab
,pdfa=true
]{hyperref}
\usepackage{array}
\usepackage{booktabs}
\usepackage{comment}
\usepackage{cleveref}
\usepackage{multirow}

\newcolumntype{L}[1]{>{\raggedright\let\newline\\\arraybackslash\hspace{0pt}}m{#1}}
\newcolumntype{C}[1]{>{\centering\let\newline\\\arraybackslash\hspace{0pt}}m{#1}}
\newcolumntype{R}[1]{>{\raggedleft\let\newline\\\arraybackslash\hspace{0pt}}m{#1}}

\usepackage{afterpage}
\usepackage{longtable}
\usepackage{capt-of}

\usepackage{siunitx}
\DeclareSIUnit\electronvolt{e\kern-.05em V}
\DeclareSIUnit\parsec{pc}

\usepackage{graphicx}
\usepackage{gensymb}
\usepackage{subfigure}
\usepackage{blindtext}

\usepackage{listings}
\usepackage{xcolor}
\usepackage[T1]{fontenc}

\definecolor{codegreen}{rgb}{0,0.6,0}
\definecolor{codegray}{rgb}{0.5,0.5,0.5}
\definecolor{codepurple}{rgb}{0.58,0,0.82}
\definecolor{backcolour}{rgb}{0.95,0.95,0.92}

\lstdefinestyle{mystyle}{
    language = Python,  
    commentstyle=\color{codegreen},
    keywordstyle=\color{magenta},
    keywordstyle=[1]\color[rgb]{0,0,0.75},
    keywordstyle=[2]\color[rgb]{0.5,0.0,0.0},
    keywordstyle=[3]\color[rgb]{0.127,0.427,0.514},
    keywordstyle=[4]\color[rgb]{0.4,0.4,0.4},
    commentstyle=\color[rgb]{0.133,0.545,0.133},
    numberstyle=\tiny\color{codegray},
    stringstyle=\color{codepurple},
    basicstyle=\ttfamily,
    breakatwhitespace=false,         
    breaklines=true,                 
    captionpos=b,                    
    keepspaces=false,                 
    numbers=none,                    
    numbersep=5pt,                  
    showspaces=false,                
    showstringspaces=false,
    showtabs=false,                  
    tabsize=2,
    morekeywords={True, False, len},
    columns=flexible
}

\newcommand\beq{\begin{alignat}{1}}
\newcommand\eeq{\end{alignat}}
\newcommand{\dhis}{\texttt{DarkHistory} }

\newcommand*\bbar[1]{%
  \vbox{%
    \hrule height 0.5pt
    \kern-0.4ex
    \hbox{%
      \kern-0.2em
      \ifmmode#1\else\ensuremath{#1}\fi
      \kern-0.1em
    }
  }
}

\begin{document}

\preprint{MIT-CTP/5113}

\title{\texttt{DarkHistory}: A code package for calculating modified cosmic ionization and thermal histories with dark matter and other exotic energy injections}

\author{Hongwan Liu}
\email{hongwan@mit.edu}
\affiliation{Center for Theoretical Physics, Massachusetts Institute of Technology, Cambridge, MA 02139, U.S.A.}
\affiliation{School of Natural Sciences, Institute for Advanced Study, Princeton, NJ 08540, U.S.A.}

\author{Gregory W. Ridgway}
\email{gridgway@mit.edu}
\affiliation{Center for Theoretical Physics, Massachusetts Institute of Technology, Cambridge, MA 02139, U.S.A.}

\author{Tracy R. Slatyer}
\email{tslatyer@mit.edu}
\affiliation{Center for Theoretical Physics, Massachusetts Institute of Technology, Cambridge, MA 02139, U.S.A.}
\affiliation{School of Natural Sciences, Institute for Advanced Study, Princeton, NJ 08540, U.S.A.}

\begin{abstract} 
	We present a new public \texttt{Python} package, \texttt{DarkHistory}, for computing the effects of dark matter annihilation and decay on the temperature and ionization history of the early universe. 
	\texttt{DarkHistory} simultaneously solves for the evolution of the free electron fraction and gas temperature, and for the cooling of annihilation/decay products and the secondary particles produced in the process. 
	Consequently, we can self-consistently include the effects of both astrophysical and exotic sources of heating and ionization, and automatically take into account backreaction, where modifications to the ionization/temperature history in turn modify the energy-loss processes for injected particles. We present a number of worked examples, demonstrating how to use the code in a range of different configurations, in particular for arbitrary dark matter masses and annihilation/decay final states. Possible applications of \texttt{DarkHistory} include mapping out the effects of dark matter annihilation/decay on the global 21cm signal and the epoch of reionization, as well as the effects of exotic energy injections other than dark matter annihilation/decay.
	The code is available at \href{https://github.com/hongwanliu/DarkHistory}{https://github.com/hongwanliu/DarkHistory} with documentation at \href{https://darkhistory.readthedocs.io/en/development/}{https://darkhistory.readthedocs.io}. Data files required to run the code can be downloaded at \href{https://doi.org/10.7910/DVN/DUOUWA}{https://doi.org/10.7910/DVN/DUOUWA}. 
\end{abstract}


\maketitle

\section{Introduction}
\label{sec:Introduction}

Dark matter annihilation or decay and other exotic sources of energy injection can significantly alter the ionization and temperature histories of the universe. In this paper we describe a new public code package, \texttt{DarkHistory}, that allows fast and accurate computation of these possible effects of exotic energy injection on astrophysical and cosmological observables.
 
In particular, we will focus on interactions that allow dark matter (DM) to decay or annihilate into electromagnetically interacting Standard Model particles. This case has been studied extensively in the literature: stringent constraints on the dark matter annihilation cross section and decay lifetime have been derived from the way these Standard Model products would distort the anisotropies of the cosmic microwave background (CMB)~\cite{Slatyer:2009yq,Slatyer:2012yq,Slatyer:2016qyl,Kanzaki:2009hf}, or increase the temperature of the Inter-Galactic Medium (IGM), consequently affecting 21-cm and Lyman-$\alpha$ line emission \cite{Liu:2016cnk, Lopez-Honorez:2016sur, Liu:2018uzy, Diamanti:2013bia}. 

\dhis facilitates the calculation of these observables and the resulting constraints. In particular, \dhis makes the temperature constraint calculations significantly more streamlined, self-consistent, and accurate. It has a modular structure, allowing users to easily adjust individual inputs to the calculation -- e.g.\ by changing the reionization model, or the spectrum of particles produced by dark matter annihilation/decay. Compared to past codes developed for such analyses~\cite{Stocker:2018avm}, \dhis has a number of important new features:
\begin{itemize}
\item the first fully self-consistent treatment of exotic energy injection. Exotic energy injections can modify the evolution of the IGM temperature $T_\text{IGM}$ and free electron fraction $x_e$, and previously this modification has been treated perturbatively, assuming the backreaction effect due to these modifications on the cooling of injected particles is negligible. This assumption can break down toward the end of the cosmic dark ages for models that are not yet excluded \cite{Liu:2016cnk}. \texttt{DarkHistory} solves simultaneously for the temperature and ionization evolution and the cooling of the injected particles, avoiding this assumption;
\item a self-contained treatment of astrophysical sources of heating and reionization, allowing the study of the interplay between exotic and conventional sources of energy injection;
\item a large speed-up factor for computation of the full cooling cascade for high-energy injected particles (compared to the code employed in e.g.\ \cite{Liu:2016cnk}), via pre-computation of the relevant transfer functions as a function of particle energy, redshift and ionization level;
\item support for treating helium ionization and recombination, including the effects of exotic energy injections; and
\item a new and more correct treatment of inverse Compton scattering (ICS) for mildly relativistic and non-relativistic electrons; previous work in the literature has relied on approximate rates which are not always accurate.
\end{itemize}
Due to these improvements, \texttt{DarkHistory} allows for rapid scans over many different prescriptions for reionization, either in the form of photoheating and photoionization rates, or a hard-coded background evolution for $x_e$. The epoch of reionization is currently rather poorly constrained, making it important to understand the observational signatures of different scenarios, and the degree to which exotic energy injections might be separable from uncertainties in the reionization model. Previous attempts to model the effects of DM annihilation and decay into the reionization epoch have typically either assumed a fixed ionization history~\cite{Stocker:2018avm} -- requiring a slow re-computation of the cooling cascade if that history is changed \cite{Liu:2016cnk} -- or made an approximation for the effect of a modified ionization fraction on the cooling of high-energy particles~\cite{Lopez-Honorez:2013lcm,Diamanti:2013bia,Poulin:2015pna,Poulin:2016anj,Lopez-Honorez:2016sur}. 

Despite our emphasis on dark matter annihilation and decay, \texttt{DarkHistory} can be used to explore the effect of other forms of exotic particle injection. Other such possible sources include Hawking radiation from black holes~\cite{Poulin:2016anj,Clark:2018ghm}, radiation from accretion onto black holes~\cite{Hektor:2018qqw}, and processes from new physics such as de-excitation of dark matter or decay of meta-stable species~\cite{Hektor:2018lec}.

In Section~\ref{sec:histories} we review the physics of the ionization and temperature evolution, in the context of the three-level-atom (TLA) approximation, including the possibility of exotic energy injections. In Section~\ref{sec:code_structure} we discuss the overall structure of \texttt{DarkHistory}, which self-consistently combines the TLA evolution of the ionization and gas temperature with the cooling of particles injected by exotic processes. This section also describes the implementation of various physical processes in the code, in particular the treatment of cooling and production of secondaries by electrons and photons. In Section~\ref{sec:modules} we relate these processes to the various modules of \texttt{DarkHistory}, before providing a number of worked examples in Section~\ref{sec:examples}. We present our conclusions and discuss some future directions in Section~\ref{sec:conclusion}. We discuss our improved treatment of ICS in detail in Appendix~\ref{app:ICS}, provide the photon spectra from positronium annihilation in Appendix~\ref{app:positronium_annihilation_spec}, discuss a series of cross checks in Appendix~\ref{app:cross_checks}, and provide a table of definitions used throughout this paper in Appendix~\ref{app:table}.

\section{Ionization and Thermal Histories}
\label{sec:histories}
\texttt{DarkHistory} computes the ionization and temperature evolution of the universe in the presence of an exotic source of energy injection, such as dark matter annihilation or decay, using a modified version of the three-level atom (TLA) model for both hydrogen and helium, based on \texttt{RECFAST}~\cite{Seager:1999km,Seager:1999bc}. The reader may refer to Ref.~\cite{AliHaimoud:2010dx} for a detailed derivation of the unmodified TLA equations with hydrogen only, and Refs.~\cite{Seager:1999km,Seager:1999bc,Wong:2007ym} for the treatment of helium recombination in \texttt{RECFAST}. In this section, we will neglect the evolution of helium for simplicity, leaving a detailed discussion of our treatment of helium to Sec.~\ref{sec:helium}. 

In the absence of any source of energy injection, the TLA model, first derived in~\cite{Peebles:1968ja,Zeldovich:1969en}, provides a pair of coupled differential equations for the matter temperature in the IGM and the hydrogen ionization fraction:
\begin{alignat}{1}
    \dot{T}_m^{(0)} &= -2 H T_m + \Gamma_C (T_\text{CMB} - T_m) \,, \nonumber \\
    \dot{x}_\text{HII}^{(0)} &= -\mathcal{C} \left[n_\text{H} x_e x_\text{HII} \alpha_\text{H} - 4(1 - x_\text{HII}) \beta_\text{H} e^{-E_{21}/T_\text{CMB}}\right] \,,
    \label{eqn:TLA}
\end{alignat}
where $H$ is the Hubble parameter, $n_\text{H}$ is the total number density of hydrogen (both neutral and ionized), 
$x_\text{HII} \equiv n_\text{HII}/n_\text{H}$ where $n_\text{HII}$ is the number density of free protons, $x_e \equiv n_e/n_\text{H}$ is the free electron fraction with $n_e$ being the free electron density, and $E_{21} = \SI{10.2}{\eV}$ is the Lyman-$\alpha$ transition energy. $T_m$ and $T_\text{CMB}$ are the temperatures of the IGM and the CMB respectively.\footnote{We follow the standard astrophysical convention in which H and H$^+$ are denoted HI and HII, while He, He$^+$ and He$^{2+}$ are denoted HeI, HeII and HeIII respectively.}
$\alpha_\text{H}$ and $\beta_\text{H}$ are case-B recombination and photoionization coefficients for hydrogen respectively,\footnote{The value of $\beta_\text{H}$ used in \texttt{DarkHistory} includes the constant and gaussian fudge factors used by version 1.5.2 of \texttt{RECFAST}.} and $\mathcal{C}$ is the Peebles-C factor that represents the probability of a hydrogen atom in the $n = 2$ state decaying to the ground state before photoionization can occur~\cite{Peebles:1968ja,AliHaimoud:2010dx}. The photoionization coefficient is evaluated at the radiation temperature, $T_\text{CMB}$, in agreement with Ref.~\cite{Chluba:2015lpa}. $\Gamma_C$ is the Compton scattering rate, given by
\begin{alignat}{1}
    \Gamma_C = \frac{x_e}{1 + \mathcal{F}_\text{He} + x_e} \frac{8 \sigma_T a_r T_\text{CMB}^4}{3 m_e} \,,
\end{alignat}
where $\sigma_T$ is the Thomson cross section, $a_r$ is the radiation constant, $m_e$ is the electron mass, and $\mathcal{F}_\text{He} \equiv n_\text{He}/n_\text{H}$ is the relative abundance of helium nuclei by number. In the absence of helium, note that $x_e = x_\text{HII}$.  The solutions to Eq.~(\ref{eqn:TLA}) --- i.e.\ without any sources of energy injection --- define what we will call the baseline temperature and ionization histories, $T_m^{(0)}(z)$ and $x_\text{HII}^{(0)}(z)$. 

Exotic sources may inject additional energy into the universe, altering the thermal and ionization evolution shown in Eq.~(\ref{eqn:TLA}). For example, the rate of energy injection from DM annihilating with some velocity averaged cross section $\langle \sigma v \rangle$, or decaying with some lifetime $\tau$ much longer than the age of the universe, is given by
\begin{alignat}{1}
    \left(\frac{dE}{dV \, dt}\right)^\text{inj} = \begin{cases} 
        \rho_{\chi,0}^2 (1 + z)^6 \langle \sigma v \rangle/m_\chi\,, & \text{annihilation}, \\
        \rho_{\chi,0} (1 + z)^3 /\tau \,, & \text{decay}, 
    \end{cases}
    \label{eqn:energy_injection}
\end{alignat}
where $\rho_{\chi,0}$ is the mass density of DM today, and $m_\chi$ is the DM mass. This injected energy, however, does not in general manifest itself instantaneously as ionization, excitation, or heating of the gas. Instead, the primary particles injected into the universe may cool over timescales significantly larger than the Hubble time, producing secondary photons that may redshift significantly before depositing their energy into the gas.

Although the primary particles injected into the universe may be any type of Standard Model particle, we will only need to consider the cooling of photons and electron/positron pairs \cite{Slatyer:2009yq}. This simplification occurs because either the primaries are stable particles like photons, electrons and positrons, neutrinos, protons and anti-protons, and heavier nuclei, or are unstable particles that resolve into these particles on time scales much shorter than the cosmological time scales under consideration.  For typical sources of energy injection we can neglect heavier nuclei because they are produced in negligible amounts, and neutrinos because they are very ineffective at depositing their energy. Protons and antiprotons generally form a subdominant component of stable electromagnetic particles across all possible Standard Model primaries~\cite{Cirelli:2010xx}, and deposit energy less effectively than electrons, positrons, and photons (although their effects are not completely negligible \cite{Weniger:2013hja}). We therefore only decompose the injection of any primary into an effective injection of photons, electrons, and positrons, in accordance with Ref.~\cite{Slatyer:2009yq} and subsequent works. Adding the contribution from protons and antiprotons may strengthen these constraints by a small amount.

A significant amount of work has been done on computing the cooling of high energy photons, electrons, and positrons~\cite{Slatyer:2009yq,Furlanetto:2009uf,Valdes:2009cq,Slatyer:2012yq,Evoli:2012zz,Galli:2013dna,Evoli:2014pva,Slatyer:2015kla,Kanzaki:2008qb,Kawasaki:2015peu}.
Once the cooling of injected primary particles is determined, the energy deposited into channel $c$ (hydrogen ionization, excitation, or heating) can be simply parametrized as

\begin{alignat}{1}
    \left(\frac{dE}{dV \, dt}\right)_c^\text{dep} = f_c(z, \mathbf{x}) \left(\frac{dE}{dV \, dt}\right)^\text{inj} \,,
    \label{eqn:fz}
\end{alignat}
with all of the complicated physics condensed into a single numerical factor that is dependent on the redshift and the ionization fractions of all of the relevant species in the gas, which we denote $\mathbf{x} \equiv (x_\text{HII}, x_\text{HeII}, x_\text{HeIII})$. When helium is neglected, the ionization dependence of these $f_c$ functions simplifies to a dependence on $x_\text{HII} = x_e$. These $f_c$ functions also depend on the energies and species of the injected particles, but for simplicity of notation we will not write these arguments explicitly. 

The effect of energy injection on the thermal and ionization history can now be captured by additional source terms,
\begin{alignat}{1}
    \dot{T}_m^\text{inj} &= \frac{2 f_\text{heat}(z, \mathbf{x})}{3(1 + \mathcal{F}_\text{He} + x_e) n_\text{H}} \left(\frac{dE}{dV\, dt}\right)^\text{inj} \,, \nonumber \\
    \dot{x}_\text{HII}^\text{inj} &= \left[\frac{f_\text{H ion}(z, \mathbf{x})}{\mathcal{R} n_\text{H}} + \frac{(1 - \mathcal{C}) f_\text{exc} (z, \mathbf{x})}{0.75 \mathcal{R} n_\text{H}}\right] \left(\frac{dE}{dV \, dt}\right)^\text{inj} \,,
\end{alignat}
where $\mathcal{R} = \SI{13.6}{\eV}$ is the ionization potential of hydrogen. 

Prior to this work, $f_c(z, \mathbf{x})$ has largely been computed assuming the standard ionization history computed by recombination codes $\mathbf{x}_\text{std} (z)$, essentially making $z$ the only independent variable of $f_c$ as a function. These calculations are therefore applicable only so long as any perturbations to the assumed ionization history (e.g.\ by additional sources of energy injection) are sufficiently small. This is generally a good approximation near recombination: at these redshifts, the ionization history is well-constrained by CMB power spectrum measurements, and therefore large perturbations to $x_e$ are highly disfavored. For $z \lesssim 100$, however, ionization levels that exceed the standard value of $x_e \sim 2 \times 10^{-4}$ by several orders of magnitude are experimentally allowed~\cite{Liu:2016cnk}. Moreover, star formation during the process of reionization rapidly ionizes and heats the universe at $z \lesssim 20$, causing the ionization and thermal history to diverge from the baseline histories.

The primary effect of an increase in ionization levels is to decrease the number of neutral hydrogen and helium atoms available to ionize, decreasing the fraction of injected power that goes into ionization of these species; on the other hand, increasing $x_e$ increases the number of charged particles available for low-energy electrons to scatter off and heat the IGM, increasing the fraction of power going into heating. Since energy injection processes generally increase $x_e$ with time, the power into heating increases at an accelerated rate at late times, making a proper calculation of $f_c(z, \mathbf{x})$ crucial for an accurate computation of the temperature history.

Computing the full $\mathbf{x}$-dependence of $f_c(z, \mathbf{x})$ also allows us to perform, for the first time, a consistent calculation of the temperature and ionization histories with both exotic energy injection processes and reionization. At the onset of reionization, stars begin to form, and the ionizing radiation emitted by these objects injects a large amount of energy into the IGM. There remains a large degree of uncertainty regarding how reionization proceeds, but given some model for the photoionization and photoheating rates, and including other important energy transfer processes such as collisional ionization and excitation, additional terms $\dot{T}_m^\text{re}$ and $\dot{x}_\text{HII}^\text{re}$ (as well as the corresponding terms for helium) can be included in Eq.~(\ref{eqn:TLA}) to model reionization. These terms are discussed in much greater detail in Sec.~\ref{sec:TLA_and_reionization}. 

To summarize, \texttt{DarkHistory} computes the ionization and thermal history in the presence of exotic sources of energy injection, with the evolution equations in the absence of helium given by

\begin{alignat}{1}
    \dot{T}_m &= \dot{T}_m^{(0)} + \dot{T}_m^{\text{inj}} + \dot{T}_m^\text{re} \,, \nonumber \\
    \dot{x}_\text{HII} &= \dot{x}_\text{HII}^{(0)} + \dot{x}_\text{HII}^{\text{inj}} + \dot{x}_\text{HII}^\text{re} \,.
    \label{eqn:TLA_DarkHistory}
\end{alignat}
In the rest of the paper, we will describe how we calculate the inputs required to integrate these equations, i.e.\ $f_c(z,\mathbf{x})$, $\dot{T}_m^\text{re}$, $\dot{x}_\text{HII}^\text{re}$ and the modifications necessary to include helium.

\section{Code Structure and Content}
\label{sec:code_structure}

In this section we discuss the structure and physics content of the \texttt{DarkHistory} package.

\subsection{Overview}
\label{sec:overview}

Fig.~\ref{fig:flowchart} shows a flowchart depicting the overall structure of \texttt{DarkHistory}. The overall goal of the code is to take in some injected spectrum of photons and electron/positron pairs at a given redshift, and partition the energy into several categories as they lose their energy over a small redshift step: 

\begin{enumerate}

    \item \textit{High-energy deposition}. This is the total amount of energy deposited into ionization, excitation and heating by any high-energy (above \SI{3}{\kilo\eV}) electron generated during any of the cooling processes;

    \item \textit{Low-energy electrons}. These are electrons that have kinetic energy below \SI{3}{\kilo\eV} where atomic cooling processes typically dominate over ICS after recombination. These electrons are separated out at each step in order to treat their energy deposition (which occurs in a timescale much shorter than the time step) more carefully;

    \item \textit{Low-energy photons}. These are photons with energies below \SI{3}{\kilo\eV} that either photoionize within the redshift step, or lie below \SI{13.6}{\eV}. Such photons either lose all their energy within the redshift step, or cool only through redshifting, and thus can be treated in a simplified manner; and

    \item \textit{Propagating photons}. These are photons that are present at the end of the redshift step and are not included in the low-energy photons category.
    
\end{enumerate}

Throughout the paper, we use the word ``electrons'' to refer to both electrons and positrons. Although the interactions of electrons and positrons with the gas differ, the ICS cross-sections are identical, and ICS dominates the energy losses down to energy scales where the positron is nonrelativistic \cite{MEDEAII}.  For nonrelativistic positrons, their mass energy is converted into photons through annihilation with electrons. Since the positron mass is much larger than the kinetic energy in this regime, neglecting differences in kinetic energy loss between electrons and positrons is unlikely to be important. In a future version of \dhis we plan to include a more sophisticated treatment of low energy electrons and positrons.

The outputs in the first three categories are used to compute the evolution of the ionization and temperature history at this redshift step, before the code moves on to the next step and performs the same calculation again. A brief description of a step in this loop is as follows:

\begin{figure*}[t!]
    \centering
    \includegraphics[scale=0.36]{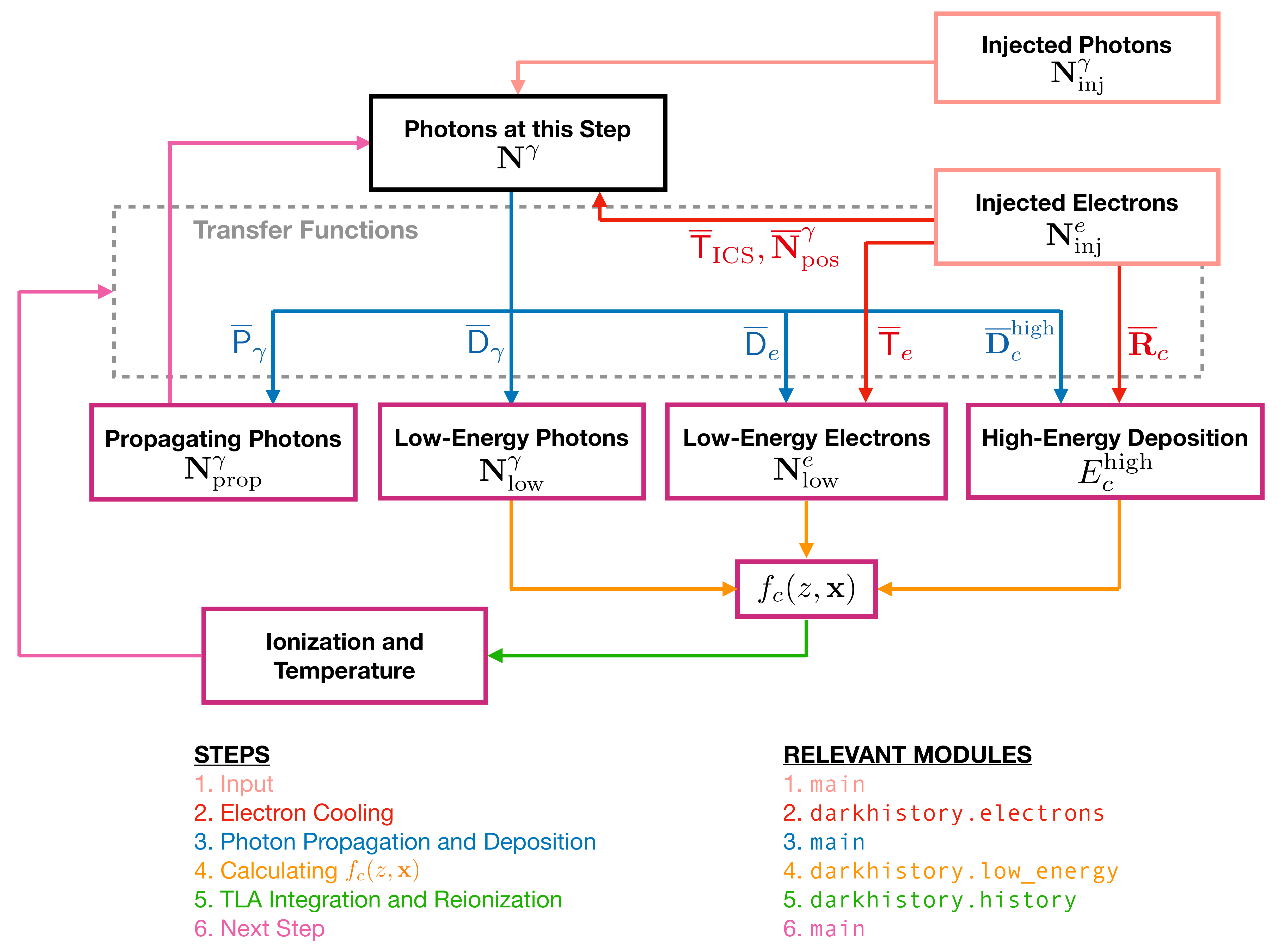}
    \caption{Flowchart showing schematically how the calculation of ionization and thermal histories in \texttt{DarkHistory} proceeds. Solid boxes represent input spectra (light pink), intermediate spectra used in calculations (black) and output spectra and quantities (purple), while arrows indicate numerical calculations that take place within the corresponding color-coded modules. The dashed grey box encloses all of the transfer functions for electron cooling (blue) and photon propagation and deposition (red), which are defined in Sec.~\ref{sec:electron_cooling} and~\ref{sec:photon_cooling} respectively. The calculation of $f_c(z)$ (orange) and the integration of the TLA (green) are explained in Sec.~\ref{sec:calculating_f} and~\ref{sec:TLA_and_reionization} respectively. Propagating photons and ionization/temperature values, which are used in calculating the transfer functions, are used as inputs for the next step (purple). All notation used here are defined in the text, and a summary table with their definitions can be found in Appendix~\ref{app:table}. Each step is outlined in Sec.~\ref{sec:overview}, and then explained in more detail in subsequent subsections within Sec.~\ref{sec:code_structure}. The modules shown here will also be outlined in Sec.~\ref{sec:modules}.}
    \label{fig:flowchart}
\end{figure*}

\begin{enumerate}
    \item \textit{Input}. Before the code begins, the user specifies a DM energy injection model or some other redshift-dependent energy injection rate, as well as the photon and $e^+e^-$ spectra produced per energy injection event. By default, \texttt{DarkHistory} starts from an initial redshift of $1+z = 3000$, ensuring that the spectra of particles present at and after recombination (at $z \sim 1000$) are accurate. Details are provided in Sec.~\ref{sec:input}. Inputs to the code are provided to the function \texttt{evolve()} found in the module \texttt{main}; some tools for obtaining spectra from an arbitrary injection of Standard Model particles can be found in the \texttt{pppc} module;

    \item \textit{Injected electron cooling}. Injected electrons (and positrons) cool through a combination of atomic processes and ICS. Transfer functions that map these injected electrons to high-energy deposition, secondary photons from ICS and positron annihilation, and low-energy electrons are computed and applied to the injected electrons. A discussion of these calculations can be found in Sec.~\ref{sec:electron_cooling} and in the \texttt{electrons} module of the code.
    
    The sum of the secondary photons produced by electron cooling, photons injected on this timestep, and propagating photons from the previous timestep are used as input to the photon cooling transfer functions, which we describe next;

    \item \textit{Photon propagation and energy deposition}. At this stage, we have a spectrum of photons that can undergo a range of  cooling processes to lose their energy over this redshift step. The effect of these cooling processes on the photon spectrum can be reduced to three transfer functions that we will describe in detail in Sec.~\ref{sec:photon_cooling}. These transfer functions have been pre-computed separately and can be downloaded at \href{https://doi.org/10.7910/DVN/DUOUWA}{https://doi.org/10.7910/DVN/DUOUWA}, together with all the other data required to run the code. These transfer functions determine how photons in this redshift step turn into propagating photons that continue on to the next redshift step, and low-energy photons and low-energy electrons that undergo further processing. All of these computations occur in the \texttt{main} module;

    \item \textit{Calculating }$f_c(z, \mathbf{x})$. The low-energy photons and low-energy electrons from this redshift step deposit their energy into ionization, heating and excitation of atoms, and the value of $f_c(z, \mathbf{x})$ at this step is computed by comparing the energy deposited in each channel to the energy injection rate for this timestep. Details of this computation are given in Sec.~\ref{sec:calculating_f}, and can be found in the \texttt{low\_energy} module;

    \item \textit{TLA integration and reionization}. With $f_c(z,\mathbf{x})$ at this step, we can now integrate the TLA across this redshift step. We can also include a reionization model, or track helium ionization, both of which add more terms to the TLA, as detailed in Sec.~\ref{sec:TLA_and_reionization}. We now know the $\mathbf{x}$ and $T_m$ that are reached at the end of this step. These calculations are done in the \texttt{history} module; and

    \item \textit{Next step}. The $\mathbf{x}$ and $T_m$ values computed above are passed to the next redshift step, so that all transfer functions at the next step can be computed at the appropriate ionization level. The propagating photons found above are also passed to the next step, and the loop repeats.
\end{enumerate}

Because $f_c(z, \mathbf{x})$ is computed by integrating the TLA at each step, and all transfer functions are evaluated at the value of $\mathbf{x}$ in the step, the backreaction of increased ionization levels is now fully accounted for. 

In the next several sections, we will describe both the physics and numerical methods that go into the loop.

\subsection{Discretization}
\label{sec:discretization}

Before describing in detail each part of \texttt{DarkHistory}, we will first describe how discretization occurs in our code, and the notation we will use throughout this paper. Typically, we will deal with some smooth spectrum of particles $dN/dE(E, A, B, \cdots)$, which is a function of the energy abscissa $E$, and several other variables that we denote here as $A, B, \cdots$. Smooth functions that are derivatives will always use `$d$' to denote differentiation, and parentheses to denote functional dependence. We shall always discretize such spectra as
\begin{alignat}{1}
    \frac{dN}{dE}(E_i, A_j, B_k, \cdots) \approx \mathsf{S}[E_i, A_j, B_k, \cdots] \,.
    \label{eqn:discretize_dNdE}
\end{alignat}
The discretized spectrum $\mathsf{S}$ is a matrix of dimension equal to the number of variables it depends on, where $i,j,k,...$ index discrete values of these variables. Throughout this paper, we will denote vectors (quantities which depend on a single variable) by a bold typeface and matrices (quantities that depend on multiple variables) by a sans-serif typeface. Discrete steps or changes are denoted by `$\Delta$', and discrete functional dependencies are written in square brackets. 

$\mathsf{S}$ times the bin width should always be regarded as a matrix of number of particles inside some bin, all with energy given by $E_i$. This matrix is mathematically defined as
\begin{multline}
    \mathsf{N}[E_i, A_j, B_k, \cdots] \equiv \mathsf{S}[E_i, A_j, B_k, \cdots] \times E_i \Delta \log E_i \,,
    \label{eqn:discretize_N}
\end{multline}
where $\Delta \log E_i$ is the log-energy bin width. We will always take $E_i \Delta \log E_i$ to be the bin width by convention. In \texttt{DarkHistory}, spectra are binned into energy values that are evenly log-spaced. $E_i$ should be regarded as the bin center, with the bin boundaries occurring at the geometric mean of adjacent energy values, and the boundaries of the first and last bin are taken to be symmetric (in log-space) about the bin centers. 

\subsection{Input}
\label{sec:input}

To initialize the loop described above, the user must specify the discretized photon and electron spectra produced per injection event, which we denote $\overline{\mathbf{N}}^\gamma_\text{inj}[E_j']$ and $\overline{\mathbf{N}}^e_\text{inj}[E_j']$. Bars denote spectra or transfer functions that have been normalized by some process or quantity, while spectra without any markings denote a number of particles per baryon from here on, unless otherwise specified.

Given the redshift-dependent rate of injection events per volume $\left(dN/dV \, dt\right)^\text{inj}$ we can determine the spectrum of particles $\mathbf{N}_\text{inj}^\alpha$ injected within a log-redshift step of width $\Delta \log (1+z)$ per baryon by
\begin{alignat}{1}
	\mathbf{N}^\alpha_\text{inj}[E_i', z] = \overline{\mathbf{N}}^\alpha_\text{inj}[E_i'] \left(\frac{dN}{dVdt}\right)^\text{inj} G(z) \, ,
  \label{eqn:injected_discretized_spec}
\end{alignat}
where $\alpha$ take on values $\gamma$ or $e$, and 
\begin{alignat}{1}
  G(z) \equiv \frac{\Delta \log(1+z)}{n_B(z) H(z)} \,,
  \label{eqn:per_baryon_to_dVdt}
\end{alignat}
where $n_B$ is the number density of baryons. $G(z)$ converts between the rate of injection events per volume and the number of injection events per baryon in the log-redshift step.

In the following sections, we will be mostly concerned with log-redshift steps, and so it is convenient to define
\begin{alignat}{1}
  y \equiv \log(1+z) \,,
\end{alignat}
and likewise $\Delta y \equiv \Delta \log (1+z)$.

\subsection{Injected Electron Cooling}
\label{sec:electron_cooling}

After specifying the injected spectra, the next step of the code is to resolve the injected electron/positron pairs, $\mathbf{N}^e_\text{inj}$. High-energy electrons and positrons cool through atomic processes (collisional ionization, collisional excitation and Coulomb heating), as well as ICS off CMB photons. After losing their kinetic energy to these processes, positrons ultimately annihilate with free electrons in the IGM, producing high-energy photons. All of these processes occur within a timescale much shorter than the timesteps considered in \texttt{DarkHistory}. Because of this, the code converts all input high-energy electrons into energy deposited into ionization, excitation, heating, scattered photons from ICS, and low-energy electrons (below \SI{3}{\kilo \eV}), which we treat separately. The photons produced from ICS are added to those that are injected promptly from the DM energy injection process, as well as propagating photons from the previous step.

We will first briefly discuss our calculation of the scattered photon and electron spectra from ICS, and then move on to describe the numerical method used to compute electron cooling.  

\subsubsection{Inverse Compton Scattering}
\label{sec:ICS}

ICS off CMB photons is an important energy loss mechanism for electrons/positrons over a large range of energies and redshifts. The efficiency of ICS as a cooling mechanism relative to atomic cooling processes has been the subject of some confusion in the literature, with some earlier studies~\cite{Valdes:2009cq,Evoli:2012zz} underestimating the cooling rate of the electrons. ICS becomes more important relative to atomic processes as the electron energy increases, but a correct treatment shows that even nonrelativistic electrons can have ICS as the main cooling mechanism in the early universe; at $z \sim 600$, for example, it is the primary energy loss mechanism for electrons with kinetic energy $\gtrsim \SI{10}{\kilo\eV}$~\cite{Galli:2013dna,Slatyer:2015kla}. Existing work on electron cooling has focused on the highly nonrelativistic regime (electron kinetic energy below \SI{3}{\kilo\eV})~\cite{Furlanetto:2006jb}, where ICS is unimportant compared to atomic cooling processes, or on the relativistic regime~\cite{Valdes:2009cq,Evoli:2012zz,Hansen:2003yj}. 

Earlier work by one of the authors~\cite{Slatyer:2012yq,Slatyer:2015kla} already incorporates ICS cooling for electrons across both the Thomson and the relativistic regimes. \texttt{DarkHistory} improves the accuracy of the calculation in the Thomson regime by using the full expression for the spectrum of scattered photons, with no further approximation. As a result, the code is able to accurately calculate the scattered photon spectrum and the energy loss spectrum of electrons. This means that we fully cover all relevant regimes for ICS for electrons of arbitrary energy scattering off the CMB at all redshifts $z \sim 10^9$ and below.\footnote{Above this redshift, photons have energies comparable to the electron mass $m_e$, and Klein-Nishina scattering can occur between photons and non-relativistic electrons, which falls outside of the two regimes considered here.} These calculations are fast and numerically stable even for nonrelativistic electrons, where conventional numerical integration can be unreliable due to the presence of catastrophic cancellations between large terms. 

We leave a full discussion of how \texttt{DarkHistory} treats ICS to Appendix~\ref{app:ICS}. In summary, the code is able to compute the scattered photon and electron spectra that are produced per unit time due to ICS off the CMB across all relevant kinematic regimes. These spectra are then taken as inputs for the numerical computation of how an electron cools taking into account all processes, which is described below.

\subsubsection{Numerical Method}
\label{sec:elec_cooling_numerical_method}

Consider an injected electron (or positron) with kinetic energy $E'$ (all quantities associated with injected particles throughout this paper will be denoted with $'$).
Let $R_c(E')$ be the energy eventually deposited into some channel $c$ by this electron, once it has lost all of its initial energy. Within a short time interval $\Delta t$ (taken to be \SI{1}{\second} in our calculation), the electron undergoes all possible cooling processes with some probability, producing the (averaged) secondary electron spectrum $dN/dE$. Within this same interval $\Delta t$, some portion of the energy $P_c(E')$ is also deposited promptly into the channel under consideration. The secondary electron spectrum then deposits its energy according to $R_c$ for energies lower than $E'$. We can thus write the following recursive equation:
\begin{alignat}{1}
    R_c(E') = \int dE \, R_c(E) \frac{dN}{dE} + P_c(E') \,.
    \label{eqn:elec_cooling_analytic}
\end{alignat}
Note that $R_c(E')$ does not include deposition to the channel $c$ via secondary photons from ICS or positron annihilation; because the cooling times of secondary photons can be much longer than a timestep, they must be treated separately. $R_c(E')$ as defined here is the ``high-energy deposition'' from electrons within the timestep, as described in Section~\ref{sec:overview}. The relevant channels are $c = \{\text{`ion'}, \text{`exc'}, \text{`heat'} \}$ for deposition into collisional ionization, collisional excitation and heating respectively. The `ion' and `exc' channels include ionization and excitation off all species.

As long as the time step $\Delta t$ is much shorter than the characteristic interaction timescale of all of the interactions, $dN/dE$ is simply the sum of all of the scattered electron spectra due to each process within $\Delta t$, normalized to a single injected electron. A detailed accounting of the relevant cross sections and secondary spectra is provided in Ref.~\cite{Slatyer:2009yq}, and these results can be used to calculate $dN/dE$ and $P_c$. We will denote the discretized version of the normalized scattered electron spectra by $\overline{\mathsf{N}}$, since it is normalized to one electron. 

Numerically, we would like to compute $\overline{\mathbf{R}}_c$, a vector containing the energy deposited into channel $c$, with each entry corresponding to a single electron with initial kinetic energy $E'$. The overline notation serves as a reminder that the quantity is normalized to one injected electron. The discretized version of Eq.~(\ref{eqn:elec_cooling_analytic}) reads
\begin{alignat}{1}
    \overline{\mathbf{R}}_c[E'_i] = \sum_j \overline{\mathsf{N}}[E'_i, E_j] \overline{\mathbf{R}}_c[E_j] + \overline{\mathbf{P}}_c[E'_i] \,,
    \label{eqn:elec_cooling_dep_tf}
\end{alignat}
where $\overline{\mathbf{P}}_c$ is the vector of the prompt energy deposition in channel $c$ per electron. This is a linear system of equations, and we can solve for each $\overline{\mathbf{R}}_c$ given $\overline{\mathsf{N}}$ and $\overline{\mathbf{P}}_c$.

A similar procedure also works for finding the ICS photon spectrum after an electron completely cools. Let the discretized spectrum be $\overline{\mathsf{T}}_{\text{ICS},0} [E_{e,i}', E_{\gamma,j}]$, where $E'_e$ is the initial electron kinetic energy, and $E_\gamma$ is the photon energy. Then the ICS photon spectrum produced after complete cooling of a single electron satisfies
\begin{alignat}{2}
    \overline{\mathsf{T}}_{\text{ICS},0} [E'_{e,i}, E_{\gamma,j}] &=&&  \sum_k\overline{\mathsf{N}}[E_{e,i}', E_{e,k}] \overline{\mathsf{T}}_{\text{ICS},0} [E_{e,k}, E_{\gamma, j}] \nonumber \\
    & && \,\, + \overline{\mathsf{N}}_\text{ICS} [E_{e,i}', E_{\gamma,j}] \,,
    \label{eqn:ics_photons}
\end{alignat}
with $\overline{\mathsf{N}}_\text{ICS}$ being the discretized version of the scattered photon spectrum defined in Eq.~(\ref{eqn:ics_scat_phot_spec}) within $\Delta t$, and indices $e$ and $\gamma$ have been inserted to clarify the difference between electron and photon energies. This spectrum consists of CMB photons that are upscattered by the injected electron; in order to be able to track energy conservation, we also need to keep track of the initial energy of the upscattered photons. We therefore also need to solve
\begin{alignat}{1}
  \overline{\mathbf{R}}_\text{CMB}[E_i'] = \sum_j \overline{\mathsf{N}}[E_i', E_j] \overline{\mathbf{R}}_\text{CMB}[E_j] + \overline{\mathbf{P}}_\text{CMB} [E_i'] \,,
  \label{eqn:elec_cooling_cmbloss}
\end{alignat}
where $\overline{\mathbf{P}}_\text{CMB}$ is the total initial energy of photons upscattered in $\Delta t$.\footnote{We do not have to track the photon spectrum, since the initial CMB photon energy is only significant for nonrelativistic injected electrons, which are always in the Thomson regime and hence scatter in a frequency-independent manner. For relativistic electrons, the initial CMB photon energy is neglected, as the photon is overwhelmingly upscattered to a much higher final energy.} At this point, we now define $\overline{T}_\text{ICS}$ to be the ICS photon spectrum with the upscattered CMB spectrum subtracted out, so that $\overline{T}_\text{ICS}$ now represents a \textit{distortion} to the CMB spectrum: 
\begin{alignat}{2}
  \overline{\mathsf{T}}_\text{ICS} [E'_{e,i}, E_{\gamma,j}] &=&& \overline{\mathsf{T}}_{\text{ICS},0} [E'_{e,i}, E_{\gamma,j}] \nonumber \\
  & && - \overline{\mathbf{R}}_\text{CMB}[E_{e,i}'] \overline{\mathbf{N}}_\text{CMB}[E_{\gamma,j}]\,,
\label{eqn:elec_cooling_ics}
\end{alignat}
where $\overline{\mathbf{N}}_\text{CMB}$ is the CMB spectrum normalized to unit total energy. The total of energy of $\overline{T}_\text{ICS}$ for each $E_{e,i}'$ therefore gives the energy lost by the incoming electron through ICS.

Finally, the low-energy electron spectrum produced is similarly given by
\begin{alignat}{2}
    \overline{\mathsf{T}}_e [E'_{e,i}, E_{e,j}] &=&& 
    \sum_k\overline{\mathsf{N}}_\text{high}[E_{e,i}', E_{e,k}] \overline{\mathsf{T}}_e [E_{e,k}, E_{e, j}] \nonumber \\
    & && \,\, + \overline{\mathsf{N}}_\text{low} [E_{e,i}', E_{e,j}] \,,
    \label{eqn:elec_cooling_lowengelec}
\end{alignat}
where $\overline{\mathsf{N}}_\text{high}$ ($\overline{\mathsf{N}}_\text{low}$) is $\overline{\mathsf{N}}$ with only high-energy (low-energy) $E_{e,k}$ included. 

In \texttt{DarkHistory}, we choose a square matrix $\overline{\mathsf{N}}$ with the same abscissa for both injected and scattered electron energies. As a result, $\overline{\mathsf{N}}$ has diagonal values that are very close to 1, since most particles do not scatter within $\Delta t$. Because of this, we find that it is numerically more stable to solve the equivalent equation
\begin{alignat}{1}
  \frac{\widetilde{E}[E_i']}{E_i'} \overline{\mathbf{R}}_c[E_i'] = \sum_j \widetilde{\mathsf{N}}[E_i', E_j] \overline{\mathbf{R}}_c[E_j] + \overline{\mathbf{P}}_c [E_i'] \,,
  \label{eqn:elec_cooling_actual}
\end{alignat}
where
\begin{alignat}{1}
  \widetilde{\mathsf{N}}[E_i', E_j] &\equiv \begin{cases}
    \overline{\mathsf{N}}[E_i', E_j] \,, & E_i' < E_j \,, \\
    0 \,, & \text{otherwise},
  \end{cases} \\ \nonumber \\
  \widetilde{E}[E_i'] &\equiv \sum_j \widetilde{\mathsf{N}} [E_i', E_j] E_j + \sum_c \overline{\mathbf{R}}_c[E_i'] \nonumber \\
  & \quad\, + \sum_j \overline{\mathsf{T}}_\text{ICS} [E_i', E_{\gamma,j}] E_{\gamma,j} \,.
\end{alignat}
The variables $\widetilde{\mathsf{N}}$ and $\widetilde{E}$ are simply the number of electrons and total energy excluding electrons that remained in the same energy bin after $\Delta t$. Eqs.~(\ref{eqn:elec_cooling_ics}) and~(\ref{eqn:elec_cooling_lowengelec}) can be similarly transformed in the same way as Eq.~(\ref{eqn:elec_cooling_actual}) and solved. Since $\widetilde{N}$ is a triangular matrix, the SciPy function \texttt{solve\_triangular()} is used for maximum speed.\footnote{The upscattering of electrons during ICS is negligible: see Appendix~\ref{app:ICS} for more details.}

Having calculated $\overline{\mathbf{R}}_c$, $\overline{\mathsf{T}}_\text{ICS}$ and $\overline{\mathsf{T}}_e$, all normalized to a single electron, the final result when an arbitrary electron spectrum $\mathbf{N}^e_\text{inj}[E'_{e,i}]$ completely cools is simply given by contracting these quantities with $\mathbf{N}^e_\text{inj}$. Note that all of these quantities are also dependent on redshift: we have simply suppressed this dependence for notational simplicity in this section.

Finally, after positrons have lost all of their kinetic energy, they are assumed to form positronium and annihilate promptly, producing a gamma ray spectrum that also gets added to the propagating photon spectrum. The positronium spectrum is given simply by
\begin{alignat}{1}
  \mathbf{N}^\gamma_\text{pos}[E_i] = \frac{1}{2} \overline{\mathbf{N}}^\gamma_\text{pos}[E_i] \sum_j \mathbf{N}^e_\text{inj}[E'_{j}] \,,
  \label{eqn:positronium_photons}
\end{alignat}
where $\overline{\mathbf{N}}^\gamma_\text{pos}$ is the positronium annihilation spectrum normalized to a single positron, shown in Appendix~\ref{app:positronium_annihilation_spec}. The factor of $1/2$ accounts for the fact that $\mathbf{N}^e_\text{inj}$ contains both electrons and positrons in equal number. 

Since all calculated quantities depend on $z$ and $\mathbf{x}$, all quantities discussed in this section have to be computed at each redshift step. This allows us to properly capture the effect of changing ionization levels on the energy deposition process. 

\subsection{Photon Propagation and Energy Deposition}
\label{sec:photon_cooling}
After resolving the injected electrons and obtaining the photons produced from their cooling, the spectrum of photons that have been newly injected per baryon per log-redshift can be discretized as
\begin{alignat}{1}
  \frac{dN^\gamma_\text{new}}{dE'_j \, dy}(E'_j) \times  E'_j \log \Delta E'_j \times \Delta y  \approx \mathbf{N}_\text{new}^\gamma[E_j'] \,,
\end{alignat}
where $\mathbf{N}^\gamma_\text{new}$ is the sum of photons injected directly by the injection event, and photons produced by the cooling of injected electrons, i.e.
\begin{alignat}{2}
  \mathbf{N}_\text{new}^\gamma[E_j'] &=&& \,\, \mathbf{N}^\gamma_\text{inj} [E'_j] + \mathbf{N}^\gamma_\text{pos}[E_j'] \nonumber \\
  & && + \sum_i \overline{\mathsf{T}}_\text{ICS} [E_{e,i}', E_j'] \mathbf{N}^e_\text{inj}[E'_{e,i}] \,.
  \label{eqn:new_inj_photons}
\end{alignat}
These photons can cool through a number of processes, including redshifting, pair production, Compton scattering and photoionization. Within a particular log-redshift step, low-energy photons and low-energy electrons are produced, and some high-energy deposition from high-energy electrons produced by $\mathbf{N}^\gamma_\text{new}$ occur. On the other hand, some part of the photon spectrum lies above \SI{13.6}{\eV} and does not photoionize within the log-redshift step; instead, these photons propagate forward to the next step.

The resulting deposition into low-energy photons and electrons was used to compute $f_c$ in Ref.~~\cite{Slatyer:2015kla}, assuming the fixed baseline ionization history. In order to capture the dependence on ionization history, however, we need to be able to calculate the propagation and deposition processes at any ionization level, redshift and injected particle energy. 

One of the main ideas of \texttt{DarkHistory} is to capture the photon cooling processes as precomputed transfer functions with injection energy, redshift and ionization levels as the dependent variables. These transfer functions then act on some incoming spectrum and produce a spectrum of propagating particles, a spectrum of deposited particles or some amount of deposited energy within a log-redshift step. These transfer functions can be evaluated at various points in injection energy, redshift, and ionization levels, and interpolated at other points. With a given injection model, we can then string together these transfer functions to work out the propagation of photons and the deposition of energy, over an extended redshift range, given any exotic source of energy injection. 

\subsubsection{Propagating Photons}
\label{sec:propagating_photons}

Consider a spectrum of photons per baryon denoted $dN^\gamma/dE'$ that is present in the universe at some log-redshift $y$. As these photons propagate, various cooling processes result in these photons being scattered into energies below \SI{13.6}{\eV}, or they may photoionize on an atom in the gas. Those particles that do not undergo either process within a redshift step are called ``propagating photons'', and continue to propagate into the next redshift step.  

We define the transfer function for propagating photons $ \overline{P}^\gamma(E', E, y', y)$ through the following relation:
\begin{alignat}{2}
  \left. \frac{dN_\text{prop}^\gamma}{dE} \right|_y &=&& \int dE' \, \overline{P}^\gamma (E', E, y', y) \left. \frac{dN^\gamma}{dE'} \right|_{y'} \,.
\end{alignat}
$\overline{P}^\gamma$ takes a spectrum of photons that are present at $y'$ and propagates them forward to a spectrum of propagating photons at $y$. $\overline{P}^\gamma(E', E, y', y)$ is exactly the number of propagating photons per unit energy that results from a single photon injected at log-redshift $y'$ with energy $E'$ cooling until log-redshift $y$. The $\overline{P}^\gamma$ functions are calculated separately using the code described in Ref.~\cite{Slatyer:2009yq,Slatyer:2015kla}. 

We distinguish between two different sources of photons between two redshifts $y'$ and $y$ (with $y' > y$): propagating photons at $y'$, $dN_\text{prop}^\gamma/dE'$, and the newly injected photons between the redshifts $y'$ and $y$, defined in discretized form in Eq.~(\ref{eqn:new_inj_photons}). With these sources, we can write the spectrum of propagating photons at $y$ as
\begin{alignat}{2}
  \left. \frac{dN_\text{prop}^\gamma}{dE} \right|_y &=&& \int dE'\, \overline{P}^\gamma(E', E, y', y) \left. \frac{dN_\text{prop}^\gamma}{dE'} \right|_{y'} \nonumber \\
  & &&+ \int dE' \int_y^{y'} d\eta \, \overline{P}^\gamma (E', E, \eta, y) \left. \frac{dN^\gamma_\text{new}}{dE'\, d\eta} \right|_\eta \,.
  \label{eqn:prop_tf_def}
\end{alignat}

We discretize this expression by defining the following discrete quantities according to the conventions set down in Eqs.~(\ref{eqn:discretize_dNdE}) and~(\ref{eqn:discretize_N}):
\begin{alignat}{2}
  \overline{\mathsf{P}}^\gamma[E'_i, E_j, y', \Delta y] E_i' \Delta \log E_i' &\approx&& \,\, \overline{P}^\gamma (E'_i, E_j, y', y' - \Delta y) \,, \nonumber \\
  \mathbf{N}^\gamma_\text{prop} [E_i', y'] &\approx&&  \left. \frac{dN_\text{prop}^\gamma}{dE'} \right|_{y'} E_i' \, \Delta \log E_i' \,,
\end{alignat}
where we have chosen some fixed value of $\Delta y$, so that the final redshift is $y = y' - \Delta y$. In \texttt{DarkHistory}, the default value is $\Delta y = 10^{-3}$, although this can be adjusted by the process of coarsening, described in Sec.~\ref{sec:coarsening}. Dropping the dependence on $\Delta y$ for simplicity, the discretized version of Eq.~(\ref{eqn:prop_tf_def}) reads
\begin{alignat}{2}
  \mathbf{N}_\text{prop}^\gamma[E_j, y] &=&& \sum_i \overline{\mathsf{P}}^\gamma [E_i', E_j, y'] \mathbf{N}^\gamma[E_i',y'] \,,
  \label{eqn:discretized_prop_tf}
\end{alignat}
where we have defined
\begin{alignat}{1}
  \mathbf{N}^\gamma[E_i', y] \equiv \mathbf{N}_\text{prop}^\gamma [E_i', y] + \mathbf{N}_\text{new}^\gamma [E_i', y] \,.
  \label{eqn:N_prop_plus_new}
\end{alignat}
%

\subsubsection{Energy Deposition}
\label{sec:energy_deposition}

Aside from $\overline{\mathsf{P}}^\gamma$, we also have three deposition transfer functions describing the energy losses of $\mathbf{N}^\gamma$ into high-energy deposition, low-energy electrons and low-energy photons, as defined in Sec.~\ref{sec:overview}. These transfer functions are defined by their action on the discretized photon spectrum, $\mathbf{N}^\gamma$, and are discretized in a similar manner.

The low-energy electron deposition transfer matrix, $\overline{\mathsf{D}}^e$, yields the low-energy electrons produced via cooling of $\mathbf{N}^\gamma$. Adding the low-energy electrons produced directly from the injected electrons $\mathbf{N}^e_\text{inj}$, we obtain the full low-energy electron spectrum $\mathbf{N}^e_\text{low}[E_j, y]$ at a particular redshift step:
\begin{alignat}{2}
	 \mathbf{N}^e_\text{low}[E_{e,j},y] &=&&  \sum_i \overline{\mathsf{D}}^e[E_{\gamma,i}', E_{e,j}, y'] \mathbf{N}^\gamma[E'_{\gamma,i}, y'] \nonumber \\
   & && + \mathbf{N}^e_\text{low,inj}[E_{e,j}, y] \,,
   \label{eqn:lowengelec_tf}
\end{alignat}
where
\begin{alignat}{1}
   \mathbf{N}^e_\text{low,inj}[E_{e,j}, y] = \sum_i \overline{\mathsf{T}}^e[E_{e,i}', E_{e,j}, y] \mathbf{N}^e_\text{inj}[E_{e,i}', y] \,,
\end{alignat}
while the deposition transfer matrix $\overline{\mathsf{D}}^\gamma$ yields the low-energy photons,
\begin{alignat}{1}
	 \mathbf{N}^\gamma_\text{low}[E_j, y] =  \sum_i \overline{\mathsf{D}}^\gamma[E_i', E_j, y'] \mathbf{N}^\gamma[E'_i, y'] \,.
   \label{eqn:lowengphot_tf}
\end{alignat}
$\mathbf{N}^\gamma_\text{low}$ is computed as a \textit{distortion} to the CMB spectrum, with $\overline{\mathsf{D}}^\gamma$ computed with the initial spectrum of upscattered CMB photons subtracted, in the same way as $\overline{T}_\text{ICS}$, as shown in Eq.~(\ref{eqn:elec_cooling_ics}). 

As the propagating photons cool over a single log-redshift step, they generate high-energy electrons along the way. These are handled in a similar manner to injected high-energy electrons as described in Sec.~\ref{sec:electron_cooling}, but instead of performing the calculation at each step, we simply provide transfer functions  $\overline{\mathbf{D}}_\text{c}^\text{high}$ that act on propagating photons and return the high-energy deposition into the channels $c = $\{`ion', `exc', `heat'\}.\footnote{For legacy reasons, \texttt{DarkHistory} actually computes the transfer function that returns the high-energy deposition per second; this is just a difference in convention.} We can then combine this with the result from electron cooling to obtain the high-energy deposition per baryon within a log-redshift step into each channel $c$:
\begin{alignat}{2}
  E_c^\text{high}[y] &=&& \sum_i \overline{\mathbf{D}}_\text{c}^\text{high}[E_{\gamma,i}', y'] \mathbf{N}^\gamma[E_{\gamma,i}', y'] \nonumber \\
  & && + \sum_i \overline{\mathbf{R}}_c [E_{e,i}', y'] \mathbf{N}^e_\text{inj}[E_{e,i}', y']  \,.
  \label{eqn:highengdep_tf}
\end{alignat}

To summarize, we have defined the following transfer functions: $\overline{\mathsf{P}}^\gamma$ for propagating photons, and $\overline{\mathsf{D}}^\gamma$, $\overline{\mathsf{D}}^e$ and $\overline{\mathbf{D}}^\text{high}_c$ for deposition into low-energy photons, low-energy electrons and high-energy deposition channels respectively. These transfer functions act on the spectrum of photons $\mathbf{N}^\gamma$ (from both the injection source and the cooling of injected electrons). Together with the transfer functions for the cooling of injected electrons, we have all the information needed to propagate injected particles and compute their energy deposition as a function of redshift. 

\subsubsection{Coarsening}
\label{sec:coarsening}

The propagating photons transfer function $\overline{\mathsf{P}}^\gamma$ can always be evaluated with the same input and output energy abscissa, so that the 2D transfer matrix at each $y$ is square. If the transfer function $\overline{\mathsf{P}}^\gamma$ does not vary significantly over redshift, then in the interest of computational speed, we can make the following approximation of Eq.~(\ref{eqn:discretized_prop_tf}) for propagation transfer matrices:
\begin{alignat}{1}
  \mathbf{N}^\gamma_\text{prop}[E_j, y - n \Delta y] \approx \left( \overline{\mathsf{P}}^\gamma_{1/2} \right)^n_{ji} \mathbf{N}_i^\gamma[y] \,,
  \label{eqn:prop_tf_coarsening}
\end{alignat}
where repeated indices are summed. $i$ and $j$ index input and output energies, and $\overline{\mathsf{P}}^\gamma_{1/2}$ is $\overline{\mathsf{P}}^\gamma$ evaluated at log-redshift $y - n \Delta y/2$ to minimize interpolation error. When making this approximation, we also have to ensure that we redefine
\begin{alignat}{1}
  \mathbf{N}_\text{inj}^\alpha[E_i', y] \to n \mathbf{N}_\text{inj}^\alpha[E_i',y]
\end{alignat}
for both channels $\alpha = e$ and $\gamma$, so that we (approximately) include all of the particles injected between $y$ and $y - n \Delta y$. 

Likewise, if both the deposition and propagation matrices do not vary significantly over redshift, we can approximate Eq.~(\ref{eqn:lowengphot_tf}) as
\begin{alignat}{2}
  \mathbf{N}^\gamma_\text{low}[E_j, y - n \Delta y] &\approx&& \left( \overline{\mathsf{D}}^\gamma_{1/2} \right)_{jk}\sum_m \left(\overline{\mathsf{P}}^\gamma_{1/2} \right)^m_{ki} \mathbf{N}_i^\gamma[y] \,,
  \label{eqn:dep_tf_coarsening}
\end{alignat}
with repeated indices once again being summed over. $\overline{\mathsf{D}}^\gamma_{1/2}$ is defined in the same manner as $\overline{\mathsf{P}}^\gamma_{1/2}$. This equation essentially applies the deposition transfer matrix at $y - n \Delta y/2$ to all $n$ steps of the propagation of the spectrum $\mathbf{N}^\gamma$ from $y$ to $y - \Delta y$, which itself is approximated by $\overline{\mathsf{P}}^\gamma_{1/2}$. In our code, we call these approximations ``coarsening'', and the number $n$ in both Eqs.~(\ref{eqn:prop_tf_coarsening}) and~(\ref{eqn:dep_tf_coarsening}) the ``coarsening factor''.

\subsubsection{Different Redshift Regimes}
\label{sec:redshift_regimes}

In \dhis we separate our transfer matrices into three redshift regimes: redshifts encompassing reionization ($z  < 50$), redshifts encompassing the times between recombination and reionization ($50 \leq z \leq 1600$), and redshifts well before recombination ($z > 1600$).  During the redshifts encompassing reionization, we allow our transfer functions to be functions of $x_\text{HII}$ and $x_\text{HeII}$, enabling the use of reionization models that evolve hydrogen and helium ionization levels separately. We only consider singly-ionized helium in the current version of \dhis since we expect $x_\text{HeIII}$ not to play an important role until $z \sim 6$.  We compute the transfer functions on a grid of $z^k$, $x^m_\text{HII}$, and $x^n_\text{HeII}$, and linearly interpolate over the grid of pre-computed transfer functions.

Between recombination and reionization, the helium ionization level lies at or below the hydrogen ionization level, since helium has a larger ionization potential at \SI{24.6}{\eV}. After recombination, current experimental constraints typically forbid a large ionization fraction, i.e.\ we expect $x_\text{HII} \lesssim 0.1$~\cite{Liu:2016cnk}. As such, setting $x_\text{HeII} = 0$ is a good approximation for the photon propagation and deposition functions: since $\mathcal{F}_\text{He} \sim 8\%$, neglecting helium ionization only results in $\lesssim 8\%$ error to $x_e$, and $\lesssim 10\%$ error in the density of neutral helium. We therefore follow the same procedure as before, except we now calculate and interpolate the transfer functions over a grid of $z^k$ and $x_\text{HII}^m$ values while holding the helium ionization level fixed to zero.

Finally, well before recombination, we expect the universe to be close to 100\% ionized and tightly coupled thermally to the CMB. Any extra source of exotic energy injection that is consistent with current experimental constraints will likely have a negligible effect on the ionization and thermal histories. We thus calculate and interpolate our transfer functions over a grid of $z^k$ values while holding the hydrogen and helium ionization levels to the baseline values provided by \texttt{RECFAST}~\cite{Wong:2007ym}.

The actual grid values $z^k$, $x^m_\text{HII}$, and $x^n_\text{HeII}$ in each of these regimes can be found in the code, and have been chosen so that interpolation errors are at the sub-10\% level when $f_c(z)$ is calculated using the same method detailed in Ref.~\cite{Slatyer:2015kla}. Our results for $f_c(z)$ without taking into account backreaction, including some improvements over Ref.~\cite{Slatyer:2015kla}, can be found in Appendix~\ref{app:cross_checks}. 

\subsection{Calculating \texorpdfstring{$f_c(z)$}{f\_c(z)}}
\label{sec:calculating_f}
The low-energy photons $\mathbf{N}^\gamma_\text{low}[E_i, z]$ and low-energy electrons $\mathbf{N}^e_\text{low}[E_i, z]$, defined in Sec~\ref{sec:overview}, 
transfer their energy into ionization and excitation of atoms, heating of the IGM, and free-streaming photons to be added to the CMB continuum.
 In \dhis we keep track of how much energy low energy photons and electrons deposit into channels c $\in $ \{`H$_\text{ion}$', `He$_\text{ion}$', `exc', `heat', `cont'\}, which represent hydrogen ionization, helium ionization, hydrogen excitation, heating of the IGM, and sub-\SI{10.2}{\eV} continuum photons respectively.  
 The energy deposition fractions $f_c(z)$ are then found by normalizing the total energy deposited into channel c within a redshift step by the total energy injected within that step according to Eq~(\ref{eqn:fz}).  We closely follow the method for computing $f_c(z)$ described in Ref.~\cite{Slatyer:2015kla}.

Before calculating $f_c(z)$ for each channel, it is instructive to see how to calculate the total amount of energy deposited per unit time and volume, $\left(dE/dV\, dt\right)^\text{dep}$. The low-energy photon and electron spectra $\mathbf{N}^\gamma_\text{low}[E_i]$ and $\mathbf{N}^e_\text{low}[E_i]$ as defined above contain a number of particles per baryon deposited within each log-redshift bin (the $z$-dependence has been suppressed since all calculations in this section occur at the same redshift step). We can convert between these and spectra containing the number of particles produced per unit volume and unit time using the conversion factor $G(z)$ introduced in Eq.~(\ref{eqn:per_baryon_to_dVdt}). For example, to obtain the total amount of energy deposited at a given redshift per unit time and volume, one simply sums over low-energy particle type and applies the conversion factor,

\begin{alignat}{1}
    \left(\frac{dE}{dVdt}\right)^\text{dep}_\text{low} = \frac{1}{G(z)} \sum_\alpha \sum_i  \,E_i' \, \mathbf{N}_\text{low}^\alpha[E'_i] \,.
\end{alignat}
To calculate the total amount of energy deposited we must also add the amount deposited by high energy electrons and photons, which we computed in Eq.~(\ref{eqn:highengdep_tf}):
\begin{alignat}{1}
	\left(\frac{dE}{dVdt}\right)_\text{high}^{\text{dep}} = \frac{1}{G(z)} \sum_c E_c^\text{high} \,.
\end{alignat}
Then the total deposited energy summed over all channels is given by
\begin{alignat}{1}
    \left(\frac{dE}{dVdt}\right)^\text{dep} = \left(\frac{dE}{dVdt}\right)^\text{dep}_\text{low} + \left(\frac{dE}{dVdt}\right)^\text{dep}_\text{high} \,.
\end{alignat}
With this example in mind, we are now ready to understand how to split the energy deposition into the different channels.

\subsubsection{Photons}
\label{sec:low_energy_phot}

We first compute $f_c(z)$ for low-energy photons, starting with energy deposition into continuum photons.  These are photons with energy below $3 \mathcal{R}/4 = \SI{10.2}{\eV}$ that are unable to effectively transfer their energy to free electrons or atoms, so they just free stream.
The energy of these photons constitutes deposition into the continuum channel, i.e.
\begin{alignat}{1}
    \left(\frac{dE^\gamma}{dV \,dt}\right)^\text{dep}_\text{cont} = \frac{1}{G(z)} \sum_{E_i=0}^{3 \mathcal{R}/4}  \,E_i\, \mathbf{N}^\gamma_\text{low}[E_i] \,.
    \label{eqn:f_cont_phot}
\end{alignat}

To calculate the total amount of energy deposited into hydrogen excitation, we make the approximation that all photons with energies between $3\mathcal{R}/4 = \SI{10.2}{\eV}$ and $\mathcal{R} = \SI{13.6}{\eV}$ deposit their energy instantaneously into hydrogen Lyman-$\alpha$ excitation, following \cite{Slatyer:2015kla}:
\begin{alignat}{1}
    \left(\frac{dE^\gamma}{dVdt}\right)^\text{dep}_\text{exc} = \frac{1}{G(z)} \sum_{E_i=3 \mathcal{R}/4}^{\mathcal{R}} \!\!\!\! E_i \, \mathbf{N}_\text{low}^\gamma[E_i] \,.
    \label{eqn:f_exc_phot}
\end{alignat}
A more complete treatment of excitation would involve keeping track of sub-\SI{13.6}{\eV} energy photons as they redshift into the Lyman-$\alpha$ transition region at \SI{10.2}{\eV}, and should also include two-photon excitation into the $2s$ state.\footnote{Two-photon $1s \to 2s$ transitions are in fact as important as Lyman-$\alpha$ transitions near recombination in determining the ionization history, due to the fact that the Lyman-$\alpha$ line is optically thick at this time.} Finally, helium excitation has been neglected, since the de-excitation of helium atoms, which occurs quickly, produces photons that can eventually photoionize hydrogen. We therefore expect almost no net deposition of energy into helium excitation. Energy injection through helium excitation would mainly affect the process of helium recombination, when the probability of ionization after excitation to a higher state is significant due to the photon bath. However, we do not track this small effect, since the change to $x_e$ would be very small. We leave a more careful treatment of excitation that can correctly take into account all of these effects to future work.

We now move on to ionization. All photons above $\mathcal{R} = \SI{13.6}{\eV}$ that are included in $N^\gamma_\text{low}$ have photoionized one of the atomic species (HI, HeI and HeII). However, after photoionizing a helium atom, the resulting ion may quickly recombine with an ambient free electron, producing an $\mathcal{R}_\text{He} = \SI{24.6}{\eV}$ or $4 \mathcal{R} = \SI{54.4}{\eV}$ photon, which may then go on to photoionize hydrogen instead.\footnote{The photoionization rate on neutral hydrogen is much faster than the Hubble rate for $x_\text{HII} \lesssim 0.9999$ for $z > 3$.}

We can handle low-energy photons with energy $E_\gamma$ that photoionize neutral helium in one of the following three ways:

\begin{enumerate}
  \item if helium is completely ignored, the photon is assumed to photoionize hydrogen, producing a low-energy electron with energy $E_\gamma - \mathcal{R}$ from photoionization and depositing $
  \mathcal{R}$ into hydrogen ionization. This is the approach used in previous calculations of $f_c(z)$~\cite{Slatyer:2015kla}, but leaves us unable to self-consistently track $x_\text{HeII}$ if desired;

  \item the photon produces a low-energy electron with energy $E_\gamma - \mathcal{R}_\text{He}$ from photoionization, depositing $\mathcal{R}$ into hydrogen ionization from the recombination photon (with energy $\mathcal{R}_\text{He}$) and producing an electron with energy $\mathcal{R}_\text{He} - \mathcal{R}$, which ultimately deposits energy into hydrogen excitation, heating and sub-\SI{10.2}{\eV} photons. This approach was previously discussed in Ref.~\cite{Galli:2013dna}, and found to result in very little difference when compared to method (1); or
  \item the photon produces a low-energy electron with energy $E_\gamma - \mathcal{R}_\text{He}$ from photoionization and deposits $\mathcal{R}_\text{He}$ into helium ionization. 
\end{enumerate}

The most accurate accounting of helium ionization lies somewhere between methods (2) and (3); however, either method will likely lead to very similar results in terms of $x_e$ and $T_m$, since the bulk of the energy is deposited by the electron from the initial photoionization for photon energies $E_\gamma \gg \mathcal{R}_\text{He}$, and the remaining energy always leads to one ionization event overall. \texttt{DarkHistory} offers the choice of these three options for implementing helium ionization.

We have checked that all three methods lead to similar ionization and temperature histories for DM models over a large range of masses decaying to both $e^+e^-$ and $\gamma \gamma$; these checks are shown in Appendix~\ref{app:cross_checks}. We recommend simply using method~(1) with helium turned off if the user is interested in ionization and temperature histories well before reionization, and using both method~(2) and~(3) with helium turned on to bracket the uncertainties associated with energy deposition on helium if the user is interested in the epoch of reionization. 

To summarize, the amount of deposited energy into hydrogen per unit time and volume is given by
\begin{alignat}{1}
    \left(\frac{dE^\gamma}{dVdt}\right)^\text{dep}_{\text{H}_\text{ion}} =  \frac{\mathcal{R} }{G(z)} \sum_{E_i > \mathcal{R}} \!\! q_\text{H}^\gamma [E_i] \mathbf{N}^\gamma_\text{low}[E_i] \,,
    \label{eqn:H_ion_dep_phot}
\end{alignat}
and into helium ionization by:
\begin{alignat}{1}
  \left(\frac{dE^\gamma}{dV \, dt}\right)^\text{dep}_{\text{He}_\text{ion}} = \frac{\mathcal{R}_\text{He}}{G(z)} \sum_{E_i > \mathcal{R}_\text{He}} \!\! q_\text{He}^\gamma [E_i] \mathbf{N}^\gamma_\text{low}[E_i] \,,
  \label{eqn:He_ion_dep_phot}
\end{alignat}
producing a low-energy electron spectrum after photoionization of
\begin{alignat}{2}
  \mathbf{N}^e_\text{ion}[E_i] &=&& \,\, q_\text{H}^e(E_i + \mathcal{R}) \mathbf{N}^\gamma_\text{low}[E_i + \mathcal{R}]  \nonumber \\
  & &&+ q_\text{He,a}^e(E_i + \mathcal{R}_\text{He}) \mathbf{N}^\gamma_\text{low}[E_i + \mathcal{R}_\text{He}] \nonumber \\
  & &&+ \delta[E_i - \mathcal{R}_\text{He} + \mathcal{R}] \sum_j  q^e_\text{He,b}(E_j) \mathbf{N}_\text{low}^\gamma[E_j] \,,
  \label{eqn:ionized_elec}
\end{alignat}
where $\delta[E_i - \mathcal{R}_\text{He} + \mathcal{R}]$ is one when the bin boundaries span the energy $\mathcal{R}_\text{He} - \mathcal{R}$ and is zero otherwise, and 
\begin{alignat}{1}
  q(E_i) \equiv \begin{cases}
  \frac{n_\text{HI} \sigma_\text{HI}(E_i)}{n_\text{HI} \sigma_\text{HI}(E_i) + n_\text{HeI} \sigma_\text{HeI} (E_i)}, & E_i > \mathcal{R}, \\
  0, & \text{otherwise},
  \end{cases}
  \label{eqn:p_def}
\end{alignat}
with the $\sigma$'s denoting the photoionization cross section of the appropriate species. $\mathbf{N}_\text{ion}^e$ is added to the low-energy electron spectrum, $\mathbf{N}^e_\text{low}$, which is then treated in the next section. The values of the $q$-coefficients depend on the method, and are shown in Table~\ref{tab:q_values}.

\setlength{\tabcolsep}{10pt}
\renewcommand{\arraystretch}{1.3}

\begin{table}
\begin{tabular}{r c c c c c}

\toprule
\hline
Method & $q_\text{H}^\gamma$ & $q_\text{H}^e$ & $q_\text{He}^\gamma$ & $q^e_\text{He,a}$ & $q^e_\text{He,b}$\\
\hline
  1 & 1 & 1 & 0 & 0 & 0 \\
  2 & 1 & $q$ & 0 & $1-q$ & $1-q$ \\
  3 & $q$ & $q$ & $1-q$ & $1-q$ & 0 \\
\botrule
\end{tabular}
\caption{List of $q$-coefficients for use in Eqs.~(\ref{eqn:H_ion_dep_phot})--(\ref{eqn:ionized_elec}). The variable $q$ is defined in Eq.~(\ref{eqn:p_def}).}
\label{tab:q_values}
\end{table}

\subsubsection{Electrons}
\label{sec:low_energy_elec}

To compute how low-energy electrons deposit their energy into the different channels, we use the results obtained by the MEDEA code \cite{Evoli:2012zz,Valdes:2009cq}, following a similar treatment to Ref. \cite{Galli:2013dna}. Although \texttt{DarkHistory} also includes a calculation of electron energy deposition, which we discussed in Sec.~\ref{sec:electron_cooling}, the MEDEA results are more accurate in the sub-\SI{3}{\kilo\eV} electron energy range, including a more detailed accounting of all possible atomic processes (such as $2s\to1s$ deexcitations) and with more up-to-date cross sections. However, at mildly nonrelativistic to mildly relativistic regimes, our calculation of ICS is more accurate, as argued in Sec.~\ref{sec:ICS}. Furthermore, the MEDEA results assume that hydrogen and helium are at similar ionization levels, which is not always a good assumption. In future versions of \texttt{DarkHistory}, an improved treatment of electrons may be a useful addition to the code.

The MEDEA code uses a Monte Carlo method to track high-energy electrons as they are injected into the IGM, and determines the fraction of the initial electron energy deposited into ionization, Lyman-$\alpha$ excitation, heating of the gas and sub-\SI{10.2}{\eV} photons. We use a table of these energy deposition fractions $p_c(E_i,x_{e,j})$~\cite{Galli:2013dna}, where $c \in $ \{`H$_\text{ion}$', `He$_\text{ion}$', `exc', `heat', `cont'\} as before, $x_{e,j}$ ranges between 0 and 1, and $E_i$ ranges between \SI{14}{\eV} and \SI{3}{\kilo\eV}, and perform an interpolation over these values. The energy deposition from electrons is then simply given by
\begin{alignat}{1}
    \left(\frac{dE^e}{dVdt}\right)^\text{dep}_\text{c} = \frac{1}{G(z)} \sum_i p_c(E_i,x_e)  E_i \, \mathbf{N}^e_\text{low}[E_i] \,,
    \label{eqn:f_elec}
\end{alignat}
keeping in mind that $\mathbf{N}^e_\text{ion}$ has already been added to $\mathbf{N}^e_\text{low}$. Between energies of \SI{10.2}{\eV} and \SI{13.6}{eV}, where collisional excitations of hydrogen are possible but not ionization, we use the result at \SI{14}{eV}, but setting the component into hydrogen ionization to zero and normalizing to unit probability. Below \SI{10.2}{eV}, electrons can only deposit energy through Coulomb heating. 

\subsubsection{High-Energy Deposition}
\label{sec:high_eng_dep}

Finally, the high-energy deposition component of the total energy deposited is given by:
\begin{alignat}{1}
  \left(\frac{dE^\text{high}}{dV \, dt}\right)^\text{dep}_c = \frac{1}{G(z)} E^\text{high}_c \,,
\end{alignat}
where $c \in $ \{ `ion', `exc', `heat' \}. Here, we add the high-energy excitation and ionization component to Lyman-$\alpha$ excitation and hydrogen ionization for simplicity, even though the high-energy deposition is computed for all atomic species. A more accurate computation of this together with a more consistent treatment of helium ionization will be a potential improvement in a future version of \texttt{DarkHistory}. 

\bigskip

With the rate of energy deposition through both low-energy photons and low-energy electrons computed, the total energy deposition rate is then straightforwardly given by
\begin{alignat}{1}
  \left(\frac{dE}{dV \, dt}\right)^\text{dep}_c = \sum_\alpha \left(\frac{dE^\alpha}{dV \, dt}\right)^\text{dep}_c \,,
\end{alignat}
where $\alpha \in \{\gamma, e, \text{high}\}$.

\subsection{TLA Integration and Reionization}
\label{sec:TLA_and_reionization}

\texttt{DarkHistory} offers several options for which set of assumptions should be used when integrating the ionization and thermal histories. In the simplest case, the user may integrate Eq.~(\ref{eqn:TLA_DarkHistory}) at each redshift step based on the $f_c(z,\mathbf{x})$ calculated above, with the reionization terms switched off. As we have discussed, including this backreaction is already a significantly better treatment compared to calculations which assume a standard recombination history, i.e.\ using $f_c(z,\mathbf{x}_\text{std}(z))$ (although backreaction can also be switched off within \texttt{DarkHistory}). 

The next significant improvement that is implemented within \texttt{DarkHistory} is the tracking of the neutral helium ionization fraction. Well before reionization, neglecting helium is a good approximation, since the number density of helium nuclei is only $\mathcal{F}_\text{He} \simeq 0.08$ of hydrogen, and we should expect only at most an 8\% correction to $x_e$ if we include helium.

However, tracking helium allows us to accomplish a self-consistent modeling of exotic energy injection and the reionization of hydrogen and neutral helium. \texttt{DarkHistory} allows users to input a model of reionization, for the first time extending the validity of these energy injection calculations into a regime where hydrogen is fully ionized and helium is singly ionized. 

\subsubsection{Helium}
\label{sec:helium}

The \texttt{DarkHistory} evolution equation governing helium without any energy injection is identical to the \texttt{RECFAST} model, and is given by~\cite{Wong:2007ym}
\begin{alignat}{2}
  \dot{x}_\text{HeII}^{(0)} &=&& \,\, \mathcal{C}^s_\text{HeI} \big(x_\text{HeII} x_e n_\text{H} \alpha^s_\text{HeI} \nonumber \\
   & && \quad - \beta^s_\text{HeI} (\mathcal{F}_\text{He} - x_\text{HeII}) e^{-E^{s,\text{He}}_{21}/T_\text{CMB}}\big) \nonumber \\
   & &&+ \mathcal{C}^t_\text{HeI} \big( x_\text{HeII} x_e n_\text{H} \alpha^t_\text{HeI} \nonumber \\
   & && \quad - 3 \beta^t_\text{HeI} (\mathcal{F}_\text{He} - x_\text{HeII}) e^{-E^{t,\text{He}}_{21}/T_\text{CMB}} \big) \,.
   \label{eqn:helium_TLA}
\end{alignat}
The singlet and triplet ground states of helium must be treated separately, and terms relevant to the singlet or triplet state are represented with a superscript $s$ or $t$ respectively. Here, $\alpha_\text{HeI}$ and $\beta_\text{HeI}$ are the recombination and photoionization for HeI, $E_{21}^\text{He}$ represents the energy difference between the corresponding $n=1$ and $n=2$ states, and finally $\mathcal{C}_\text{HeI}$ is the analog to the Peebles-C coefficient found in Eq.~(\ref{eqn:TLA}), representing the probability of a helium atom in the $n=2$ state decaying to either the singlet or triplet ground state before photoionization can occur. The reader should refer to Refs.~\cite{Wong:2007ym,Kholupenko:2008gb,Kholupenko:2007qs} for details on the numerical values of the coefficients, as well as how to compute $\mathcal{C}_\text{HeI}$. 

We emphasize that although we have implemented all of the modifications to the standard TLA in Eq.~(\ref{eqn:TLA}), our code should not be used for high-precision cosmology, given that it has not been tested extensively, e.g.\ with different cosmological parameters from the central values used in \texttt{DarkHistory}. We find that our code agrees to within 3\% of the \texttt{RECFAST} $x_e$ values for the cosmological parameters used here, which is sufficient for computing the effects of exotic energy injection at this stage.

In the presence of exotic sources of energy injection, low-energy photons and electrons can also change the helium ionization level. Once again, we express the energy injection source term as
\begin{alignat}{1}
  \dot{x}^\text{inj}_\text{HeII} = \frac{f_\text{He ion}(z,\mathbf{x})}{\mathcal{R}_\text{He} n_\text{H}} \left(\frac{dE}{dV \, dt}\right)^\text{inj} \,,
\end{alignat}
where $\mathcal{R}_\text{He} = \SI{24.6}{\eV}$ is the ionization potential of neutral helium. As we discussed in Sec.~\ref{sec:low_energy_phot}, there are three different methods available to evaluate $f_{\text{He}_\text{ion}}$ which bracket the uncertainties involved in helium ionization.

To summarize, the user may opt to track the change in helium ionization levels. This means that in addition to Eq.~(\ref{eqn:TLA_DarkHistory}), we also include
\begin{alignat}{1}
  \dot{x}_\text{HeII} = \dot{x}_\text{HeII}^{(0)} + \dot{x}_\text{HeII}^\text{inj} + \dot{x}_\text{HeII}^\text{re} \,,
  \label{eqn:TLA_helium_darkhistory}
\end{alignat}
where $\dot{x}_\text{HeII}^\text{re}$ is the contribution from processes that are active during reionization.

\subsubsection{Reionization}

The evolution equations shown in Eqs.~(\ref{eqn:TLA_DarkHistory}) and~(\ref{eqn:TLA_helium_darkhistory}) can be integrated with all reionization terms switched off if the user is primarily interested in temperatures or ionization levels well before reionization starts at $z \sim 20$. In this regime, turning off helium is also a reasonable approximation. 

With reionization however, the helium ionization level should be solved as well for complete consistency. We solve the TLA differential equations shown in Eqs.~(\ref{eqn:TLA_DarkHistory}) and~(\ref{eqn:TLA_helium_darkhistory}) in two separate redshift regimes. Prior to some user-defined reionization redshift $1 + z_\text{re}$ ($z_\text{re} \leq 50$), we set $\dot{T}_m^\text{re}$, $\dot{x}_\text{HII}^\text{re}$ and $\dot{x}_\text{HeII}^\text{re}$ to zero. Once reionization begins, we set $\dot{x}_\text{HII}^{(0)}$ and $\dot{x}_\text{HeII}^{(0)}$ to zero for $z < z_\text{re}$ instead, switching over to the specified reionization model with its own photoionization and recombination rates.\footnote{We do not set $\dot{T}_m^{(0)} = 0$, since both adiabatic cooling and Compton scattering off the CMB remain active during reionization.} We also begin tracking doubly-ionized helium $x_\text{HeIII}$, which is always assumed to be zero before reionization. 

The $\dot{T}_m^\text{re}$, $\dot{x}_\text{HII}^\text{re}$ and $\dot{x}_\text{HeII}^\text{re}$ terms depend on the details of how reionization proceeds, which is still relatively uncertain. However, choosing a model for the formation of stars and active galactic nuclei (AGNs) and the associated photoionization and photoheating rates, these terms can be evaluated. \texttt{DarkHistory} by default includes the Puchwein+ model of Ref.~\cite{Puchwein:2018arm}. We also demonstrate how to implement the older Madau and Haardt model~\cite{Haardt:2011xv} in Example 8. Both models provide a photoionization rate $\Gamma_{\gamma X}^\text{ion}(z)$ and a photoheating rate $\mathcal{H}_{\gamma X}^\text{ion}(z)$ as a function of redshift and species $X$. 

Along with these energy injection rates, we must also include other relevant processes that alter the ionization fraction of each species. Since these processes generally convert kinetic energy to atomic binding energy, cooling or heating of the gas due to these processes must also be included in $\dot{T}_m^\text{re}$. The processes we include are:
\begin{enumerate}

    \item collisional ionization, occuring at a rate $\Gamma_{eX}^\text{ion}$ for each species $X$, and an associated cooling rate $-\mathcal{H}_{eX}^\text{ion}$;  

    \item case-A recombination, described by a rate coefficient $\alpha_{A,X}$ for each species $X$, and an associated cooling rate $-\mathcal{H}_X^\text{rec}$;

    \item collisional excitation cooling, with a rate $-\mathcal{H}_{eX}^\text{exc}$; and

    \item bremsstrahlung cooling, with a rate $-\mathcal{H}^\text{br}$. 

\end{enumerate}

The cooling rates here have been defined with a negative sign so that all quantities denoted by $\mathcal{H}$ contribute positively to any temperature change. Expressions for all of these rates can be found in Ref.~\cite{Bolton:2006pc}. They are explicitly dependent on the ionization fraction of all three of the relevant species, namely $x_\text{HI}$, $x_\text{HeI}$ and $x_\text{HeII}$.  The full expressions for the evolution of each of these fractions are as follows:
\begin{alignat}{2}
\label{eqn:tla_reion_xe}
    \dot{x}_\text{HII} &=&&\, x_\text{HI} \left(\Gamma_{\gamma \text{HI}}^\text{ion} + n_e \Gamma_{e \text{HI}}^\text{ion} \right) - x_\text{HII} n_e \alpha_{A, \text{HI}} \,, \nonumber \\
    \dot{x}_\text{HeII} &=&&\, x_\text{HeI} \left(\Gamma_{\gamma \text{HeI}}^\text{ion} + n_e \Gamma_{e \text{HeI}}^\text{ion}\right) + x_\text{HeIII} n_e \alpha_{A,\text{HeIII}} \nonumber \\
    & && - x_\text{HeII} \left(\Gamma_{\gamma \text{HeII}}^\text{ion} + n_e \Gamma_{e \text{HeII}}^\text{ion} + n_e \alpha_{A,\text{HeII}}\right) \,, \nonumber \\
    \dot{x}_\text{HeIII} &=&&\, x_\text{HeII} \left(\Gamma_{\gamma \text{HeII}}^\text{ion} + n_e \Gamma_{e \text{HeII}}^\text{ion} - x_\text{HeIII} n_e \alpha_{A,\text{HeIII}} \right) \,,
\end{alignat}
with the temperature evolution given by
\begin{alignat}{2}
\label{eqn:tla_reion_T}
    \dot{T}_m^\text{re} &=&& \, \frac{2}{3(1 + \mathcal{F}_\text{He} + x_e) n_\text{H}} \nonumber \\
    & && \times \sum_X \left(\mathcal{H}_{eX}^\text{ion} + \mathcal{H}_X^\text{rec} + \mathcal{H}_{eX}^\text{exc} + \mathcal{H}^\text{br} \right) \, .
\end{alignat}

Instead of specifying a full reionization model, the user may also choose the simpler alternative of fixing the value of $x_\text{HII}$ and $x_\text{HeII}$ as a function of redshift once reionization begins, and integrate the temperature evolution alone instead. We note that this approach is not self-consistent, since fixing the ionization levels forces us to neglect any additional contribution to ionization from exotic energy injection sources. However, if the contribution to ionization is known to be small, this can serve as a useful approximation.

\subsubsection{Numerical Integration}

To ensure that ionization fractions always remain appropriately bounded during integration, we introduce the variable
\begin{alignat}{1}
  \zeta_i \equiv \text{arctanh} \left[ \frac{2}{\chi_i} \left(\frac{n_i}{n_\text{H}} - \frac{\chi_i}{2}\right) \right] \,,
  \label{eqn:yi_variable}
\end{alignat}
where $\chi_i = 1$ for HI and $\chi_i = \mathcal{F}_\text{He}$ for HeI and HeII. This transformed equation is then integrated using the standard \texttt{odeint} integrator provided by SciPy.

At early times, the equations we are integrating are very stiff, and solving them directly with numerical integration can often run into difficulties. We therefore assume that when $x_\text{HII} > 0.99$ or $x_\text{HeII} > 0.99 \mathcal{F}_\text{He}$, either variable follows their Saha equilibrium values. 

In Sec.~\ref{sec:example_reionization}, we will show several thermal and ionization histories that showcase \texttt{DarkHistory}'s capabilities in tracking the helium ionization level, exotic energy injection and reionization all at the same time.

\section{Modules}
\label{sec:modules}
In this section we summarize the main modules in \texttt{DarkHistory}. We will pay particular attention to the modules shown in the flow chart in Fig.~\ref{fig:flowchart}, 
and as far as possible provide links between the code and the text.
Keep in mind that this is not a complete list and that it is subject to change in future versions of the code. There is more thorough documentation in \texttt{DarkHistory} itself that will be periodically updated at \href{https://darkhistory.readthedocs.io/en/development/}{https://darkhistory.readthedocs.io}, and will contain a more complete explanation of the code. In the interest of space, we only provide the full path of each module in the code when it is mentioned for the first time.

\subsection{Data}
\label{sec:data}

First, the user must download the data files found at \href{https://doi.org/10.7910/DVN/DUOUWA}{https://doi.org/10.7910/DVN/DUOUWA}. These files contain the photon propagation transfer function $\overline{\mathsf{P}}_\gamma$ and deposition transfer functions $\overline{\mathsf{D}}_\gamma$, $\overline{\mathsf{D}}_e$ and $\overline{\mathbf{D}}_c^\text{high}$, which have all been precomputed as discussed above. They also contain transfer functions for ICS calculations discussed in Appendix~\ref{app:ICS}, structure formation annihilation boost factors computed in Ref.~\cite{Liu:2016cnk}, the baseline thermal and ionization histories, data from \textsc{pppc4dmid}~\cite{Cirelli:2010xx} and $f_c(z)$ computed without backreaction for DM annihilation and decay, where photons and $e^+e^-$ are injected at a fixed set of energies. 

\subsection{\texttt{config}}
\label{sec:module_config}

The \texttt{config} module contains the code required to access the downloaded data, and to store them in memory for use. Users should ensure that the variable \texttt{data\_path} points to the directory containing the data files.

\subsection{\texttt{main}}
\label{sec:module_main}

The \texttt{main} module contains the function that implements the loop shown in Fig.~\ref{fig:flowchart}, \texttt{evolve()}. The usage of this function will be discussed in great detail in Sec.~\ref{sec:examples}. 

\subsection{\texttt{darkhistory.physics}}
\label{sec:module_physics}

This module contains physical constants and useful functions found in cosmology, particle physics and atomic physics. We use units of \SI{}{cm} for length, \SI{}{s} for time and \SI{}{\eV} for energy, mass and temperature. Some examples of functions that are included in this module include the Hubble parameter as a function of redshift, \texttt{physics.hubble()}, and the Peebles-C factor $\mathcal{C}$ found in Eq.~(\ref{eqn:TLA}), \texttt{physics.peebles\_C()}. Constants provided in this module are taken from central values of the Planck 2018 TT,TE,EE+lowE results~\cite{Aghanim:2018eyx} and the Particle Data Group review of particle physics~\cite{Tanabashi:2018oca}.

\subsection{\texttt{darkhistory.electrons}}
\label{sec:module_electrons}

The \texttt{electrons} module contains all of the functions necessary to perform the electron cooling calculation. The \texttt{positronium} submodule contains functions that return the spectrum of photons obtained during positronium annihilation, which we denoted as $\overline{\mathbf{N}}_\gamma^\text{pos}$ in Eq.~(\ref{eqn:positronium_photons}); Example 7 demonstrates how to use this module. The \texttt{ics} submodule contains all of the machinery necessary to compute the ICS scattered photon and electron spectra; for more details on how to use this submodule, refer to Example 4 in the code. 

\texttt{elec\_cooling} contains the code necessary to compute the transfer functions $\overline{\mathbf{R}}_c$, $\overline{\mathsf{T}}_\text{ICS}$ and $\overline{\mathsf{T}}_e$, as defined in Eqs.~(\ref{eqn:elec_cooling_dep_tf}),~(\ref{eqn:elec_cooling_ics}) and~(\ref{eqn:elec_cooling_lowengelec}) respectively; Example 6 shows how this module is used.

\subsection{\texttt{darkhistory.history}}
\label{sec:module_history}

This module contains our implementation of the TLA and reionization.
The submodule \texttt{tla} corresponds to Sec.~\ref{sec:histories} 
where the function \texttt{get\_history} implements the TLA, including all of the terms discussed in Eqs.~(\ref{eqn:TLA_DarkHistory}) and Eqs.~(\ref{eqn:TLA_helium_darkhistory})--(\ref{eqn:tla_reion_T}). 
The submodule \texttt{reionization} contains the Puchwein+ reionization model, and contains all of the coefficients found in Eqs.~(\ref{eqn:tla_reion_xe}) and~(\ref{eqn:tla_reion_T}). 

\subsection{\texttt{darkhistory.low\_energy}}
\label{sec:module_low_energy}

This module calculates $f_c(z)$. 
The \texttt{lowE\_photons} and \texttt{lowE\_electrons} submodules correspond 
to Sec.~\ref{sec:low_energy_phot} and Sec.~\ref{sec:low_energy_elec}, respectively, implementing Eqs.~(\ref{eqn:f_cont_phot})--(\ref{eqn:ionized_elec}) and Eq.~(\ref{eqn:f_elec}) respectively. 
The \texttt{lowE\_deposition} submodule then combines the energy deposited by photons, electrons (including photoionized electrons) and high-energy deposition to make $f_c(z, \mathbf{x})$.

\subsection{\texttt{darkhistory.spec}}
\label{sec:module_spec}

This module contains functions for handling and generating spectra and transfer functions. All one dimensional spectra in the code can be handled using the class \texttt{Spectrum}, which stores not just the data of the spectrum, but also the abscissa, and other relevant information like redshift or the injection energy of the particle that produced the spectrum. This class includes many convenience functions, such as the ability to rebin the spectrum into a new binning while conserving total number and energy, or the ability to quickly obtain the total number of particles within some energy range. Example 1 in our code gives a quick introduction to this class.

The user may also want to store closely related spectra in one object. This may be desirable for spectra of the same particle type over different redshifts, or if they correspond to spectra from the same injected particle but at different injection energies. The class \texttt{Spectra} has been written to do exactly this. Example 2 provides a good overview of what this class can do. 

\subsection{\texttt{darkhistory.spec.pppc}}
\label{sec:module_pppc}

Within the \texttt{spec} module, a dedicated submodule \texttt{pppc} has been written to calculate the electron and photon spectra from the injection of any arbitrary Standard Model particle, based on the \textsc{pppc4dmid} results. The function \texttt{pppc.get\_pppc\_spec()} is the main function to use for this end. See Example 4 for more information on how to use this function.

\section{Using the Code}
\label{sec:examples}

We will now apply \dhis to perform a variety of calculations in order to highlight the key functionalities of the code. Each of the subsections corresponds to an example Jupyter notebook that has been provided as part of the code; the user should refer to these examples for a deeper look at the full capability of the code, as well as to the online documentation. In this paper, we will simply highlight capabilities and interesting physics results. 

Within the code and in this section, the word ``redshift'' and variables that represent redshift (usually called \texttt{rs} in the code) refer to the quantity $1+z$, since this is the physically relevant quantity in many cosmological calculations.

\subsection{A Simple Model: \texorpdfstring{$\chi \chi \to b\bar{b}$}{DM DM -> b bbar}}

As a first example, we will demonstrate how to compute the ionization and thermal history of a simple annihilation model. Consider a \SI{50}{\giga\eV} Majorana fermion DM particle that undergoes $s$-wave annihilation to a pair of $b \overline{b}$ quarks, with an annihilation cross section $\langle \sigma v \rangle = \SI{2e-26}{\centi\meter\cubed\per\second}$, close to the required thermal freezeout cross section for the correct relic abundance. Similar models have been considered as a possible dark matter explanation for the galactic center excess~\cite{Calore:2014nla} and the AMS-02 antiproton excess~\cite{Cui:2016ppb,Cuoco:2016eej}. We perform the calculation in a relatively simplified setting: with no reionization, no backreaction included, but with a boost to the annihilation rate from structure formation.  For more details, see Example 9 in the code.

The function that we use to compute histories is \texttt{main.evolve()}. There are many keyword parameters that can be used with this function, and the user should refer to the example notebooks and the online documentation for more information. To find the thermal history for this model, \texttt{evolve()} can be called in the following fashion:

\begin{lstlisting}{language=python}
    import main
    import darkhistory.physics as phys

    bbbar_noBR = main.evolve(
        DM_process='swave', mDM=50e9, 
        sigmav=2e-26, 
        primary='b', start_rs=3000.,
        coarsen_factor=32, backreaction=False,
        struct_boost=phys.struct_boost_func()
    )
\end{lstlisting}
\begin{figure*}[t!]
\begin{tabular}{c}
\includegraphics[scale=.52]{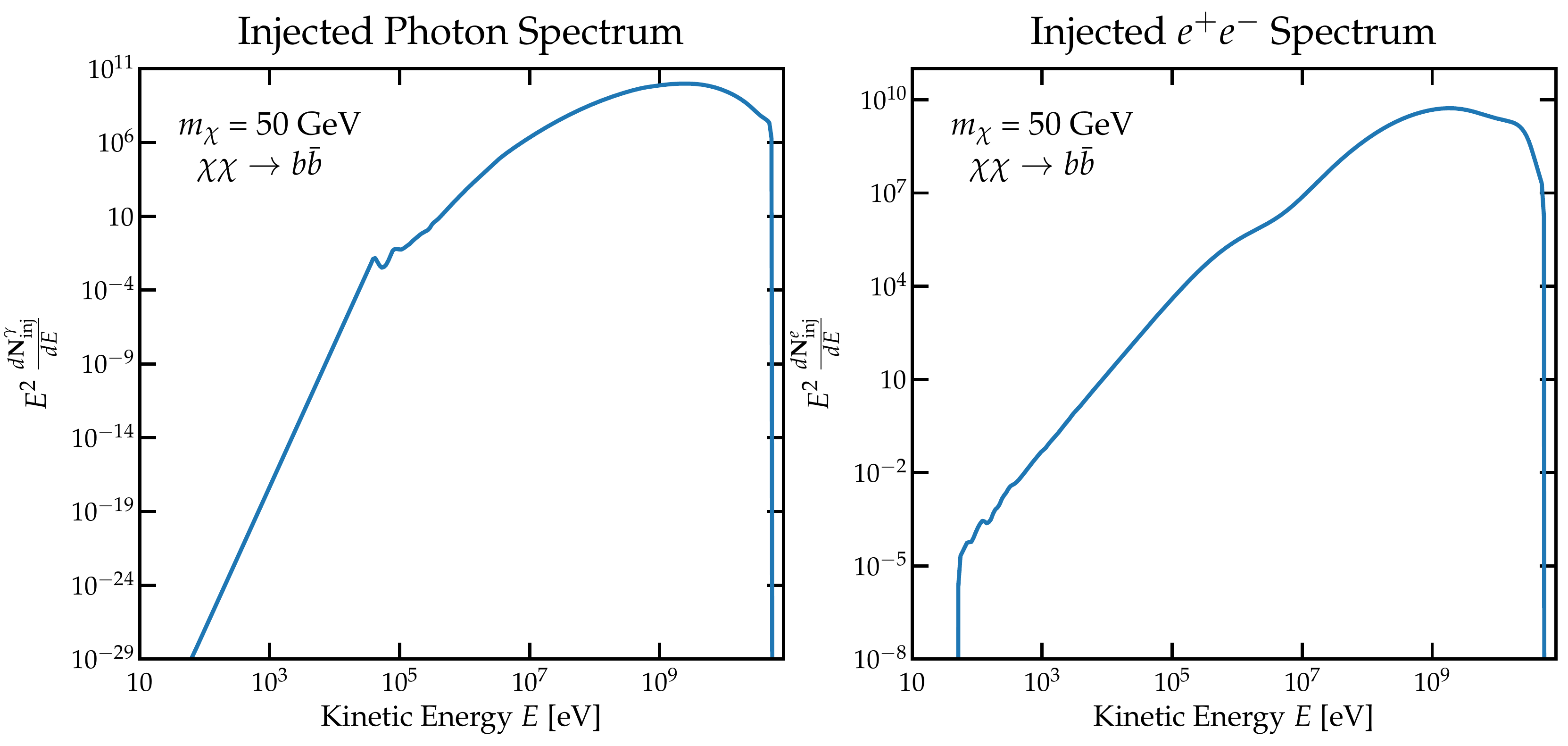}
\end{tabular}
\caption{Photon (left) and $e^+e^-$ (right) spectra produced by a single annihilation event, $\chi \chi \to b \overline{b}$, with $m_\chi = \SI{50}{\giga\eV}$. These spectra are based on the raw data provided by \textsc{pppc4dmid}.}
\label{fig:bbbar_spectra}
\end{figure*}

The keyword parameters are as follows:

\begin{enumerate}
 \item \lstinline|DM_process='swave'| -- specifies the DM process of interest. Currently, \texttt{DarkHistory} can handle $s$-wave annihilating and decaying DM models (\lstinline|DM_process='decay'|) with this keyword; 

 \item \lstinline|mDM=50e9| -- specifies the DM mass, in \SI{}{\eV}; 

 \item \lstinline|sigmav=2e-26| -- specifies the velocity averaged annihilation cross section, in \SI{}{\centi\meter\cubed\per\second}; 

 \item \lstinline|primary='b'| -- specifies the annihilation channel. The options include all of those offered by \textsc{pppc4dmid}~\cite{Cirelli:2010xx}, and the spectra are extracted from the raw data provided by the cookbook. The $e^+e^-$ and photon spectra from the showering of a single $b \overline{b}$ pair are shown in Fig.~\ref{fig:bbbar_spectra}. These are proportional to the injection spectra $\mathbf{N}_\text{inj}^\alpha$ defined in Sec.~\ref{sec:input}, and can be generated using the function \texttt{pppc.get\_pppc\_spec()}; 

 \item \lstinline|start_rs=3000| -- the redshift at which to start the evaluation. $1+z = 3000$ is the highest redshift at which we have produced the photon cooling transfer functions, and represents the highest redshift that should be specified here. In this example, \texttt{start\_rs} fixes the initial conditions of the TLA in Eq.~(\ref{eqn:TLA}) at the baseline ionization and temperature values at this redshift; 

 \item \lstinline|coarsen_factor=32| -- the coarsening factor, defined in Sec.~\ref{sec:coarsening}. For a comparison between solutions with different coarsening factors, see Appendix~\ref{app:cross_checks}; 

 \item \lstinline|backreaction=False| -- this turns backreaction on and off; and

 \item \lstinline|struct_boost=phys.struct_boost_func()| -- the structure formation prescription to use. Once dark matter halos start to collapse, the annihilation rate gets enhanced by the factor 
 \begin{alignat}{1}
   1 + \mathcal{B}(z) \equiv \frac{\langle \rho_\chi^2 \rangle}{\langle \rho_\chi \rangle^2} 
   \label{eqn:boost_factor}
 \end{alignat}
 compared to the smooth annihilation rate shown in Eq.~(\ref{eqn:energy_injection}). Here, \texttt{struct\_boost} is a function that takes redshift as the argument, and returns $1 + \mathcal{B}(z)$. The user can make use of the structure formation boosts that are saved by default in \texttt{DarkHistory} in the \texttt{physics} module, which include the boost factors computed in Ref.~\cite{Liu:2016cnk}, and is used as the default boost factor by \lstinline{struct_boost_func()}. 

\end{enumerate}

By default, the solver integrates the equations down to $1+z = 4$, and will not evolve the helium ionization levels. These choices can of course be changed with other keyword parameters. Note that the function is not limited to DM processes or \textsc{pppc4dmid} spectra; other keyword parameters allow the user to specify their own injection rates as a function of redshift (see the documentation for the keyword parameters \texttt{rate\_func\_N} and \texttt{rate\_func\_eng}), along with the spectra of photons and $e^+e^-$ injected (see the documentation for the keyword parameters \texttt{in\_spec\_elec} and \texttt{in\_spec\_phot}). 

\begin{figure*}[t!]
\begin{tabular}{c}
\includegraphics[scale=.55]{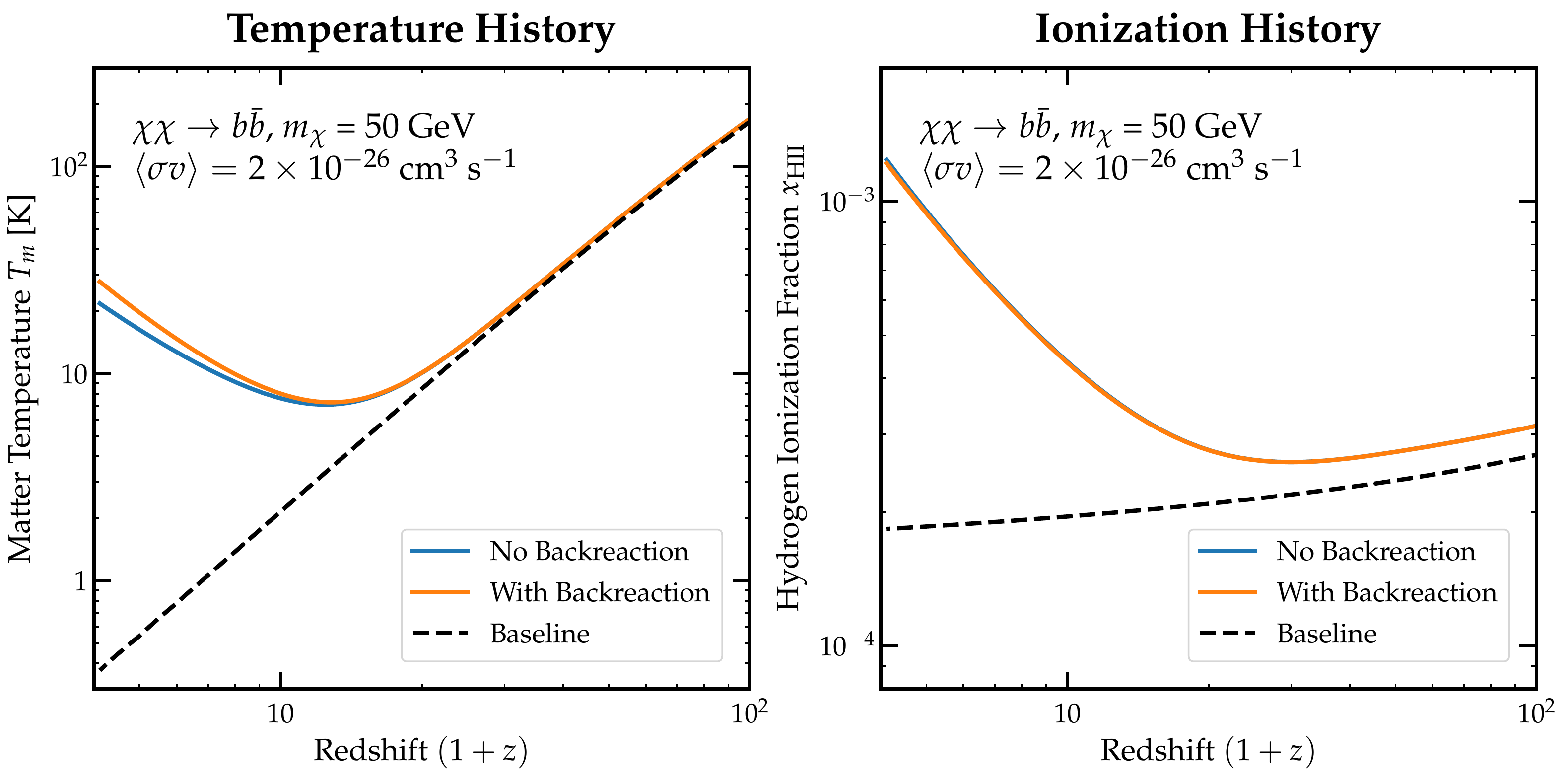}
\end{tabular}
\caption{Matter temperature $T_m$ (left) and hydrogen ionization fraction $x_\text{HII}$ (right) solved in the presence of dark matter annihilation into $b \bar{b}$ pairs using \dhis.  Eq.~\ref{eqn:TLA_DarkHistory} is solved without dark matter energy injection to produce the baseline histories (black, dashed), with energy injection but without backreaction (blue), and with dark matter annihilation and backreaction (orange).  We assume a dark matter mass of $50$ GeV and a velocity averaged annihilation cross section of $2 \times 10^{-26}$ cm$^3$ s$^{-1}$.
}
\label{fig:bbbar}
\end{figure*}

The output of \texttt{evolve()}, stored in \texttt{bbbar\_noBR}, is a dictionary containing the redshift abscissa of the solutions, the ionization and temperature solutions, the propagating photon, low-energy photon and low-energy electron spectra, and the computed value of $f_c(z)$. To access the redshift, ionization and temperature, we can simply do: 
\begin{lstlisting}{language=python}
  # Redshift abscissa.
  rs_vec   = bbbar_noBR['rs']   
  # Matter temperature in eV.
  Tm_vec   = bbbar_noBR['Tm']
  # Ionization fraction. 
  # Stored as 1+z by {xHII, xHeII, xHeIII}.
  xHII_vec = bbbar_noBR['x'][:,0] 
\end{lstlisting}

In Fig.~\ref{fig:bbbar} we plot $T_m$ and $x_\text{HII}$ as a function of redshift for the $\chi\chi \to b\bar{b}$ model. For DM masses above $\gtrsim \SI{10}{\giga\eV}$, values of $\langle \sigma v \rangle$ required for thermal freezeout are unconstrained by the CMB anisotropy power spectrum energy injection constraints: the ionization fraction, which changes by approximately 25\% only at high redshifts, does not change enough to affect the power spectrum significantly. The sudden increase in ionization and temperature at $z \sim 30$ corresponds to an increase in the boost factor used (halos with an Einasto profile with halo substructure boost included~\cite{Liu:2016cnk}, found in \texttt{physics.struct\_boost\_func()}). 

We also show in Fig.~\ref{fig:bbbar} for completeness the effect of turning on backreaction, i.e.\ including the effect of the increased ionization level on the evolution of the ionization and thermal histories. This is conveniently done by setting \texttt{backreaction=True}. In this particular example, the effect of backreaction is small, but we will show more scenarios where backreaction has large effect on $T_m$, and explain why this can be significant in the next example. 

%

\subsection{Backreaction}
\label{sec:backreaction}

\begin{figure*}
 \centering
 \includegraphics[scale=0.55]{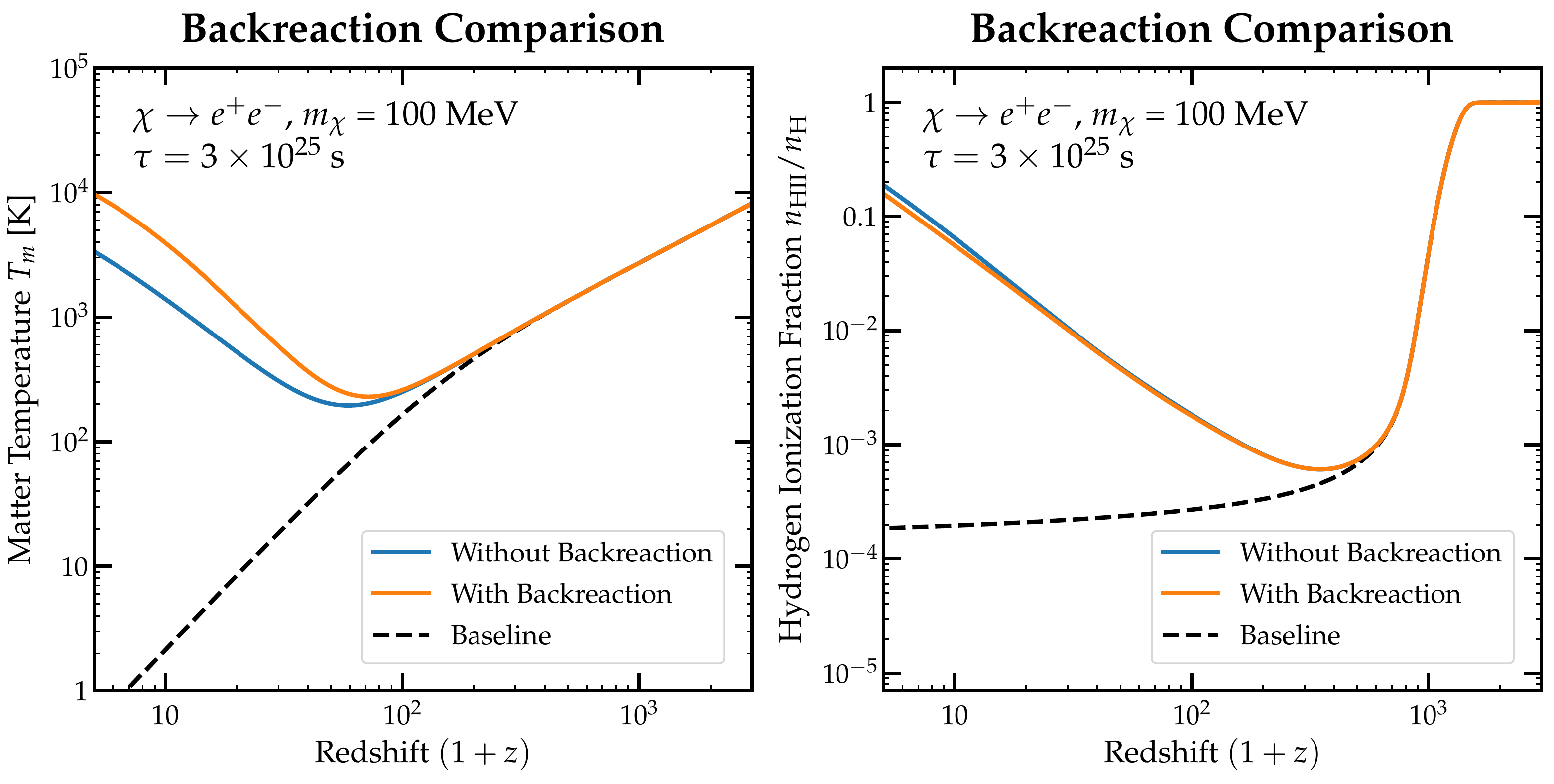}
 \caption{Temperature (left) and ionization (right) histories including the effects of dark matter decay to electrons and positrons.  We choose a lifetime of \SI{3e25}{\second}, which is consistent with the CMB constraints from Ref.~\cite{Slatyer:2016qyl}. We plot the baseline histories (black, dashed), the histories including dark matter energy injection but not backreaction (blue), and the histories including energy injection and backreaction (orange). These plots are a single vertical slice of the contour plots in Fig.~\ref{fig:backreact_mass_scan}. Additionally, these plots constitute a cross-check on \texttt{DarkHistory}, as they agree well with similar results obtained in Ref.~\cite{Liu:2016cnk}.}
 \label{fig:single_decay}
\end{figure*}

Let us explore the effects of backreaction a bit more using some of the code found in Example 10 of \texttt{DarkHistory}. As was described in Sec~\ref{sec:histories}, one of \texttt{DarkHistory}'s main improvements 
to ionization and temperature history calculations is its ability to include the effects of back-reaction. To see its importance, consider the example of \SI{100}{\mega\eV} dark matter decaying to a pair of $e^+e^-$, with a lifetime of $\tau = \SI{3e25}{\second}$, a value that is close to the minimum lifetime allowed by constraints from the CMB power spectrum~\cite{Slatyer:2016qyl}. The ionization and thermal histories can be evaluated in this way: 
\begin{lstlisting}{language=python}
  decay_BR = main.evolve(
    DM_process='decay', mDM=1e8, lifetime=3e25, 
    primary='elec_delta', start_rs=3000.,
    coarsen_factor=16, backreaction=True
  )
\end{lstlisting}
The new keywords here are:
\begin{enumerate}

 \item \lstinline|DM_process='decay'| -- specifies the DM process of interest to be decays; 
 \item \lstinline|lifetime=3e25| -- specifies the decay lifetime in seconds; and
 \item \lstinline|primary='elec_delta'| -- the \texttt{primary} channel options \lstinline|'elec_delta'| and \lstinline|'phot_delta'| can be used to inject an $e^+e^-$ and $\gamma \gamma$ pair respectively, with no electroweak corrections applied. 

\end{enumerate}

To do the calculation without backreaction, we can simply set \lstinline|backreaction=False|. However, with \lstinline|primary='elec_delta'| or \lstinline|'phot_delta'|, \texttt{DarkHistory} can instead rely on tabulated results of $f_c(z)$ for these two channels, using the same method based on results from Ref.~\cite{Slatyer:2015kla}, to calculate the ionization and thermal histories without evolving the input spectrum, leading to a significant speed-up. This can be done using the function \texttt{tla.get\_history()}:

\begin{lstlisting}{language=python}
  import numpy as np
  from darkhistory.tla import get_history
  # get_history takes a redshift vector:
  rs_vec = np.flipud(np.arange(5, 3000, 0.1)) 

  result = get_history(
     rs_vec, baseline_f=True, mDM=1e8, 
     lifetime=3e25, DM_process='decay',
     inj_particle='elec_delta' 
  )
\end{lstlisting}
with the following parameters:

\begin{enumerate}

 \item \lstinline|rs_vec| -- the redshift vector, ordered from high to low, over which the temperature and ionization histories are to be evaluated; 
 \item \lstinline|baseline_f=True| -- this tells the code to use the baseline $f_c(z)$ computed by \texttt{DarkHistory} without backreaction. As we discussed in Sec.~\ref{sec:calculating_f}, these $f_c(z)$ agree with those computed in Ref.~\cite{Slatyer:2015kla} to within 10\%, and
 \item \lstinline|inj_particle='elec_delta'| --  used to specify one of two options \lstinline|'elec_delta'| or \lstinline|'phot_delta'|.

\end{enumerate}

The output \texttt{result} is an array of shape \lstinline|(len(rs_vec), 4)|, with the second dimension indexing $\{T_m, x_\text{HII}, x_\text{HeII}, x_\text{HeIII}\}$. The temperature (in \SI{}{\eV}) can be accessed through \lstinline|T_m = results[-1,0]|. 

Although only the $f_c(z)$ values for the injection for an $e^+e^-$ and $\gamma \gamma$ pair have been saved for use with \texttt{DarkHistory}, the $f_c(z)$ for any arbitrary channel can be computed from a weighted average of the electron and photon results~\cite{Slatyer:2015kla}. We stress once again, however, that this can only be done assuming no backreaction.

The histories are shown in Fig~\ref{fig:single_decay}, with and without backreaction turned on. First, even though the ionization level at $z \sim 10$ is three orders of magnitude larger than the baseline, such a scenario is actually still consistent with the CMB power spectrum constraints, owing to the fact that the ionization build-up occurs relatively late: the CMB constraints are sensitive to changes in $x_e$ near recombination, and become less sensitive at later times. 

Comparing the temperature histories with and without backreaction, we see that the main effect of this increase in $x_e$ on the energy deposition processes is to increase energy deposition into heating. Ionization and excitation rates depend on the neutral fraction, which is still close to 100\% even with energy deposition from DM. However, the energy rate into Coulomb heating is proportional to $x_e$, so taking into account the significantly elevated $x_e$ values leads to higher temperature levels. By about $z \sim 10$, $T_m$ with backreaction is larger than without backreaction by a factor of $\sim 4$, with the difference continuing to grow. Neglecting backreaction therefore leads to a severe underestimate of $T_m$, and including this effect consistently will certainly be important in understanding what measurements of $T_m$ at $z \simeq 20$ through the 21-cm signal or the Lyman-$\alpha$ power spectrum can tell us about exotic sources of energy injection.

\begin{figure*}
\begin{tabular}{cc}
 \includegraphics[scale=0.455]{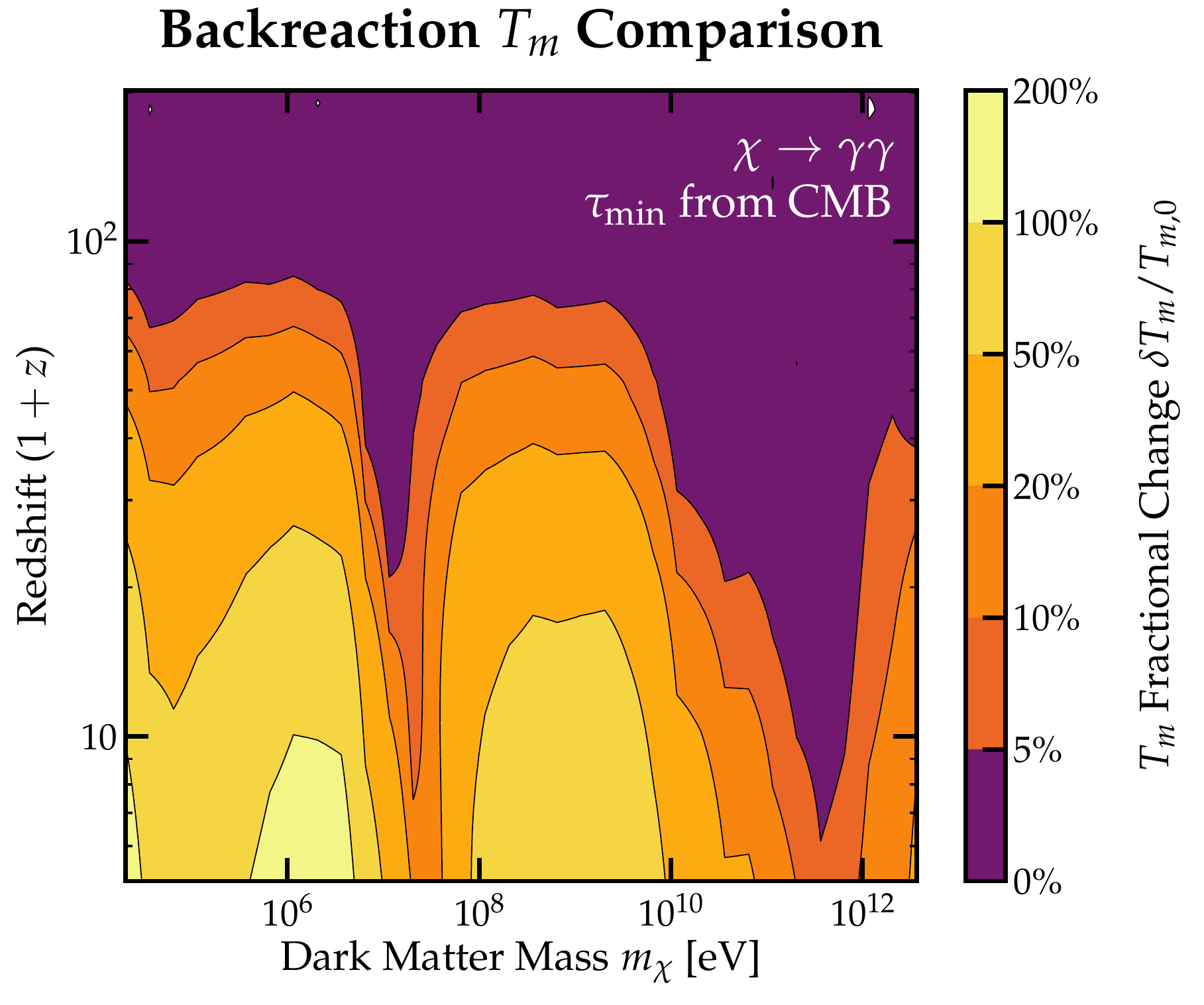} &
 \includegraphics[scale=0.455]{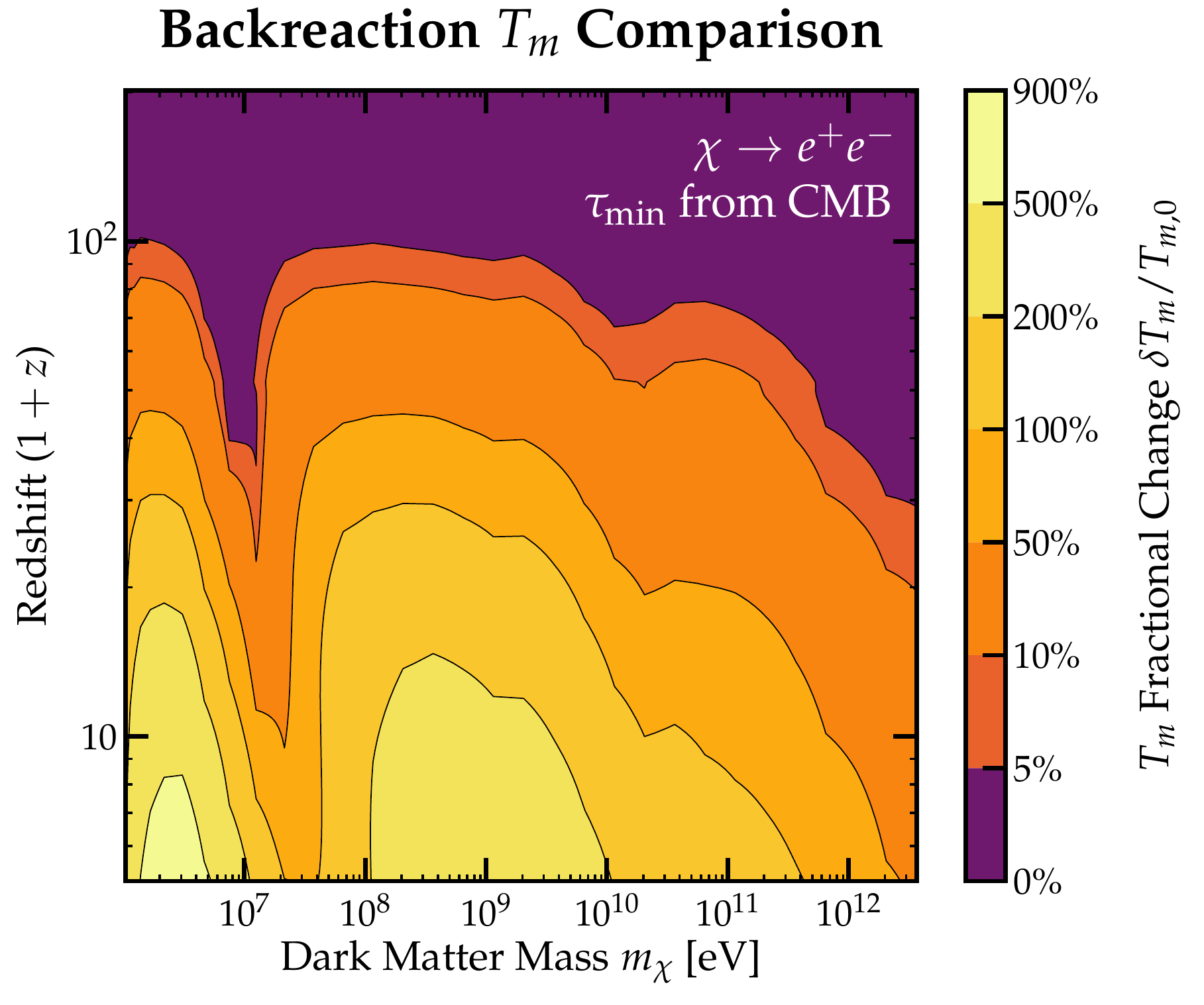} \\
 \includegraphics[scale=0.455]{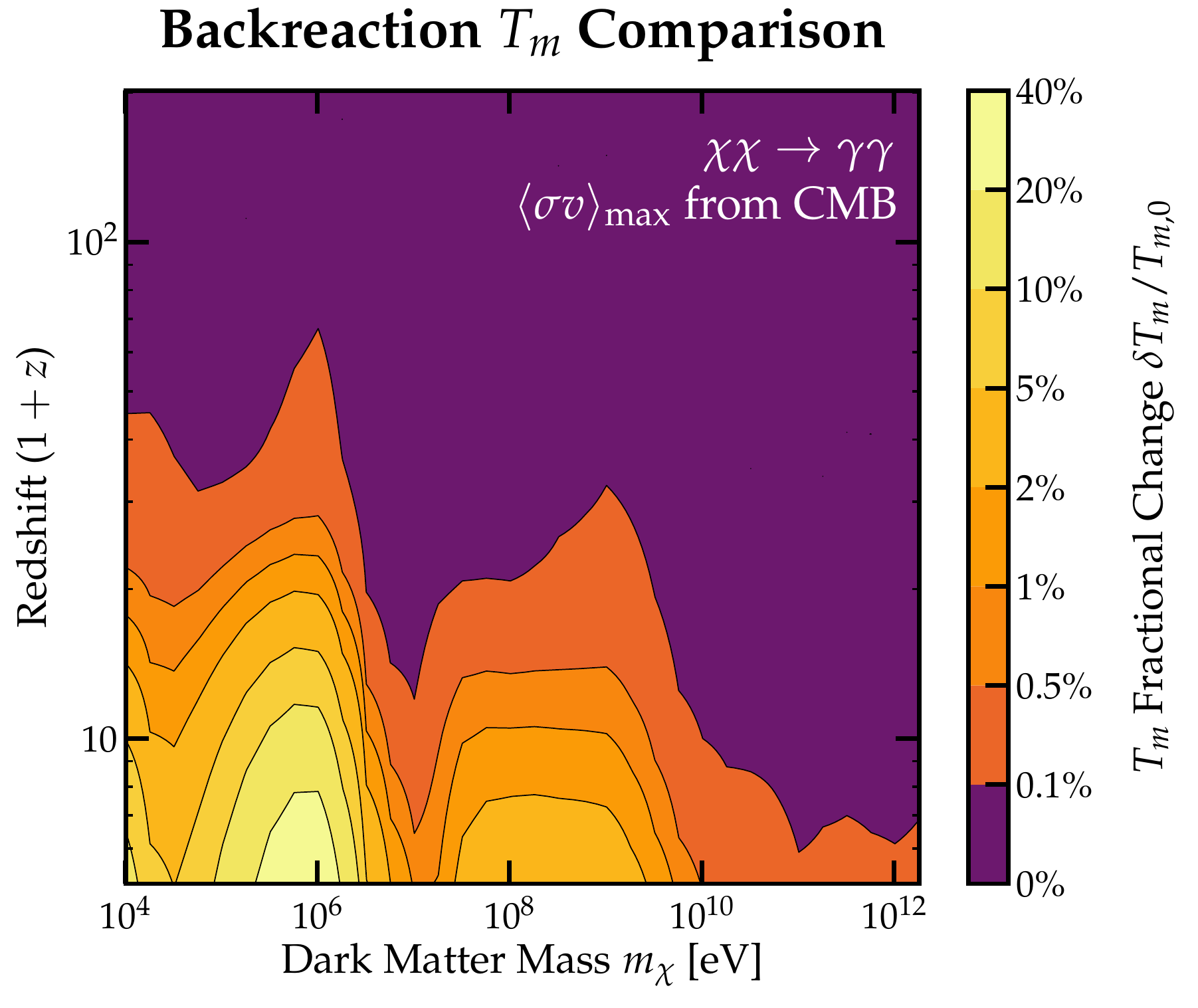} &
 \includegraphics[scale=0.455]{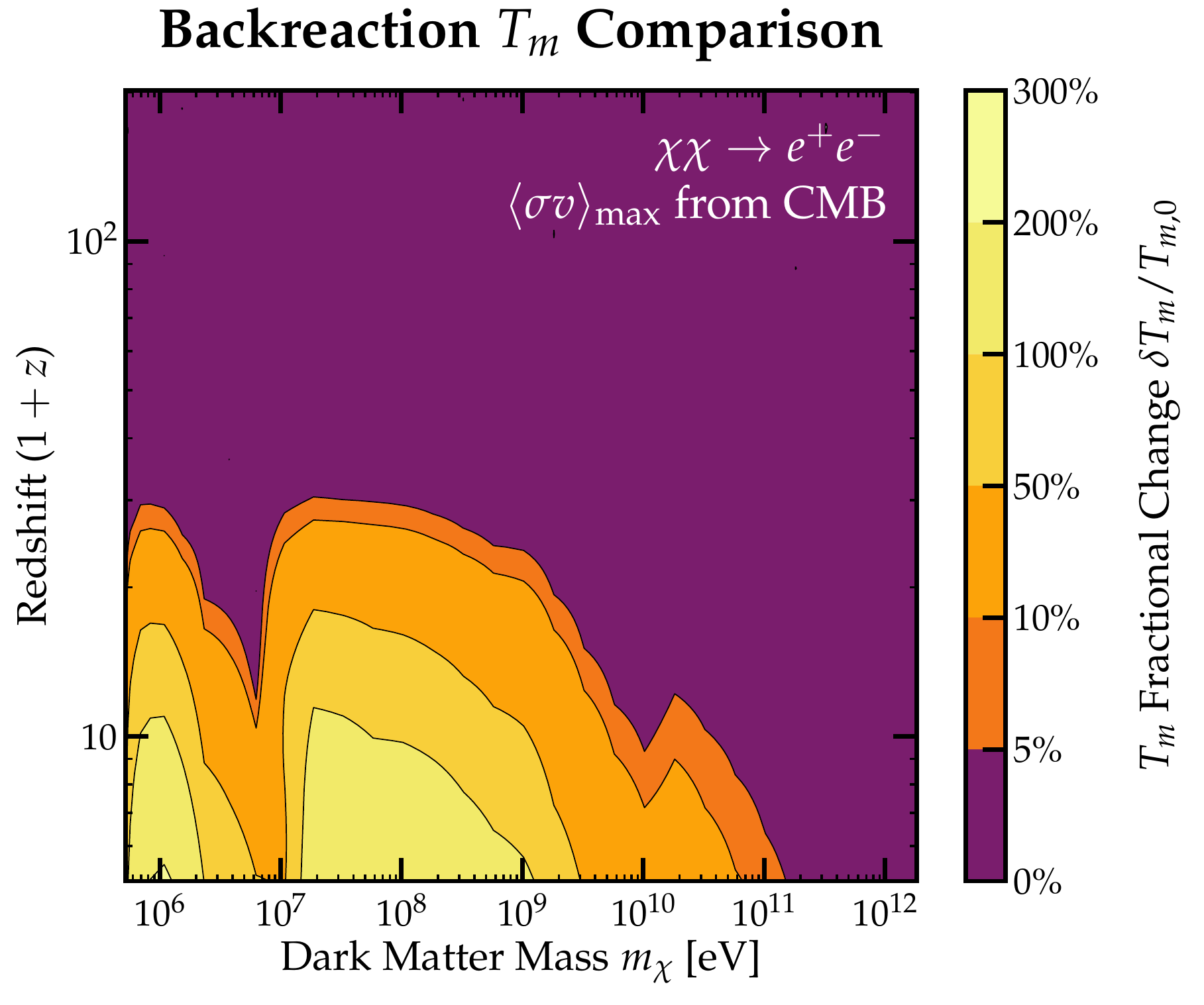} \\
\end{tabular}
\caption{Contour plots of the fractional change in temperature $\delta T_m/T_{m,0}$ caused by including the effects of backreaction, as a function of dark matter mass and redshift (See Eq.~(\ref{eqn:fractional_Tm_change})). For each dark matter mass, we choose the minimum $\tau$ or maximum $\langle \sigma v \rangle$ allowed by current CMB power spectrum constraints~\cite{Slatyer:2016qyl,Slatyer:2015kla}. 
}
\label{fig:backreact_mass_scan}
\end{figure*}

We can perform the calculation over a range of DM masses by looping over values of \texttt{mDM}. For each value of $m_\chi$, we select the minimum lifetime $\tau$ which is consistent with the CMB power spectrum constraints, and compare the difference between the temperature history with backreaction ($T_{m,\text{BR}}$) and without ($T_{m,0}$) by computing the fractional change in temperature,
\begin{alignat}{1}
 \frac{\delta T_m}{T_{m,0}}(m_\chi, z) = \frac{T_{m,\text{BR}}(m_\chi, z) -T_{m,0}(m_\chi, z)}{T_{m,0}(m_\chi, z)} \,.
 \label{eqn:fractional_Tm_change}
\end{alignat}
In Fig~\ref{fig:backreact_mass_scan} we plot this variable over a range of redshifts and dark matter masses for this particular channel ($\chi \to e^+e^-$), but also for decay and annihilation into $e^+e^-$ and $\gamma \gamma$, taking the maximum $\langle \sigma v \rangle$ again allowed by the CMB power spectrum constraints. At a redshift of $z \sim 17$ near the end of the cosmic dark ages, $\delta T_m/T_{m,0} \sim 100\%$ (i.e.\ $T_m$ with backreaction is a factor of 2 larger than without) or more can easily be obtained. Even larger deviations are possible at lower redshifts, depending on the channel under consideration.

%

\subsection{21-cm Sensitivity}
\label{sec:21_cm_sensitivity}

\begin{figure*}
\begin{tabular}{cc}
 \quad \includegraphics[scale=.55]{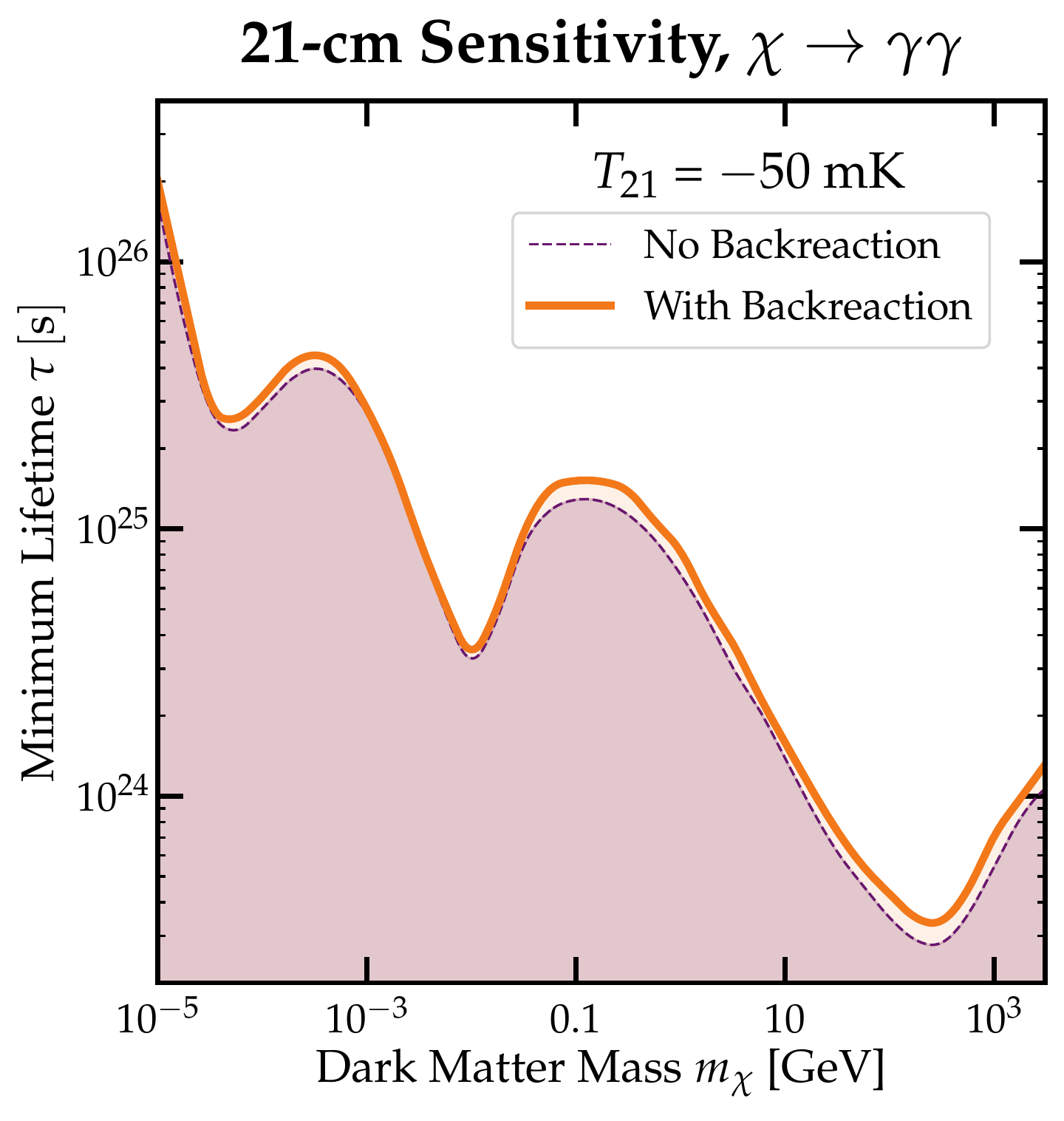}&  \
 \quad \includegraphics[scale=.55]{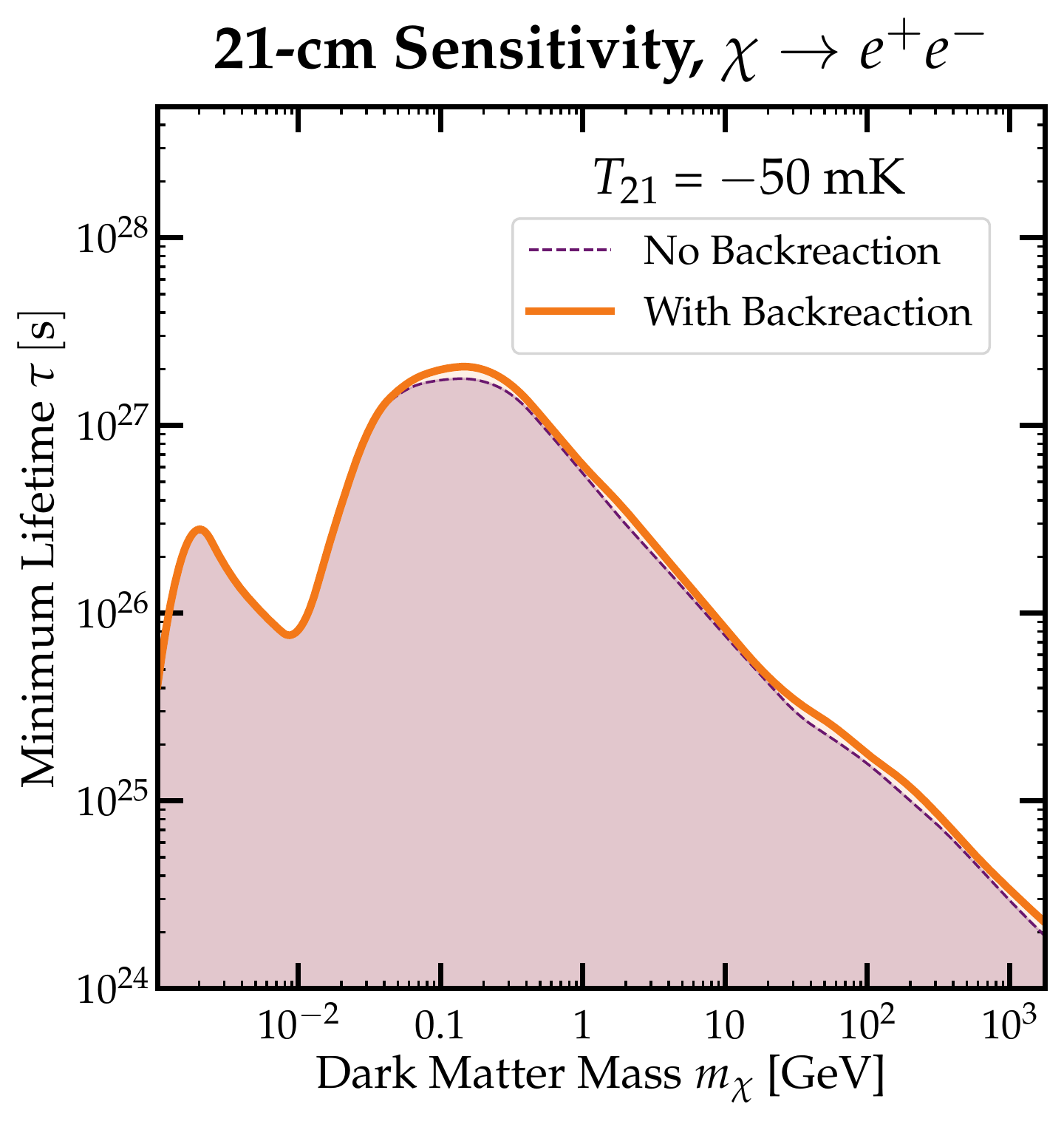} \\[6pt]
 \includegraphics[scale=.55]{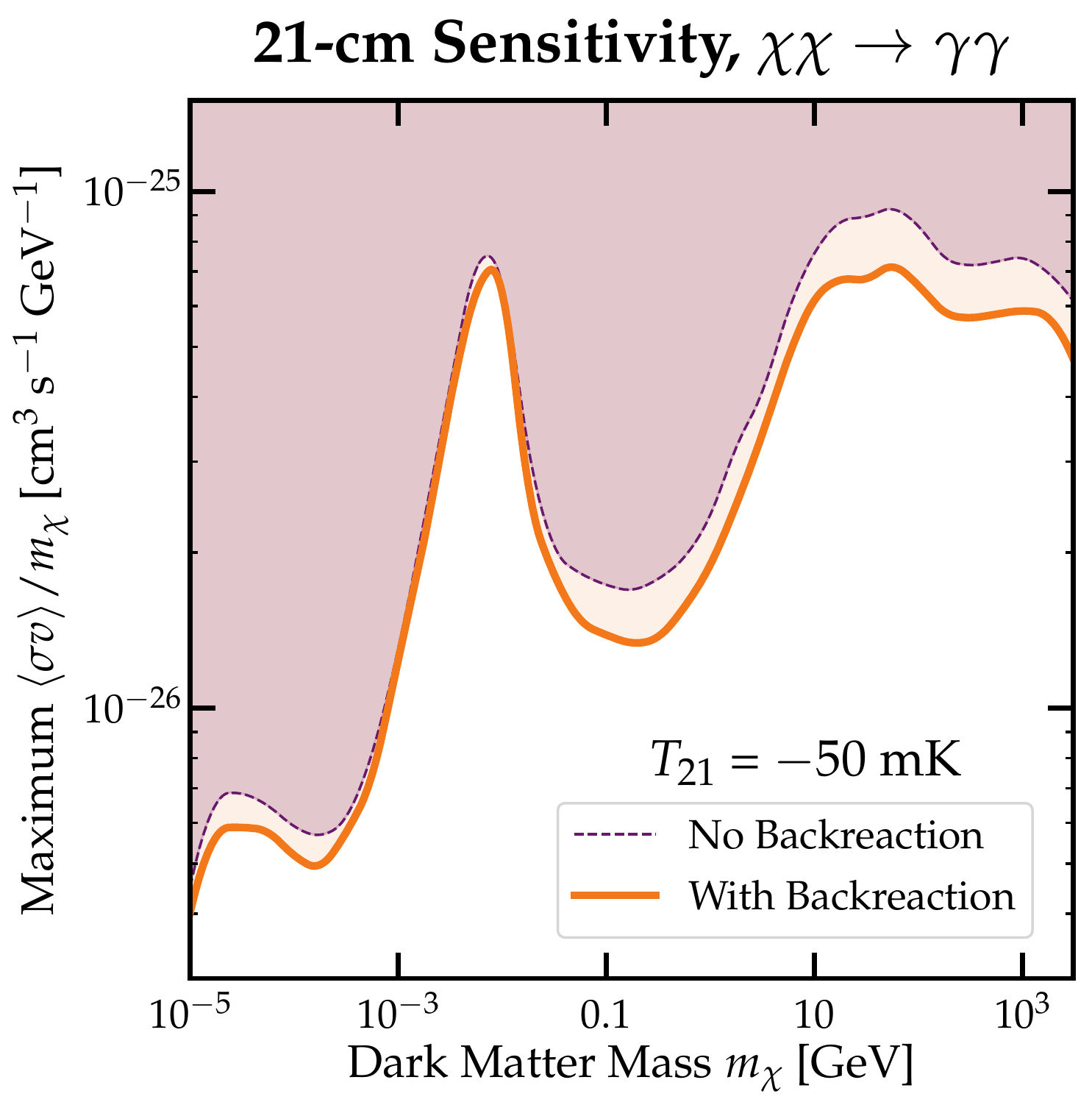} & \
 \includegraphics[scale=.55]{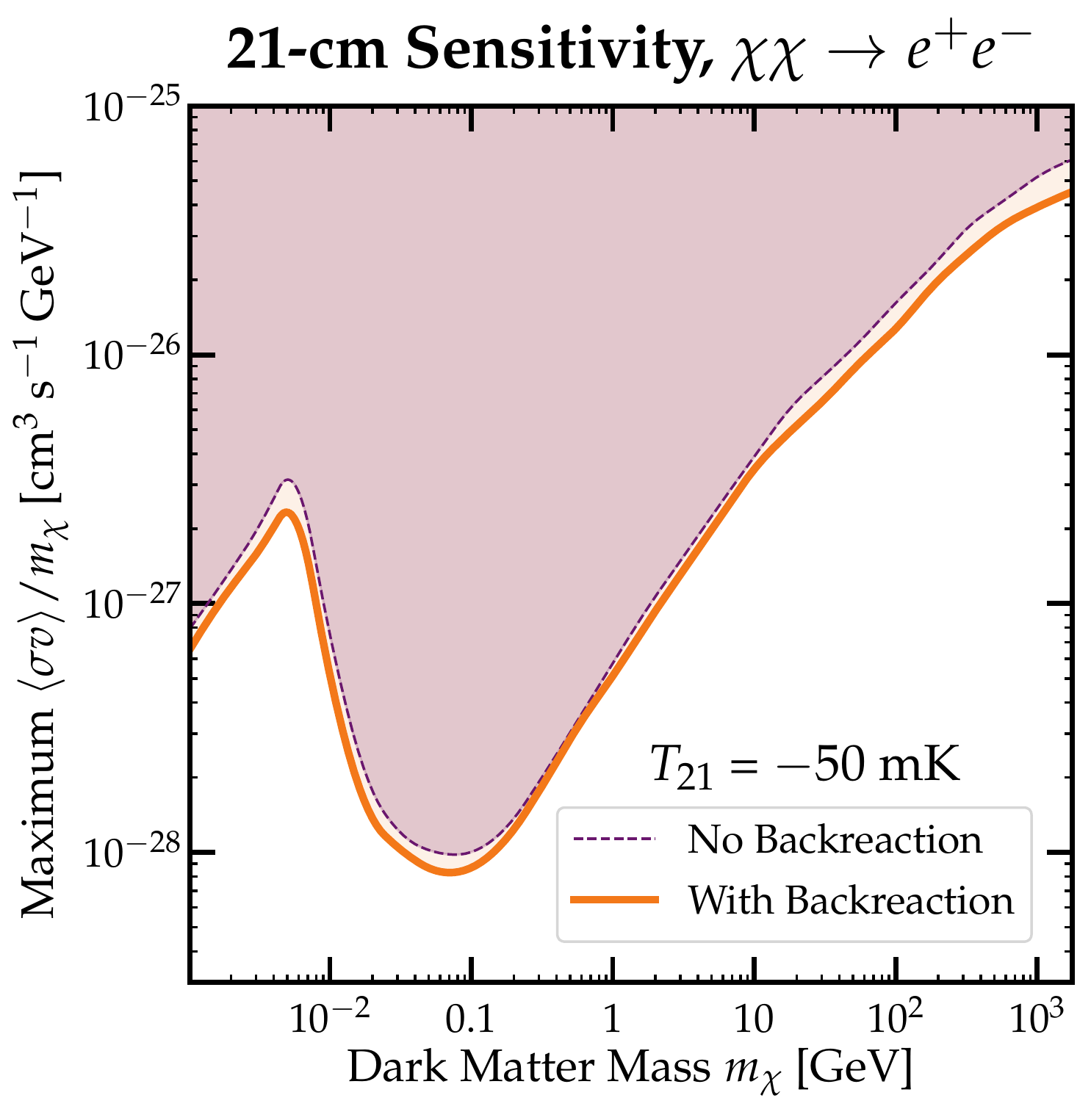} \\[6pt]
\end{tabular}
\caption{The minimum dark matter decay lifetime (top row) and maximum annihilation cross section (bottom row) bounds, derived from the global 21-cm signal. We assume a differential 21-cm brightness temperature of $T_\text{21} = -\SI{50}{\milli\kelvin}$, corresponding to a maximum $T_m$ of about \SI{20.3}{\kelvin} at $z \sim 17$. We consider decay and annihilation into $\gamma \gamma$ (left column) and $e^+e^-$ (right column) and compute the bounds with (orange, solid) and without (purple, dashed) backreaction.}
\label{fig:21cm}
\end{figure*}

The global 21-cm signal is a measurement of the sky-averaged differential brightness temperature $T_{21}$ with respect to the background radiation. Measurements of this signal would open a window into the ionization and temperature histories of the universe at the cosmic dawn (see e.g.\ Ref.~\cite{Pritchard:2011xb} for a review of 21-cm cosmology). A first claim of such a measurement has already been made by the EDGES collaboration~\cite{Bowman:2018yin}. The brightness temperature of the 21-cm hydrogen absorption line relative to the background radiation temperature is given by~\cite{Pritchard:2011xb}:
\begin{alignat}{2}
	T_\text{21} &\approx&& \,\, x_\text{HI}(z) 
	 \left( \frac{0.15}{\Omega_m}\right)^{1/2} 
	 \left( \frac{\Omega_b h}{0.02}\right) \nonumber \\
	 & &&\times \left( \frac{1+z}{10} \right)^{1/2} 
	 \left[ 1 - \frac{T_R(z)}{T_S(z)} \right]  \; \SI{23}{\milli \K} \,,
	\label{eqn:T21}
\end{alignat}
where $\Omega_b$ is the baryon energy density today as a fraction of the critical density, $h$ is the Hubble parameter today in \SI{}{\kilo\meter \per \second \per \mega\parsec}, $T_R$ is the background radiation temperature (typically assumed to be the CMB temperature) and $T_S$ is the spin temperature of neutral hydrogen as a function of redshift, which determines the relative population of neutral hydrogen in the two hyperfine states. Due to the presence of an intense Lyman-$\alpha$ radiation field once stars begin to form, it is expected that $T_S \approx T_m$ at the cosmic dawn. This fact allows us to turn the 21-cm global signal into a limit on $T_m$ itself, assuming that $T_R = T_\text{CMB}$. 

We will focus on $1+z \approx 18$, roughly the central value of the absorption trough measured by EDGES~\cite{Bowman:2018yin}. At this redshift, almost all hydrogen is neutral, i.e.\ $x_\text{HI} \approx 1$, and we can invert Eq.~(\ref{eqn:T21}) to find $T_S$ as a function of $T_\text{21}$. Since $T_m < T_S$, this yields the bound
\begin{alignat}{1}
	T_m(z=17) < \left( 1 - \frac{T_\text{21}}{\SI{35}{\milli\kelvin}} \right)^{-1} \SI{49}{\kelvin} \,.
	\label{eqn:Tm_bound}
\end{alignat}
This temperature bound in turn puts a limit on the DM decay lifetime or cross-section 
because too much dark matter decay/annihilation would heat up $T_m$ past this point.

In contrast to the CMB power spectrum energy injection bounds, which is most sensitive to changes in $x_e$ around the time of recombination, the 21-cm global signal constraints are more sensitive to energy injection processes that are more active at late times, and are dependent primarily on $T_m$ instead. Since $T_m$ is significantly impacted by including the effects of backreaction, the calculation performed by \dhis becomes important for setting accurate constraints using the 21-cm global signal.

\begin{figure*}[t]
   \centering
 \includegraphics[scale=0.55]{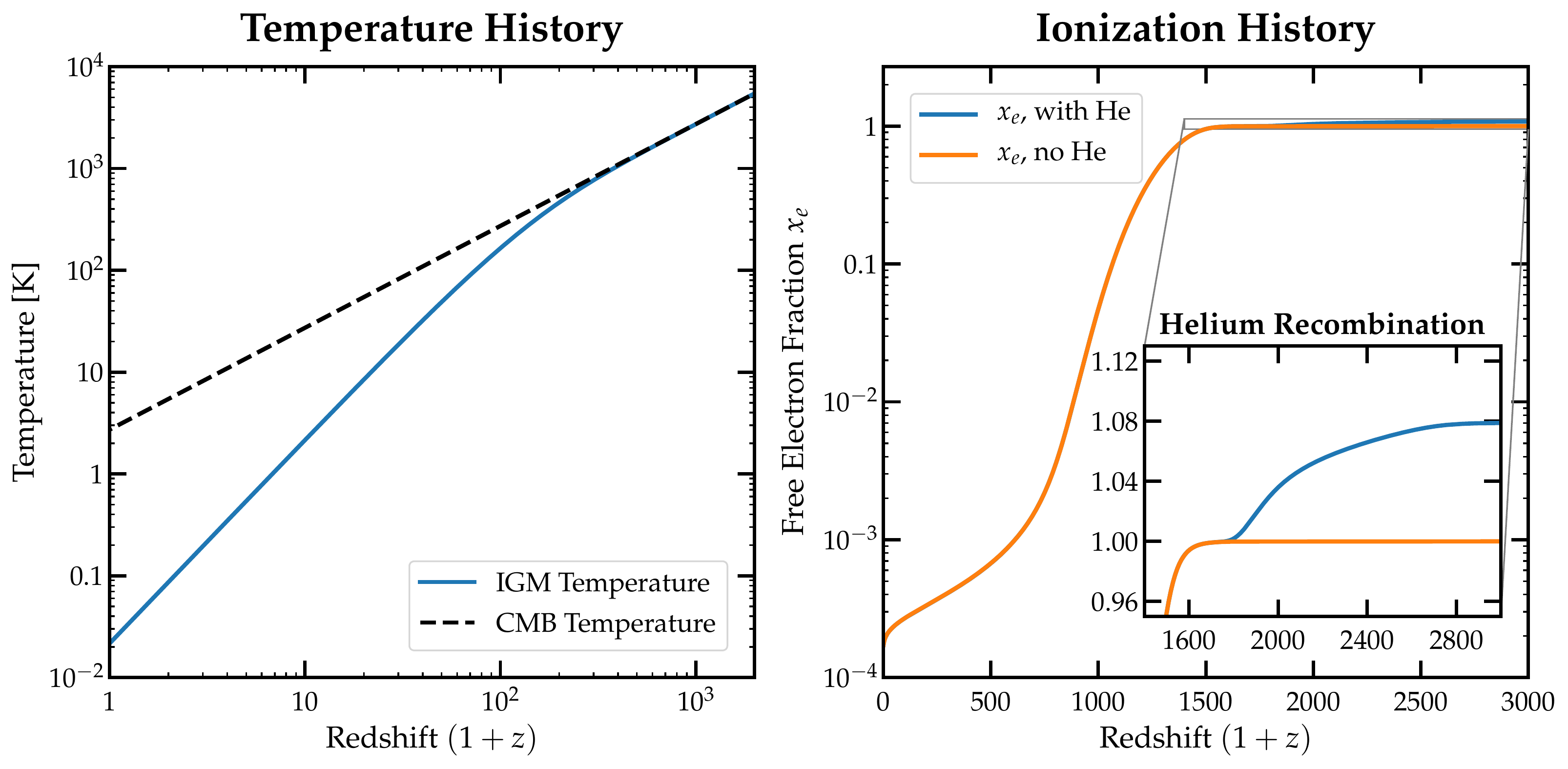}
 \caption{Temperature (left) and free electron fraction $x_e$ (right) as a function of redshift. $x_e$ is solved in \texttt{DarkHistory} with (blue) and without (orange) helium; both options lead to a similar temperature history (blue). With helium, helium recombination is correctly computed (inset). The CMB temperature is shown (black, dashed) for reference.}
 \label{fig:He_recomb}
\end{figure*}

To illustrate this, we perform a simple sensitivity study by obtaining the constraints for a measured $T_\text{21}$ of \SI{-50}{\milli\kelvin}, and compare the constraints with and without backreaction taken into account. Although this value of $T_{21}$ is inconsistent with the EDGES experiment, it is impossible to interpret the EDGES result without proposing new physics that may be at play during the cosmic dark ages~\cite{Liu:2018uzy}, which is a more complicated task and less relevant to helping users understand the code.  The following analysis is worked out in more detail within the code in Example 11.

$T_{21} = \SI{-50}{\milli\kelvin}$ means that we require $T_m < \SI{20.3}{\kelvin}$ according to Eq.~(\ref{eqn:Tm_bound}). We once again scan over a grid of dark matter masses and lifetimes/cross-sections decaying/annihilating into $e^+e^-$ and $\gamma\gamma$, using \texttt{get\_history()} for the case with no backreaction and \texttt{evolve()} for the case with backreaction, as explained in the previous section, to find where in parameter space dark matter energy injection leads to a violation of Eq.~(\ref{eqn:Tm_bound}).

The resulting exclusion plots are shown in Fig.~\ref{fig:21cm}.  
We see that in each case the calculation with backreaction can be between 10\%-50\% stronger than without backreaction, which we would expect because backreaction leads to larger temperatures. We emphasize that this is the result for just one chosen value of $T_{21}$; for larger (less negative) $T_{21}$, we expect that the importance of backreaction will increase, since the energy injection is less constrained, allowing for larger values of $x_e$.

\subsection{Helium, Dark Matter and Reionization}
\label{sec:example_reionization}

\begin{figure*}[t]
   \centering
 \includegraphics[scale=0.55]{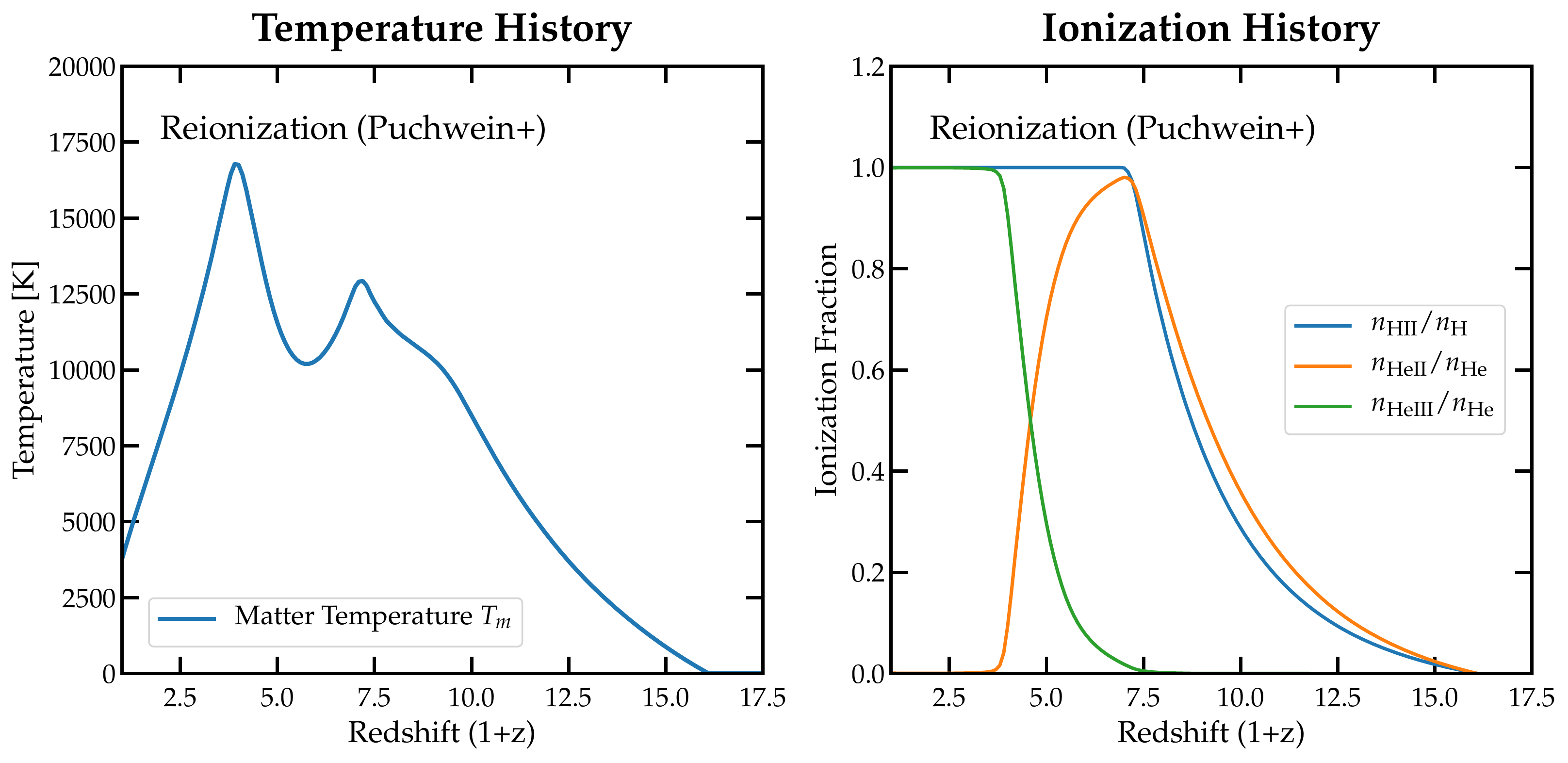}
 \caption{Temperature (left) and free electron fraction (right) as a function of redshift, solved in \texttt{DarkHistory} with the default Puchwein+ reionization model~\cite{Puchwein:2018arm}. The IGM temperature (blue) is shown on the left, while the ionization fractions $n_\text{HII}/n_\text{H}$ (blue), $n_\text{HeII}/n_\text{He}$ (orange) and $n_\text{HeIII}/n_\text{He}$ (green) are shown as well. These results agree very well with the same plots shown in Ref.~\cite{Puchwein:2018arm}.}
 \label{fig:Puchwein_reion}
\end{figure*}

\begin{figure*}[t]
 \centering
 \includegraphics[scale=0.55]{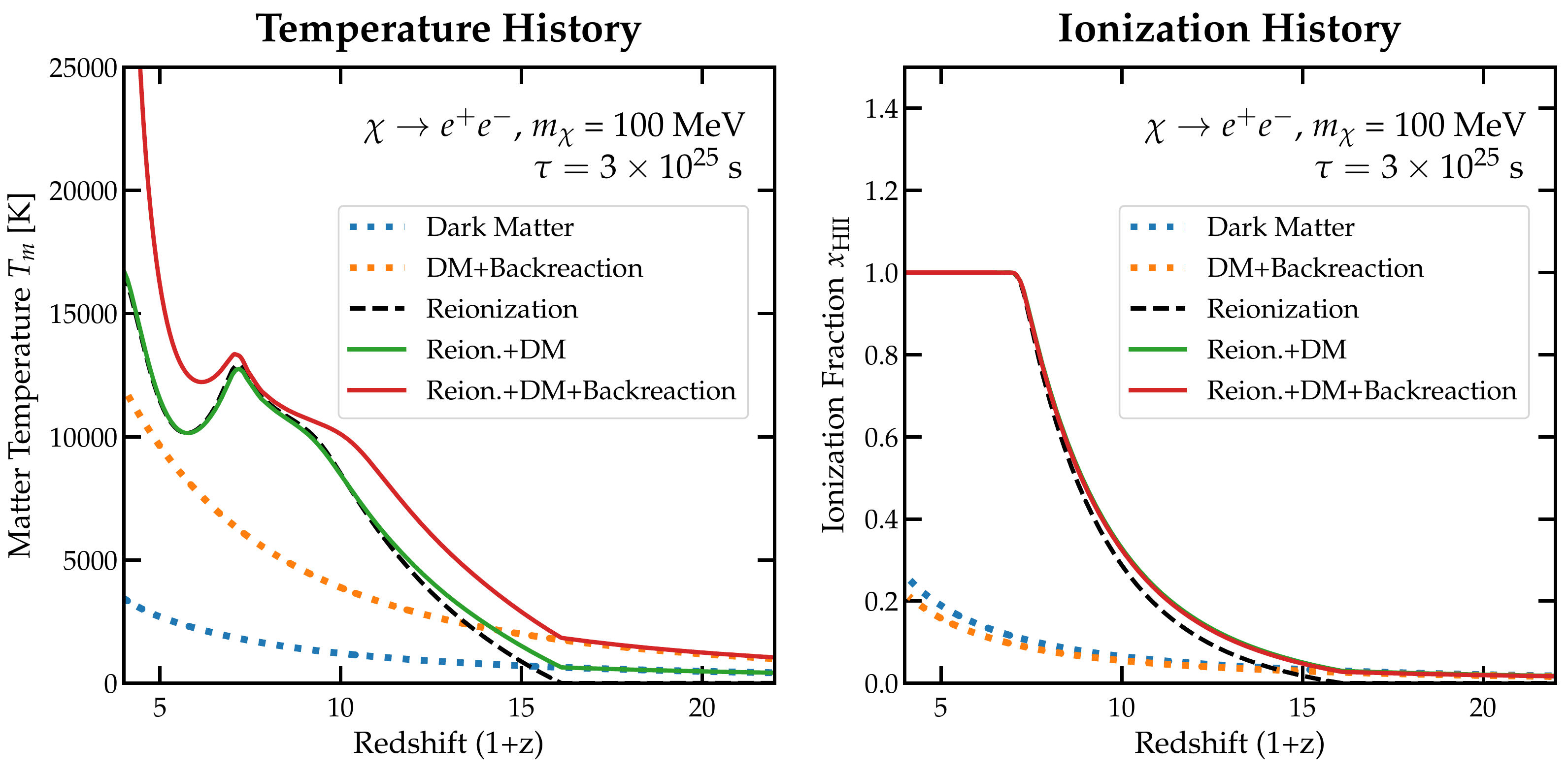}
 \caption{Temperature (left) and hydrogen ionization (right) history of the universe with DM decay and the default reionization model. The DM has a mass of $m_\chi = \SI{100}{\mega\eV}$ and decays to $e^+e^-$ with a lifetime of \SI{3e25}{s}. The temperature and ionization with DM decay alone is shown without (blue, dotted) and with (orange, dotted) backreaction included. The combined effect of DM decay and reionization without (green) and with (red) backreaction can be compared to the reference reionization model (black, dashed).}
 \label{fig:DM_reion}
\end{figure*}

Finally, we will take a closer look at the different options one can use within the code to evaluate temperature and ionization histories. Throughout this section, we will demonstrate these different options mostly using \texttt{get\_history()}, but similar options are also available in \texttt{evolve()}, which calls \texttt{get\_history()} with all of the relevant options provided. We refer the reader to the online documentation and to Example 8 in the code for more details.

Without any exotic energy injection or any reionization, the function \texttt{get\_history()} accepts a redshift vector, and simply returns the baseline ionization and temperature histories, obtained by solving Eq.~(\ref{eqn:TLA_DarkHistory}):  
\begin{lstlisting}[language=python]
  import numpy as np
  from darkhistory.tla import get_history

  # Redshift vector in decreasing order. 
  rs_vec = np.flipud(np.arange(1., 3000., 0.1))
  soln_baseline = get_history(rs_vec)
\end{lstlisting}
Turning on helium evolution within \texttt{get\_history()} is controlled by the flag \texttt{helium\_TLA}, i.e.
\begin{lstlisting}
  soln_He = get_history(rs_vec,helium_TLA=True)
\end{lstlisting}
Fig.~\ref{fig:He_recomb} shows the solution to Eq.~(\ref{eqn:TLA}) with just the ``$(0)$'' terms, i.e.\ without any energy injection or reionization, and compares that solution to one with Eq.~(\ref{eqn:TLA_helium_darkhistory}) added as well. This is simply the standard ionization history with helium recombination ($z \sim 1800$) and hydrogen recombination ($z \sim 1100$), eventually leading to the residual ionization fraction at redshifts well below hydrogen recombination of about $x_e \sim 2 \times 10^{-4}$. The inset of Fig.~\ref{fig:He_recomb} shows that \dhis is able to correctly reproduce helium recombination; the entire ionization history agrees with \texttt{RECFAST} results at the central cosmological parameters used by \dhis to within $\sim 3\%$. We recommend that helium ionization levels are tracked when used in combination with reionization.

The next important option is whether to include the effects of reionization. This option is controlled by the flag \texttt{reion\_switch}: 
\begin{lstlisting}
  soln_default_reion = get_history(
   rs_vec, helium_TLA=True, reion_switch=True
  )
\end{lstlisting}
With no other options set, setting \texttt{reion\_switch} to \lstinline|True| causes \dhis to use the standard reionization model, which is based on the photoionization and photoheating rates provided in~\cite{Puchwein:2018arm}. Fig.~\ref{fig:Puchwein_reion} shows the IGM temperature as well as the ionization levels of the different atomic species as a function of redshift. Both of these results agree well with the same result shown in Ref.~\cite{Puchwein:2018arm}. Reionization of hydrogen and neutral helium is complete by about $z \sim 6$; soon after, HeII starts to become doubly ionized, leading to a decrease in $n_\text{HeII}$ and a corresponding increase in $n_\text{HeIII}$. Dips in $T_m$ correspond to a decrease in photoheating rates once a species becomes completely ionized and the production of high-energy electrons from photoionization off these species ceases.

Aside from the default reionization model, the user may also supply their own reionization models in two different ways: by either providing their own photoionization and photoheating rates on each atomic species (e.g. based on a model that is different from the default, e.g.~\cite{Haardt:2011xv}), or by fixing the ionization history below a certain redshift, e.g.\ with a $\tanh$ model~\cite{Lewis:2008wr,Adam:2016hgk}. We leave a discussion of how to use these options to Example 8 in the code. 

With the ability to include both helium and reionization, we can now add a new source of energy injection and compute the effects on ionization and temperature levels. We remind the reader that this means we are solving Eq.~(\ref{eqn:TLA_DarkHistory}) together with Eq.~(\ref{eqn:TLA_helium_darkhistory}). This is accomplished in the code with both \texttt{reion\_switch} and \texttt{helium\_TLA} set to \lstinline|True|, and supplying the same keyword parameters used to inject energy from DM shown in Sec.~\ref{sec:backreaction}. We can add decaying DM with mass \SI{100}{\mega\eV} into an $e^+e^-$ pair with a lifetime of \SI{3e25}{\second}, like so (using \texttt{evolve()} in this example):
\begin{lstlisting}
  main.evolve(
    DM_process='decay', mDM=1e8, 
    lifetime=3e25, primary='elec_delta',
    start_rs=3000., coarsen_factor=1, 
    backreaction=True, helium_TLA=True, 
    reion_switch=True
  )
\end{lstlisting}
By turning on and off the various flags \texttt{backreaction}, \texttt{helium\_TLA} and \texttt{reion\_switch}, we can produce histories including or excluding these various effects. 

The results from different combinations of these switches are summarized in Fig.~\ref{fig:DM_reion}. The dashed lines shows the histories with DM decay only, and illustrates the significant difference that can arise after taking into account backreaction, which we have already seen in Fig.~\ref{fig:single_decay}. Combining the DM energy injection with the reionization model gives the solid lines in Fig.~\ref{fig:DM_reion}. These curves should be compared to the default reionization model temperature and ionization histories, shown in the black, dashed lines. When computing the DM energy deposition without taking into account backreaction, we find that the amount of energy deposited into heating from DM is much smaller than heating from reionization processes once they begin in earnest, and so adding the DM decays on top of reionization produces only a small perturbation in the temperature history relative to $T_m$ for just the reionization model alone. In some cases, the addition of DM actually decreases $T_m$: this can happen due to reionization proceeding at a faster rate, leaving fewer atoms to photoionize and thus suppressing photoheating. 

It is clear, however, that neglecting backreaction leads to a severe underestimation of the energy deposition into heating. Performing the full calculation with DM, reionization and backreaction correctly accounted for produces the line in red, which shows that the addition of DM significantly increases $T_m$ compared to both the reionization model and the case where DM energy deposition is added without backreaction. Reionization greatly enhances the energy deposition rate into heating of the IGM by increasing the number of free charged particles available for Coulomb heating, and properly accounting for backreaction using \dhis is critical to predicting the IGM temperature growth due to energy injection once reionization begins. 

\section{Conclusion}
\label{sec:conclusion}

We have developed and made public a new code package for mapping out the effects of arbitrary exotic energy injections --- including dark matter annihilation and decay to arbitrary Standard Model final states --- on the temperature and ionization history of the early universe. \texttt{DarkHistory} is capable of self-consistently including the effects of conventional astrophysical sources of ionization and heating, and of including feedback effects that can significantly enhance the degree of heating. Additionally, the ICS module can be employed independently of the rest of the code, as an accurate and efficient numerical calculator of ICS across a very wide range of electron and photon energies. We have outlined here a number of worked examples, and provide more examples with the online code at \href{https://github.com/hongwanliu/DarkHistory}{https://github.com/hongwanliu/DarkHistory}.

\texttt{DarkHistory} has a modular framework and can in the future be improved in several different directions, while keeping the same underlying structure. In this first version we have focused on the homogeneous signal, and neglected the possible effect of new radiation backgrounds and/or gas inhomogeneities on the cascade of secondaries produced by injected high-energy particles. Such effects may become important in the late cosmic dark ages and the epoch of reionization. The spectrum of low-energy photons produced by energy injection, and the resulting distortion to the spectrum of the CMB, is a possible observable in its own right; the current version of \texttt{DarkHistory} provides only a partial calculation of this spectral distortion, due to our approximate treatment of low-energy electrons, but we intend to improve this aspect in future work. The effects of other new physics on the temperature/ionization evolution -- in particular, scattering between baryons and DM -- can be incorporated within the same framework. We also intend to explore the possibility of interfacing DarkHistory with existing public codes for computing the recombination history, perturbations to the CMB, and 21cm signals. 

The tools we have developed in this work can be used to understand the visible imprints of exotic energy injections that could appear in the CMB and the 21cm line of neutral hydrogen, and hence to place precise constraints on dark matter annihilation and decay. We hope they will help pave the way for a comprehensive description of the ways in which dark matter interactions, and other physics beyond the Standard Model, could reshape the early history of our cosmos.

\section{Acknowledgement}
 The authors would like to thank Shi-Fan Stephen Chen for a fruitful early collaboration. We would also like to thank Yacine Ali-Ha\"{i}moud, Anastasia Fialkov, Katherine Mack, Vivian Poulin, Nicholas L. Rodd, Sarah Sch\"{o}n and Cannon Vogel for helpful discussions. We acknowledge the use of the following Python packages: NumPy~\cite{van2011numpy}, SciPy~\cite{jones_scipy_2001}, Matplotlib~\cite{Hunter:2007}, as well as \href{https://github.com/ml31415/numpy-groupies}{\texttt{numpy\_groupies}} for fast rebinning of spectra.

This work was supported by the Office of High Energy Physics of the U.S. Department of Energy under Grant No.~DE-SC00012567 and DE-SC0013999. This work was also partly supported by the MIT Research Support Committee (NEC Corporation Fund for Research in Computers and Communications). GR is supported by a National Science Foundation Graduate Research Fellowship. TRS is partially supported by a John N. Bahcall Fellowship.

\appendix

\section{Inverse Compton Scattering}
\label{app:ICS}

In this appendix, we discuss in detail the methods used to compute the spectra of photons that are produced by the cooling of electrons through ICS. We restore $\hbar$, $c$ and $k_B$ in this appendix, since the exact numerical value of these spectra is important. 

\subsection{Scattered Spectra}

We begin with some preliminaries that will be important in understanding our subsequent discussion of ICS. The goal is to determine the secondary photon spectrum produced on average by multiple scatterings of a single electron.

Consider an electron with energy $E_e$ and corresponding Lorentz factor $\gamma$ incident on some distribution of photons $n(\epsilon)$ with initial energy $\epsilon$ in the comoving frame. Since we are only interested in ICS off the CMB, we will only consider an isotropic photon bath in the co-moving frame, distributed as a blackbody with some temperature $T$. The electron has some probability per unit time of scattering the photons into some outgoing energy $\epsilon_1$, with some probability distribution $dN_\gamma/(d\epsilon \, d\epsilon_1 \, dt)$, which we call the ``differential scattered photon spectrum''. This quantity is proportional to the number density per unit energy of the photon bath $n(\epsilon)$, so that integrating over $\epsilon$ also integrates over the distribution of these photons. This can be interpreted as a normalized scattered photon spectrum for ICS by many electrons with the same energy. Integrating the differential scattered photon spectrum with respect to $\epsilon$ gives us the ``scattered photon spectrum'', 
\begin{alignat}{1}
    \frac{dN_\gamma}{d \epsilon_1 \, dt} (E_e, T, \epsilon_1) = \int_{\epsilon_\text{min}}^{\epsilon_{\text{max}}} d\epsilon \frac{dN_\gamma}{d\epsilon \, d\epsilon_1 \, dt} (E_e, T, \epsilon, \epsilon_1) \,,
    \label{eqn:ics_scat_phot_spec}
\end{alignat}
with $\epsilon_\text{min}$ and $\epsilon_\text{max}$ determined by the kinematics of ICS.

We further define the ``scattered photon energy loss spectrum'',
\begin{alignat}{1}
    \frac{dN_\gamma}{d\Delta \, dt} (E_e, T, \Delta) &= \int d\epsilon \frac{dN_\gamma}{d\epsilon \, d\epsilon_1 \, dt} (E_e, T, \epsilon, \epsilon_1 = \epsilon + \Delta),
  \label{eqn:scattered_phot_engloss_spec}
\end{alignat}
where $\Delta$ is the change in energy of a photon scattering by a single electron. This is simply the distribution of scattered photons as a function of the energy gained or lost by the photon during the scattering. 

Now, consider some arbitrary injection spectrum of electrons $d \tilde{N}_e/dE_1$. The tilde serves to remind the reader that this is a distribution of electrons, and not a normalized quantity. From the definition of Eq.~(\ref{eqn:scattered_phot_engloss_spec}), we define the ``scattered electron spectrum'' as
\begin{alignat}{1}
  \frac{d \tilde{N}_e}{dE_1\, dt} = \int_0^\infty dE \, \frac{d \tilde{N}_e}{dE} \frac{dN_\gamma}{d \Delta \, dt} (E, T, \Delta = E - E_1) \,,
  \label{eqn:scattered_elec_spec_exact}
\end{alignat}
where $E_1$ is the energy of the scattered electron. However, this result allows some electrons to gain energy after scattering, significantly complicating our calculations. Intuitively, we expect electrons that upscatter from $E \to E_1$ to partially cancel with downscatters from $E_1 - E$, justifying an approximate treatment where we simply cancel out photons that downscatter (and upscatters an electron) with photons that upscatter (and downscatters an electron). We leave a full justification of this to the end of this section, but for now, we will accordingly define the ``scattered electron net energy loss spectrum'',
\begin{alignat}{1}
    \frac{dN_e}{d \Delta \, dt} (\beta, T, \Delta) = \frac{dN_\gamma}{d\Delta\, dt}(\beta, T, \Delta) - \frac{dN_\gamma}{d\Delta \, dt}(\beta, T, -\Delta),
    \label{eqn:electron_eng_loss_spec}
\end{alignat}
with $\Delta \geq 0$ in the expression above. For relativistic electrons, the average energy lost due to an upscattering a photon is much larger than the average energy gained due to downscattering a photon, and it is therefore a good approximation to consider only scattering events where electrons lose their energy~\cite{Blumenthal:1970gc}. The upscattered photons also have outgoing energy $\epsilon_1 \gg \epsilon$, and so a reasonable approximation to make in the relativistic limit is
\begin{alignat}{1}
    \left. \frac{dN_e}{d\Delta \, dt} \right|_{\beta \to 1} \approx \left. \frac{dN_\gamma}{d\epsilon_1 \, dt}\right|_{\beta \to 1} \, .
    \label{eqn:engloss_approx}
\end{alignat}

Now we can turn our attention to justifying the approximation laid out in Eq.~(\ref{eqn:electron_eng_loss_spec}). First, we split the exact integral in Eq.~(\ref{eqn:scattered_elec_spec_exact}) into an integral from 0 to $E_1$, and from $E_1$ to $\infty$. The first integral can be rewritten as (dropping the $T$ dependence for clarity)
 \begin{multline}
  \int_0^{E_1} dE \, \frac{d \tilde{N}_e}{dE} \frac{dN_\gamma}{d \Delta \, dt} (E, \Delta = E - E_1) \\
  = - \int_{E_1}^{2E_1} dx \frac{d \tilde{N}}{dx} \frac{dN_\gamma}{d\Delta \, dt} (E = 2E_1 - x, \Delta = E_1 - x) \,,
 \end{multline}
where we have simply made the substitution $x = 2E_1 - E$. In this part of the integral, we are dealing with upscattered electrons and downscattered photons, and so we know that $dN_\gamma / (d \Delta \, dt)$ only has support when $E - E_1 \sim T_\text{CMB} \ll E, E_1$, since ICS is only included for electrons with $E > \SI{3}{\kilo\eV}$~\cite{Slatyer:2015kla}. This implies that the integral only has support near $x = E_1$, and we can therefore make the following approximation:  
 \begin{multline}
  \int_0^{E_1} dE \, \frac{d \tilde{N}_e}{dE} \frac{dN_\gamma}{d \Delta \, dt} (E, \Delta = E - E_1) \\
  \approx - \int_{E_1}^{\infty} \, dx \frac{d \tilde{N}_e}{dx} \frac{dN_\gamma}{d\Delta \, dt} (E = x, \Delta = E_1 - x) \\
  = - \int_{E_1}^\infty dE \, \frac{d\tilde{N}_e}{dE} \frac{dN_\gamma}{d \Delta \, dt} (E, \Delta = E_1 - E) \,,
 \end{multline}
where in the last step we have trivially relabeled $x \to E$. We have therefore shown that
\begin{alignat}{2}
  \frac{d \tilde{N}_e'}{dE_1 \, dt} &\approx&& -\int_{E_1}^\infty dE \, \frac{d \tilde{N}_e}{dE} \frac{dN_\gamma}{d \Delta \, dt} (E, \Delta = E_1 - E) \nonumber \\
  & && \,\, + \int_{E_1}^\infty dE \, \frac{d\tilde{N}_e}{dE} \frac{dN_\gamma}{d \Delta \, dt} (E, \Delta = E - E_1) \,,
\end{alignat}
and that is a good approximation due to the relatively low temperature of the CMB.

With these definitions in mind, we are now ready to understand how to compute these scattered spectra when the electron is in two limits. For $\gamma > 20$, the spectra are computed in the relativistic limit, while below that, scattering with the CMB at all relevant redshifts lie well within the Thomson regime. Together, they cover all relevant kinematic regimes that we consider in our code.

\subsubsection{Relativistic Electrons}

The differential upscattered photon spectrum produced by ICS between an electron and the CMB blackbody spectrum in the relativistic regime ($\gamma \gg 1$) is given by~\cite{Blumenthal:1970gc}
\begin{alignat}{2}
    \frac{dN_\gamma}{d\epsilon \, d\epsilon_1 \, dt} &=&& \frac{2\pi r_0^2 c}{\gamma^2} \frac{n(\epsilon, T)}{\epsilon} \bigg[ 2q \log q + (1 + 2q)(1 - q) \nonumber \\
    & && + \frac{1}{2} \frac{(\Gamma(\epsilon) q)^2}{1 + \Gamma(\epsilon) q}(1 - q) \bigg],
    \label{eqn:ICS_phot_spec_rel}
\end{alignat}
 where $r_0$ is the classical electron radius, $m_e$ is the electron mass, $\epsilon$ is the incident photon energy in the comoving frame, and $\epsilon_1$ is the scattered photon energy in the same frame, and we have defined
\begin{alignat}{1}
    \Gamma(\epsilon) = \frac{4 \epsilon \gamma}{m_e c^2}\, , \quad q = \frac{\epsilon_1}{\gamma m_e c^2 - \epsilon_1} \frac{1}{\Gamma(\epsilon)} \, .
\end{alignat}
We stress that Eq.~(\ref{eqn:ICS_phot_spec_rel}) is strictly only correct when photons are upscattered by the incoming electron, which corresponds to the kinematic regime $\epsilon \leq \epsilon_1 \leq 4 \epsilon \gamma^2/(1 + 4 \epsilon \gamma/m)$. In the opposite regime where $\epsilon/(4\gamma^2) \leq \epsilon_1 < \epsilon$ and photons get downscattered, we have~\cite{Jones:1968zza}
 \begin{alignat}{1}
   \frac{dN_\gamma}{d\epsilon \, d\epsilon_1 \, dt} = \frac{\pi r_0^2 c}{2 \gamma^4 \epsilon} \left(\frac{4\gamma^2 \epsilon_1}{\epsilon} - 1 \right) n(\epsilon, T) \,.
   \label{eqn:Jones_correction}
 \end{alignat}
For ICS off CMB photons, the $n(\epsilon)$ is the number density of photons per unit energy; for a blackbody, this is
\begin{alignat}{1}
    n_\text{BB}(\epsilon, T) = \frac{1}{\pi^2 \hbar^3 c^3} \frac{\epsilon^2}{\exp(\epsilon/k_B T) - 1} \, ,
\end{alignat}
where $T$ is the temperature of the CMB. 

The complete upscattered photon spectrum for ICS off the CMB is therefore obtained by performing the integral in Eq.~(\ref{eqn:ICS_phot_spec_rel}) over $\epsilon$, with the kinematic limits given by $1/4\gamma^2 \leq q \leq 1$~\cite{Blumenthal:1970gc}. Since the CMB photons at $z \lesssim 3000$ have energies less than \SI{1}{\eV}, the amount of energy transferred by an electron is always completely dominated by Eq.~(\ref{eqn:ICS_phot_spec_rel}). Furthermore, one can check that at $q = 1/4\gamma^2$, $\epsilon \gg T$. We can therefore make the approximation that Eq.~(\ref{eqn:ICS_phot_spec_rel}) gives the full ICS spectrum while neglecting Eq.~(\ref{eqn:Jones_correction}), and take the integral limits to be $0 \leq q \leq 1$ instead. This assumption is made in the ICS transfer functions provided as part of the downloaded data, but options are available in the \texttt{ics} module to turn these various assumptions off.

The quantity $\Gamma(\epsilon)$ separates the two kinematic regimes of Compton scattering: $\Gamma \gg 1$ for the Klein-Nishina regime, where Compton scattering in the electron rest frame is highly inelastic, and $\Gamma \ll 1$ for the Thomson regime, where it is almost elastic instead.\footnote{Although the scattering process is almost elastic in the initial electron rest frame, it is certainly not elastic in the co-moving frame. In the co-moving frame, the electron loses a small fraction of its energy per collision, but each collision can upscatter a CMB photon by a significant factor.} Eq.~(\ref{eqn:ICS_phot_spec_rel}) applies to both regimes, with the only assumption being $\gamma \gg 1$.

To avoid computing the scattered photon spectrum repeatedly in the code, we use the following relation between spectra at different temperatures:
\begin{alignat}{1}
    \frac{dN_\gamma}{d\epsilon_1 \, dt} (E_e, yT, \epsilon_1) = y^4 \frac{dN_\gamma}{d\epsilon_1 \, dt}(yE_e, T, y\epsilon_1) \,,
    \label{eqn:rel_temp_relation}
\end{alignat}
for any real positive number $y$, even if $yE_e$ is unphysical.\footnote{This trick can only be performed by integrating over $0 \leq q \leq 1$, and is the key reason for making such an approximation.} In \texttt{DarkHistory}, we evaluate the scattered photon spectrum at $1+z = 400$, and use this relation to compute the subsequent spectra at lower redshifts by a straightforward interpolation.

\subsubsection{Thomson Regime}

In the Thomson regime, the rate at which photons are scattered is given by~\cite{Blumenthal:1970gc}
\begin{alignat}{1}
    \frac{dN_\gamma}{dt} = \sigma_T c N_\text{rad},
    \label{eqn:thomson_scattering_rate}
\end{alignat}
where $N_\text{rad}$ is the total number density of incident photons, with $\sigma_T = 8 \pi r_0^2/3$ being the Thomson cross section. Note that the scattering rate is independent on the incident photon energy. The energy loss rate of the electron is~\cite{Blumenthal:1970gc}
\begin{alignat}{1}
    \frac{dE_e}{dt} = \frac{4}{3} \sigma_T c \gamma^2 \beta^2 U_\text{rad},
    \label{eqn:thomson_energy_loss_rate}
\end{alignat}
where $\beta$ is the velocity of the electron, with $U_\text{rad}$ being the total energy density of the incident photons.

While Eqs.~(\ref{eqn:thomson_scattering_rate}) and~(\ref{eqn:thomson_energy_loss_rate}) are well-known, the actual spectrum of scattered photons in the Thomson regime is much less so. The complete expression for the differential scattered photon spectrum with no further assumptions is, as far as the authors know, first given in Ref.~\cite{Fargion:1996xr}, and we reproduce their final result here for completeness. For $(1-\beta)\epsilon_1/(1+\beta) < \epsilon < \epsilon_1$, we have
\begin{multline}
    \left. \frac{dN_\gamma}{d \epsilon \,d\epsilon_1\, dt}  (\beta, T, \epsilon, \epsilon_1) \right|_{\epsilon < \epsilon_1} = \frac{\pi r_0^2 c n(\epsilon, T)}{4 \beta^6 \gamma^2 \epsilon} \Bigg\{ \frac{1}{\gamma^4} \frac{\epsilon}{\epsilon_1} - \frac{1}{\gamma^4} \frac{\epsilon_1^2}{\epsilon^2} \\
    + (1 + \beta) \left[\beta(\beta^2 + 3) + \frac{1}{\gamma^2}(9 - 4\beta^2) \right] \\
    + (1-\beta) \left[\beta(\beta^2 + 3) - \frac{1}{\gamma^2}(9 - 4\beta^2) \right] \frac{\epsilon_1}{\epsilon} \\
    - \frac{2}{\gamma^2}(3 - \beta^2) \left(1 + \frac{\epsilon_1}{\epsilon}\right) \log \left(\frac{1+\beta}{1-\beta} \frac{\epsilon}{\epsilon_1} \right) \Bigg\},
    \label{eqn:thomson_scattered_diff_phot_spec}
\end{multline}
and for $\epsilon_1 < \epsilon < (1+\beta) \epsilon_1 / (1-\beta)$, 
\begin{alignat}{1}
    \left. \frac{dN_\gamma}{d\epsilon \, d\epsilon_1 \, dt} (\beta, T, \epsilon, \epsilon_1) \right|_{\epsilon \geq \epsilon_1} = -\left. \frac{dN_\gamma}{d\epsilon \, d\epsilon_1 \, dt} (-\beta, T, \epsilon, \epsilon_1) \right|_{\epsilon < \epsilon_1} \!\!\!\!\!.
    \label{eqn:thomson_spectrum_upp_low_relation}
\end{alignat}
All other values of $\epsilon$ outside of the ranges specified are kinematically forbidden, and so to find the spectrum, we need to integrate over $\epsilon$ with $n(\epsilon) = n_\text{BB}(\epsilon)$ in the finite range specified above, i.e.\
\begin{multline}
    \frac{dN_\gamma}{d \epsilon_1 \, dt}(\beta, T, \epsilon_1) = \int_{\frac{1-\beta}{1+\beta}\epsilon_1} ^{\epsilon_1} d\epsilon \left. \frac{dN_\gamma}{d\epsilon \, d\epsilon_1 \, dt} (\beta, T, \epsilon, \epsilon_1) \right|_{\epsilon < \epsilon_1} \\
    - \int_{\epsilon_1}^{\frac{1+\beta}{1-\beta} \epsilon_1} d\epsilon \left. \frac{dN_\gamma}{d\epsilon \, d\epsilon_1 \, dt} (-\beta, T, \epsilon, \epsilon_1) \right|_{\epsilon < \epsilon_1} \!\!\!\!\!.
    \label{eqn:thomson_scattered_phot_spec}
\end{multline}
The relationship between spectra at different temperatures is given by
\begin{alignat}{1}
    \frac{dN_\gamma}{d\epsilon_1 \, dt}(\beta, yT, \epsilon_1) = y^2 \frac{dN_\gamma}{d\epsilon_1 \, dt}(\beta, T, \epsilon_1/y) \,.
    \label{eqn:thomson_temp_relation}
\end{alignat}

The scattered photon energy loss spectrum $dN_\gamma/(d\Delta \, dt)$ is similarly given by
 \begin{multline}
    \frac{dN_\gamma}{d\Delta dt}(\beta, T, \Delta) \\
    = \begin{cases} 
        \left. \int_{\frac{1-\beta}{2\beta}\Delta}^\infty d \epsilon \, \frac{dN_\gamma}{d \epsilon \, d\epsilon_1 \, dt}(\beta, T, \epsilon, \epsilon + \Delta) \right|_{\epsilon < \epsilon_1}, & \Delta > 0, \\
        \left. \int_{-\frac{1+\beta}{2\beta}\Delta}^\infty d \epsilon \, \frac{dN_\gamma}{d \epsilon \, d\epsilon_1 \, dt}(\beta, T, \epsilon, \epsilon + \Delta) \right|_{\epsilon \geq \epsilon_1}, & \Delta \leq 0.
    \end{cases}
 \end{multline}
The relation shown in Eq.~(\ref{eqn:thomson_temp_relation}) between scattered photon spectra of different temperatures also holds for the energy loss spectrum, with $\epsilon_1 \to \Delta$. 

\subsection{Numerical Methods}

Computationally, to evaluate all of the scattered spectra, we need to perform numerical quadrature over a large range of electron and scattered photon energies; using a standard grid of $5000 \times 5000$ energy values, the grid would take the standard \texttt{numpy} integrator over a day to populate. While a substantial speed-up may be obtained by using packages like \texttt{Cython}~\cite{behnel2010cython}, numerical quadrature for ICS in the Thomson regime is also subject to significant numerical errors when the electron is nonrelativistic due to the existence of catastrophic cancellations. A semi-analytic approach provides both a faster method and a way to avoid such errors in a robust manner. 

\subsubsection{Thomson and Relativistic Regime: Large \texorpdfstring{$\beta$}{beta}}

For $\beta \gtrsim 0.1$, we can obtain the scattered photon spectrum in Eq.~(\ref{eqn:thomson_scattered_diff_phot_spec}) in the Thomson regime or Eq.~(\ref{eqn:ICS_phot_spec_rel}) in the relativistic regime, as well as the scattered electron energy loss spectrum in the Thomson regime in Eq.~(\ref{eqn:electron_eng_loss_spec}), by direct integration.

The problem of integrating these expressions reduces to obtaining an expression for indefinite integrals over the Bose-Einstein distribution of the form
\begin{alignat}{1}
    P_{f} (y) \equiv \int \frac{f(y)\, dy}{e^y - 1} \, .
    \label{eqn:planck_integral}
\end{alignat}
Throughout this appendix, we ignore the constant of integration for such indefinite integrals, since we will ultimately be taking differences of such expressions to find definite integrals. For $f(y) \equiv y^n$ with integer $n \geq 0$, the indefinite integral is known explicitly:
\begin{alignat}{1}
    P_{y^n} (x) = -n! \sum_{s=0}^n \frac{x^s}{s!} \text{Li}_{n-s+1}(e^{-x}) \quad (n = 0, 1, 2, \cdots),
\end{alignat}
where $\text{Li}_{m}(z)$ is the polylogarithm function of order $m$ with argument $z$ (see Appendix~\ref{app:ICS_integrals_series} for the definition). Note however that NumPy does not have a numerical function for the polylogarithm of order $m > 2$, and so the semi-analytic method that we describe below is still necessary for $P_{y^n}$, $n \geq 2$ due to this limitation. 

For other functions $f(y)$, closed-form solutions do not exist. However, an expression for the indefinite integral as an infinite series can be obtained~\cite{Zdziarski:2013gza}. Importantly, more than one series expression exists for all of the integrals $P_{f}(x)$ of interest in both the relativistic and nonrelativistic regimes, so that it is always possible to find a series expression that converges quickly for any integration limit. We tabulate the series expressions already found in Ref.~\cite{Zdziarski:2013gza} for completeness, together with the many new series expressions derived in this paper required for the nonrelativistic limit in Appendix~\ref{app:ICS_integrals_series}. 

\subsubsection{Thomson Regime: Small \texorpdfstring{$\beta$}{beta}}

In the Thomson regime for $\beta \lesssim 0.1$, catastrophic cancellations between terms in the integral make even the method described above insufficient. After integrating Eq.~(\ref{eqn:thomson_scattered_phot_spec}) over $\epsilon$ to get the scattered photon spectrum, for example, the final result must be $\mathcal{O}(\beta^0)$, even though the prefactor in Eq.~(\ref{eqn:thomson_scattered_diff_phot_spec}) is $\mathcal{O}(\beta^{-6})$. The integrals of all of the terms in the curly braces of Eq.~(\ref{eqn:thomson_scattered_diff_phot_spec}) and their analog from Eq.~(\ref{eqn:thomson_spectrum_upp_low_relation}) must therefore cancel among themselves to 1 part in $\beta^{-6}$; such a computation is impossible to perform for $\beta \lesssim 0.003$ due to floating point inaccuracy, even with double precision. 

We avoid this problem by expanding the scattered photon spectrum in Eq.~(\ref{eqn:thomson_scattered_phot_spec}) and the mean electron energy loss spectrum in Eq.~(\ref{eqn:electron_eng_loss_spec}). Eq.~(\ref{eqn:thomson_scattered_phot_spec}) can be expanded straightforwardly in $\beta$, but Eq.~(\ref{eqn:electron_eng_loss_spec}) must be expanded in both $\beta$ and $\xi \equiv \Delta/T$, since catastrophic cancellations occur when either variable is small. In \texttt{DarkHistory}, we expand these expressions up to $\mathcal{O}(\beta^6)$ and $\mathcal{O}(\xi^6)$, but the precision of this calculation is systematically improvable by adding more terms to the code as desired. The exact expressions for the expansions, details of their derivations and several consistency checks for these expressions can be found in Appendix~\ref{app:ICS_integrals_series}. 

\subsection{Results}

\begin{figure}
    \centering
    \includegraphics[scale=0.45]{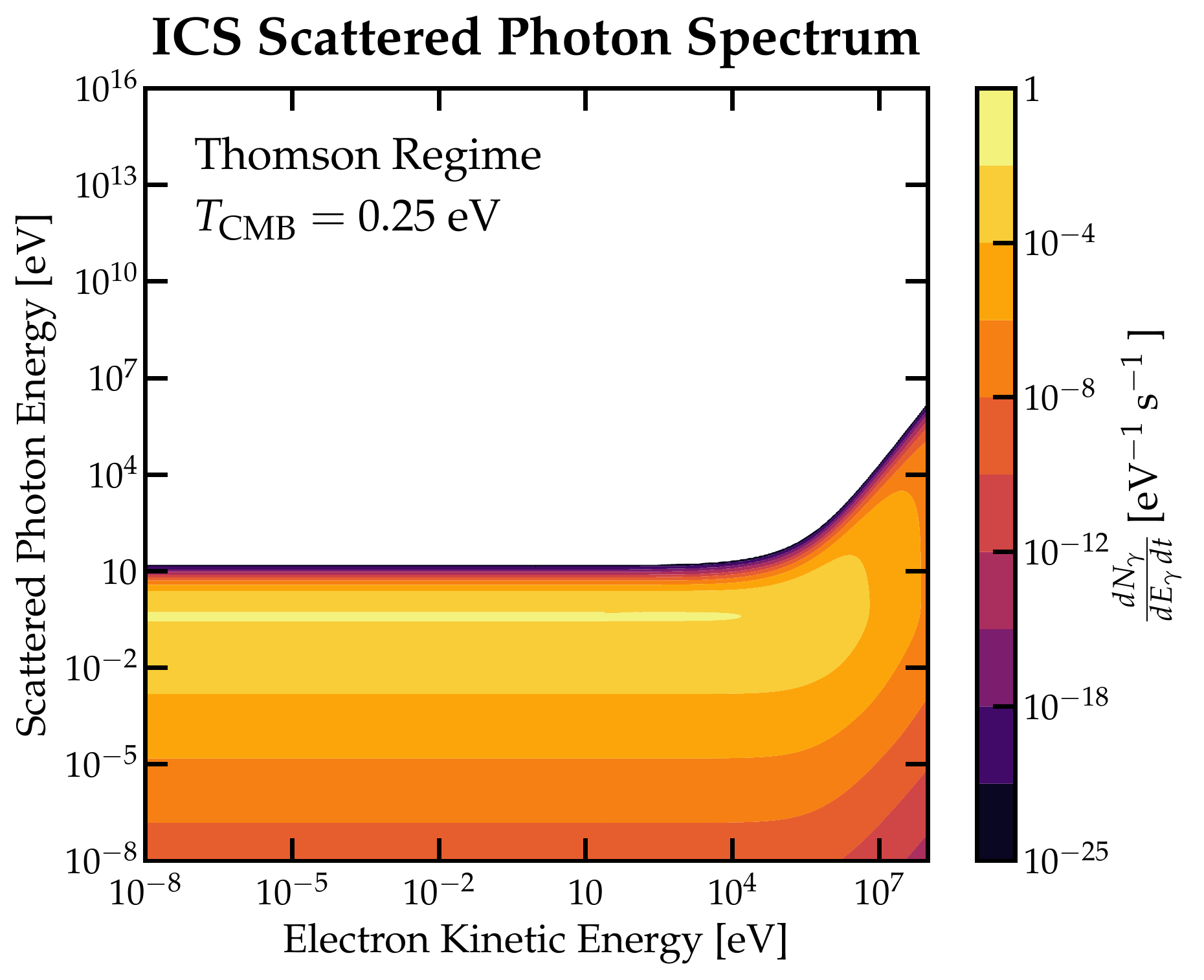}
    \caption{The ICS scattered photon spectrum in the Thomson regime, with $T_\text{CMB}$ = \SI{0.25}{\eV}. }
    \label{fig:ics_thomson_scattered_phot_spec}
\end{figure}

\begin{figure}
    \centering
    \includegraphics[scale=0.45]{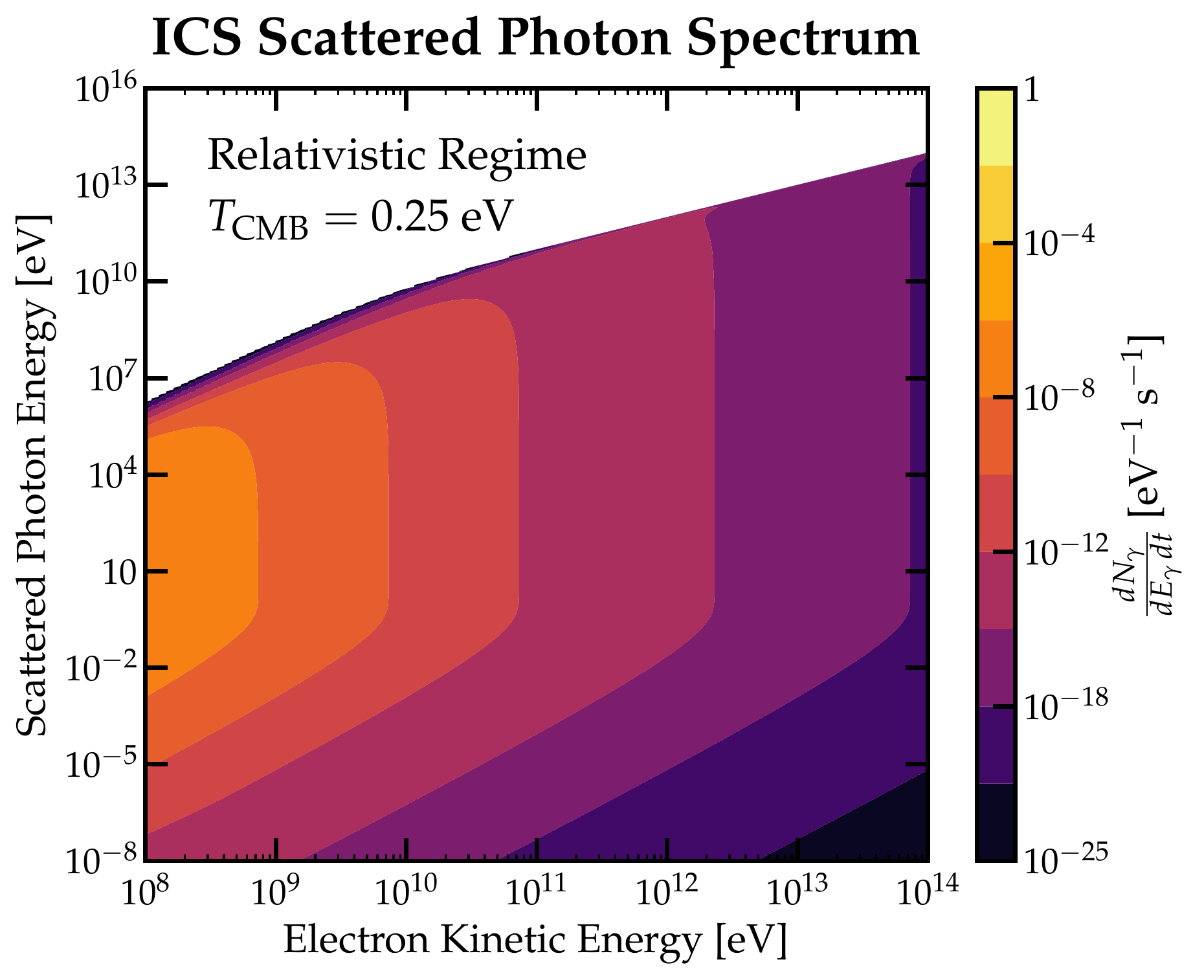}
    \caption{The ICS scattered photon spectrum in the relativistic regime, with $T_\text{CMB}$ = \SI{0.25}{\eV}. }
    \label{fig:ics_rel_scattered_phot_spec}
\end{figure}

\begin{figure}
    \centering
    \includegraphics[scale=0.45]{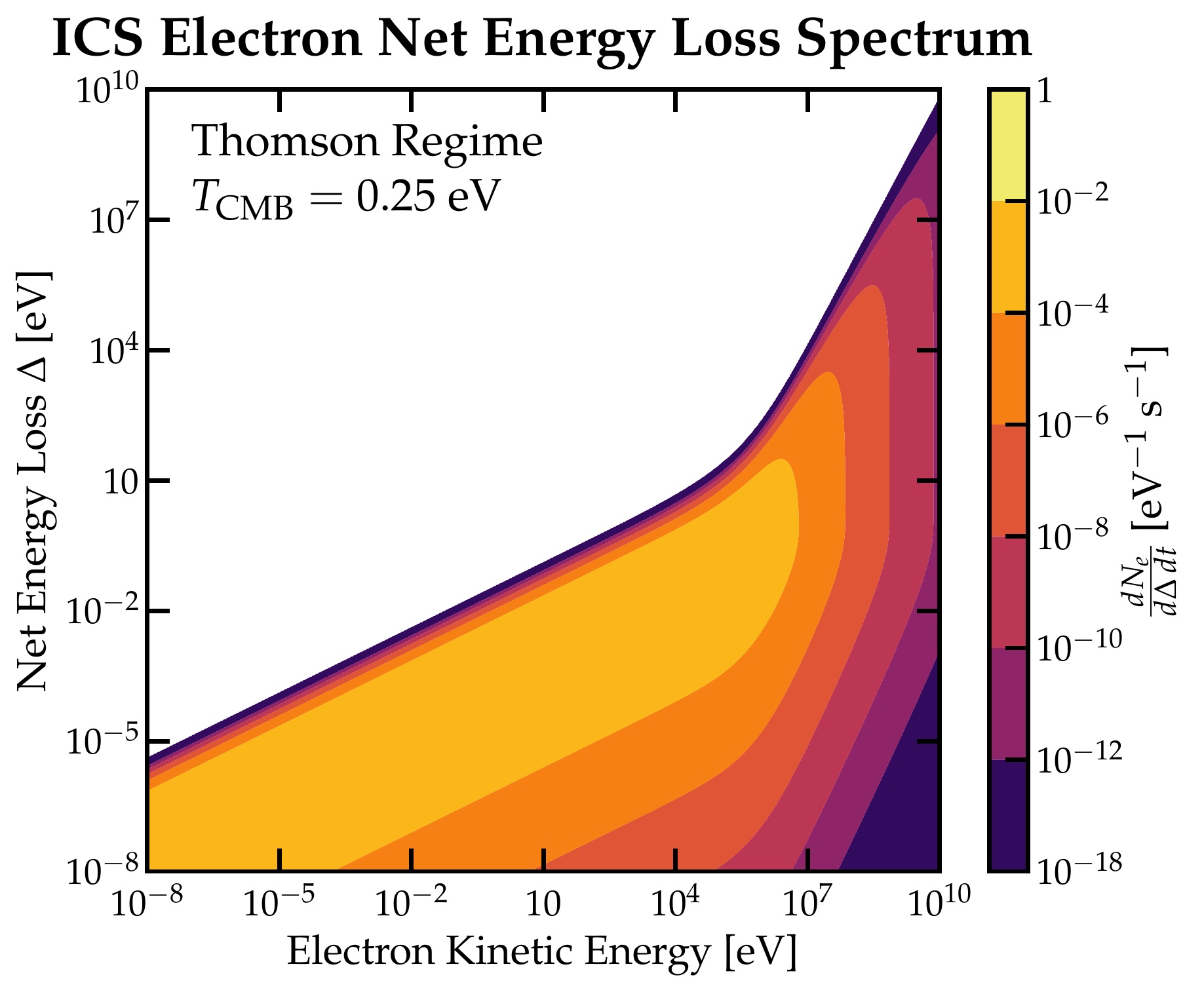}
    \caption{The ICS mean electron energy loss spectrum in the Thomson regime, with $T_\text{CMB}$ = \SI{0.25}{\eV}. }
    \label{fig:ics_thomson_engloss_spec}
\end{figure}

\begin{figure}
   \centering
   \includegraphics[scale=0.45]{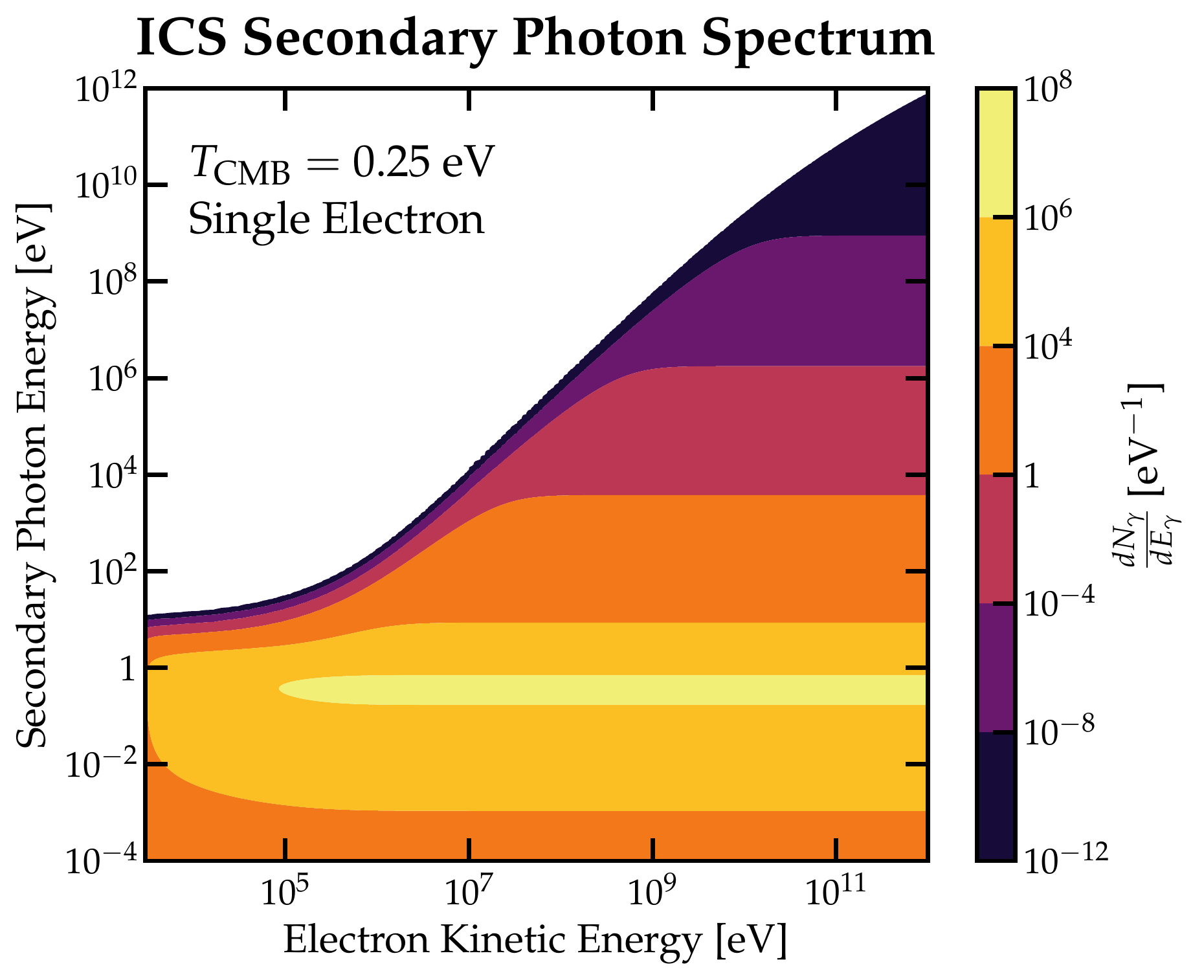}
   \caption{The ICS secondary photon spectrum after complete cooling of a single electron, with $T_\text{CMB}$ = \SI{0.25}{\eV}.}
   \label{fig:ics_sec_phot_spec}
\end{figure}

Figs.~\ref{fig:ics_thomson_scattered_phot_spec} and~\ref{fig:ics_rel_scattered_phot_spec} show the scattered photon spectrum in the Thomson and relativistic regimes respectively as a function of electron energy, at a CMB temperature of \SI{0.25}{\eV}, corresponding to a redshift of $z \approx 1065$ that is near recombination. By default, \texttt{DarkHistory} transitions between these two limits at $\gamma = 20$. Fig.~\ref{fig:ics_thomson_engloss_spec} shows the mean electron energy loss spectrum in the Thomson regime. Above $\gamma = 20$, \texttt{DarkHistory} uses the approximation shown in Eq.~(\ref{eqn:engloss_approx}). Finally, the computed secondary photon spectrum after completely cooling of all electrons and positrons through ICS is shown in Fig.~\ref{fig:ics_sec_phot_spec}. 

All results shown here are computed using a $500 \times 500$ grid of electron and photon energies/energy loss, and each can be completed under ten seconds on a typical personal computer.

\subsection{Integrals and Series Expansions}
\label{app:ICS_integrals_series}

We are now ready to detail the integrals and series expansions used in the numerical methods described above.

\subsubsection{Bose Einstein Integrals}

Each $P_f(x)$ that is of interest has a series that converges quickly for small values of $x$, and another that converges quickly for large values of $x$. \texttt{DarkHistory} by default chooses $x = 2$ as the value to switch between the two expressions. 

Suppose we approximate the indefinite integral in Eq.~(\ref{eqn:planck_integral}) $P_f(x)$ by the first $N$ terms of its series expression, which we denote $S_N(x)$. Let $S_N^s(x)$ and $S_N^l(x)$ be the series expressions we obtain for $x < 2$ and $x \geq 2$ respectively. In all of the cases we are interested in, $S^l_{N \to \infty}(x \to \infty) = 0$ (with the constant of integration taken to be zero) due to the exponential function in the denominator of the original integral, and so 
\begin{alignat}{1}
	S^l_{N\to \infty}(b > 2) = -\int_b^\infty \frac{f(y) \, dy}{e^y - 1} \,.
\end{alignat}
Then defining $\Delta S_N^{s,l} (a, b) = S_N^{s,l}(b) - S_N^{s,l}(a)$, the definite integral is evaluated as
\begin{multline}
	\int_a^b \frac{f(y) \, dy}{e^y - 1} \\
	= \begin{cases}
		\Delta S_{N\to\infty}^s(a, b), & a < 2, b < 2; \\
		\Delta S_{N\to\infty}^s(a, 2) + \Delta S_{N\to\infty}^l(2, b), & a < 2, b \geq 2; \\  
		\Delta S_{N\to\infty}^l(a, b), & a \geq 2, b \geq 2.
	\end{cases}
\end{multline}
Terms are added sequentially until the next contribution to the full integral falls below a given relative tolerance; the default value for this tolerance used by \texttt{DarkHistory} is $10^{-10}$. 

Before listing the series expressions, we must first introduce some notation that will be relevant. The numbers and analytic functions defined below are all well-known, but are often defined with different normalizations or given different names. We explicitly define all relevant functions used here for clarity.

$B_n$ are the Bernoulli numbers, defined through the following exponential generating function:
\begin{alignat}{1}
    \frac{x}{e^x - 1} \equiv \sum_{n=0}^\infty \frac{B_n x^n}{n!},
    \label{eqn:bernoulli_numbers}
\end{alignat}
with $B_{0,1,2,\cdots} = 1, -1/2, 1/6, \cdots$. Note that $B_{2j+1} = 0$ for all integers $j> 0$. 

Next, we define the generalized exponential integrals
\begin{alignat}{1}
	E_n(x) \equiv \int_1^\infty \frac{e^{-xt}}{t^n} \, dt
\end{alignat}
and the closely related incomplete gamma function
\begin{alignat}{1}
    \Gamma(n, x) \equiv \int_x^\infty t^{n-1} e^{-t}\, dt.
\end{alignat}
The polylogarithm of order $m$, denoted $\text{Li}_m(z)$, is defined as
\begin{alignat}{1}
    \text{Li}_m(z) = \sum_{k=1}^\infty \frac{z^k}{k^m} \,.
\end{alignat}
Finally, we define $_2F_1(a,b;c;z)$, the Gaussian hypergeometric function, as
\begin{alignat}{1}
    _2F_1(a, b; c; z) &\equiv 1 + \frac{ab}{1! c} z + \frac{a(a+1) b(b+1)}{2!c(c+1)} z^2 + \cdots \nonumber\\
    &=\sum_{n=0}^\infty \frac{(a)_n (b)_n}{(c)_n} \frac{z^n}{n!},
\end{alignat}
where $(x)_n \equiv x(x+1)\cdots(x+n-1)$ is the Pochhammer symbol. This function only appears in the form $R(n, x) \equiv \Re[_2 F_1(1, n+1; n+2; x)]$, where $\Re$ denotes the real part; to avoid the slow evaluation of the \texttt{hyp2f1} function in NumPy, we use instead the following relation:
\begin{alignat}{2}
    R(n, x) &\equiv&& \,\, \Re[_2 F_1(1, n+1; n+2; x)] \nonumber \\
    &=&& - (n+1) x^{-(n+1)} \log \left(\left| 1 - x \right| \right) \nonumber \\
    & &&- \sum_{j=1}^n \frac{n+1}{j} x^{j-n-1} \,.
\end{alignat} 

The list of all of the series expressions that we use, including those already derived in~\cite{Zdziarski:2013gza}, are shown in Tables~\ref{tab:series_expressions_low} and~\ref{tab:series_expressions_high} for $x < 2$ and $x\geq 2$ respectively.

\renewcommand{\arraystretch}{3}

\setlength{\tabcolsep}{15pt}

\begin{table*}[t!]
\begin{tabular}{
        >{$\displaystyle}c<{$}   
        >{$\displaystyle}c<{$}}

\toprule
\hline
f(y) & P_{f} ,\, x < 2 \\
\hline
y^n, \, n \geq 1 
    & \sum_{k=0}^\infty \frac{B_k x^{k+n}}{k!(k+n)} \\
y \log y
    & x \log x - x + \sum_{k=1}^\infty \frac{B_k x^{k+1}}{k!(k+1)} \left[\log x - \frac{1}{k+1} \right] \\
y \log(y+a),\, a > -x
    & \sum_{k=0}^\infty \frac{B_k x^{k+1}}{k!(k+1)} \left[\log(x + a) + \frac{R(k, -x/a)}{k+1} - \frac{1}{k+1} \right] \\
1
    & \log(1 - e^{-x}) \\
\log y 
    & \frac{1}{2}\log^2 x + \sum_{k=1}^\infty \frac{B_k x^k}{k! k} \left[ \log x - \frac{1}{k} \right] \\
\log(y+a), \, a > 0
    & \log x \log a - \text{Li}_2(-x/a) + \sum_{k=1}^\infty \frac{B_k x^k}{k! k} \left[\log(x + a) - \frac{x}{a(k+1)} R(k, -x/a) \right] \\
\log(y+a), \, -x < a < 0
    & \log(-x/a) \log(x+a) - \text{Li}_2(1+x/a) + \sum_{k=1}^\infty \frac{B_k x^k}{k! k} \left[\log(x+a) - \frac{x}{a(k+1)} R(k, -x/a) \right] \\
\frac{1}{y+a}, \, a > -x
    & \frac{1}{a} \log \left(\frac{x}{x+a}\right) + \sum_{k=1}^\infty \frac{B_k x^k}{k! k} \left[\frac{1}{a} - \frac{k x }{(n+1)a^2} R(k, -x/a) \right]\\
y^{-n},\, n \geq 1 
    & \sum_{k=0}^{n-1} \frac{B_k}{k!} \frac{x^{k-n}}{k-n} + \frac{B_n}{n!} \log x + \sum_{k=1}^\infty \frac{B_{k+n}}{(k+n)!} \frac{x^k}{k} \vspace{0.3cm} \\
    
\botrule
\end{tabular}
\caption{Series expressions for the relevant indefinite integrals of the form shown in Eq.~(\ref{eqn:planck_integral}). Here, $y$ is the integration variable, and $x$ denotes the integration limit of interest. These expressions are used for $x < 2$.}
\label{tab:series_expressions_low}
\end{table*}

\renewcommand{\arraystretch}{3}

\setlength{\tabcolsep}{15pt}

\begin{table*}[t]
\begin{tabular}{
        >{$\displaystyle}c<{$}    
        >{$\displaystyle}c<{$}}

\toprule
\hline
f(y) & P_{f},\, x \geq 2 \\
\hline
(y+a)^n, \, \forall n \in \mathbb{Z},\, a > -x
    & \sum_{k=1}^\infty \frac{e^{ka} \Gamma \big(n+1, k(x+a) \big)}{k^{n+1}} = \sum_{k=1}^\infty \frac{e^{ka} E_{-n}\big( k(x+a) \big)}{(x+a)^{-k-1}} \\
y \log(y+a), \, a > -x
    &  \sum_{k=1}^\infty \frac{e^{ka}}{k^2} \left[(1+kx) e^{-k(x+a)} \log(x+a) + (1+kx)E_1\big(k(x+a)\big) + E_2\big(k(x+a)\big) \right] \\
\log(y+a), \, a > -x
    &  \sum_{k=1}^\infty \frac{e^{ka}}{k} \left[e^{-k(x+a)} \log(x + a) + E_1 \big( k(x+a) \big) \right] \vspace{0.3cm} \\
    
\botrule
\end{tabular}
\caption{Series expressions for the relevant indefinite integrals of the form shown in Eq.~(\ref{eqn:planck_integral}). Here, $y$ is the integration variable, and $x$ denotes the integration limit of interest. These expressions are used for $x \geq 2$.}
\label{tab:series_expressions_high}
\end{table*}

\subsubsection{Nonrelativistic Thomson Limit: Small Parameter Expansion}

The expression for the scattered photon spectrum in the Thomson limit, shown in Eq.~(\ref{eqn:thomson_scattered_phot_spec}), can be expanded in the small $\beta$ limit, to obtain
\begin{alignat}{1}
    \frac{dN_\gamma}{d\epsilon_1 dt_1} = \frac{3 \sigma_T k_B^2 T^2}{32 \pi^2 \hbar^3 c^2} \sum_{n=0}^\infty \sum_{j=1}^{2n}\frac{A_n \beta^{2n} x_1^3 P_{n,j}(x_1) e^{-jx_1}}{(1 - e^{-x_1})^{2n+1}} \, ,
    \label{eqn:thomson_scattered_phot_spec_expansion}
\end{alignat}
where $x_1 = \epsilon_1/T$, $A_n$ is a constant, and $P_{j,n}(x_1)$ is some rational or polynomial function in $x_1$. These quantities are as follows. For $n = 0$ i.e.\ $\mathcal{O}(\beta^0)$, 
\begin{alignat}{1}
    A_0 = \frac{32}{3}, \quad P_{0, 1}(x) = \frac{1}{x} \,.
\end{alignat}
For $n = 1$, i.e.\ $\mathcal{O}(\beta^2)$,
\begin{alignat}{1}
    A_1 = \frac{32}{9}, \quad P_{1,1}(x) = x - 4, \quad P_{1,2}(x) = x + 4 \, .
\end{alignat}
For $n = 2$, i.e.\ $\mathcal{O}(\beta^4)$, 
\begin{alignat}{1}
    A_2 &= \frac{16}{225}, \nonumber \\
    P_{2,1}(x) &= 7x^3 - 84x^2 + 260x - 200, \nonumber \\
    P_{2,2}(x) &= 77x^3 - 252x^2 - 260x + 600, \nonumber \\
    P_{2,3}(x) &= 77x^3 + 252x^2 - 260x - 600, \nonumber \\
    P_{2,4}(x) &= 7x^3 + 84x^2 + 260x + 200,
\end{alignat}
and finally for $n = 3$, i.e.\ $\mathcal{O}(\beta^6)$, 
\begin{alignat}{2}
    A_3 &=&& \,\, \frac{16}{4725}, \nonumber \\
    P_{3,1}(x) &=&& \,\, 11x^5 - 264x^4 + 2142x^3 \nonumber \\
    & && \qquad\quad- 7224x^2 + 9870x - 4200, \nonumber \\
    P_{3,2}(x) &=&& \,\, 3(209x^5 - 2200x^4 + 6426x^3 \nonumber \\
    & && \qquad\quad-2408x^2 - 9870x + 7000), \nonumber \\
    P_{3,3}(x) &=&& \,\, 2(1661x^5 - 5280x^4 - 10710x^3 \nonumber \\
    & && \qquad\quad+28896x^2 + 9870x - 21000), \nonumber \\
    P_{3,4}(x) &=&& \,\, 2(1661x^5 + 5280x^4 - 10710x^3 \nonumber \\
    & && \qquad\quad-28896x^2 + 9870x + 21000), \nonumber \\
    P_{3,5}(x) &=&& \,\, 3(209x^5 + 2200x^4 + 6426x^3 \nonumber \\
    & && \qquad\quad+2408x^2 - 9870x - 7000), \nonumber \\
    P_{3,6}(x) &=&& \,\, 11x^5 + 264x^4 + 2142x^3 \nonumber \\
    & && \qquad\quad+7224x^2 + 9870x + 4200.
\end{alignat}
Furthermore, when $x_1$ is small, it becomes numerically advantageous to expand Eq.~(\ref{eqn:thomson_scattered_phot_spec_expansion}) in $x_1$ as well, leaving a simple polynomial in $x_1$ and $\beta$, i.e.\
\begin{alignat}{1}
    \frac{dN_1}{d\epsilon_1 dt_1} = \frac{3 \sigma_T k_B^2 T^2}{32 \pi^2 \hbar^3 c^2} \sum_{n=0}^\infty \sum_{j=1}^\infty C_{n,j} \beta^{2n} x_1^j \, .
    \label{eqn:thomson_scattered_phot_spec_expansion2} 
\end{alignat}
The values of $C_{n,j}$ are shown in Table~\ref{tab:C_n_j}.
\renewcommand{\arraystretch}{2}

\setlength{\tabcolsep}{7pt}

\begin{table*}[t!]
\begin{tabular}{c c c c c c c c}

\toprule
\hline
$C_{n,j}$ & $x_1$ & $x_1^2$ & $x_1^3$ & $x_1^5$ & $x_1^7$ & $x_1^9$ & $x_1^{11}$ \\
\hline
    $\beta^0$ & 32/3 & -16/3 & 8/9 & -2/135 & 1/2835 & -1/113400 & 1/4490640 \\
    $\beta^2$ & -64/9 & 0 & 32/27 & -4/45 & 8/1701 & -1/4860 & 1/124740 \\
    $\beta^4$ & -256/225 & 0 & 32/27 & -296/1125 & 1208/42525 & -64/30375 & 389/3118500 \\
    $\beta^6$ & -832/1575 & 0 & 32/27 & -1828/3375 & 31352/297675 & -10669/850500 & 10267/9355500 \vspace{0.2cm}\\
\botrule
\end{tabular}
\caption{List of coefficients $C_{n,j}$ for use in Eq.~(\ref{eqn:thomson_scattered_phot_spec_expansion2}).}
\label{tab:C_n_j}
\end{table*}

Three checks can be performed to verify that this is indeed the correct expansion in $\beta$. First, taking $\beta \to 0$, the scattered photon spectrum simply becomes $dN_\gamma/(d\epsilon_1 \, dt_1) = n_{\text{BB}}(\epsilon_1, T) \sigma_T c$, which is exactly the expected result for Thomson scattering in the rest frame of the electron: all photons simply scatter elastically at a rate governed by the Thomson scattering cross section, thus remaining in a blackbody distribution. Second, a more non-trivial check is to integrate Eq.~(\ref{eqn:thomson_scattered_phot_spec_expansion}) with respect to $\epsilon_1$, giving the total Thomson scattering rate given in Eq.~(\ref{eqn:thomson_scattering_rate}). Since the scattering rate is independent of $\beta$, the $\mathcal{O}(\beta^0)$ term in the series should integrate to exactly $\sigma_T c N_\text{rad}$ where $N_\text{rad}$ is the number density of the blackbody photons, while the other higher order terms should integrate to exactly zero. This is indeed the case for the series expansion shown here. Lastly, one can check that Eq.~(\ref{eqn:thomson_scattered_phot_spec_expansion}) agrees with the energy loss expression Eq.~(\ref{eqn:thomson_energy_loss_rate}), by noting that
\begin{alignat}{1}
  \int \frac{dN_\gamma}{d\epsilon_1 \, dt_1} \epsilon_1 \, d\epsilon_1 = \sigma_T c u_\text{BB}(T) + \frac{4}{3} \sigma_T c \beta^2 \gamma^2 u_\text{BB}(T) \,,
\end{alignat}
where $u_\text{BB}(T)$ is the blackbody energy density with temperature $T$, i.e.\ the produced secondary photon spectrum must have the same energy as the upscattered CMB photons plus the energy lost from the scattering electron. This check has also been performed for the series expansions shown here.

For the scattered electron energy loss spectrum shown in Eq.~(\ref{eqn:electron_eng_loss_spec}), the small $\beta$ and $\xi$ expansion can be written as
\begin{alignat}{1}
    \frac{dN_e}{d\Delta \, dt} = \frac{3 \sigma_T k_B^2 T^2}{32 \pi^2 \hbar^3 c^2}  \sum_{n=0}^\infty \left[ \sum_{j=1}^{2n} \frac{A^{j+1} Q_{n,j}(e^{-A}) }{(1-e^{-A})^j \beta^{-2n}} + R_n(A) \right],
    \label{eqn:electron_engloss_spec_expansion}
\end{alignat}
where $Q_{n,j}(x)$ is a polynomial, $A \equiv \Delta/(2\beta T) = \xi/(2 \beta)$, and $R_n(A)$ is a sum of integrals of the form
\begin{alignat}{1}
    P_k(A) = A^{k+1} \int_A^\infty \frac{x^{-k} \,dx}{e^x - 1}\, .
\end{alignat}
These integrals can be evaluated using the same methods detailed in Appendix~\ref{app:ICS_integrals_series}. The list of polynomials $Q_{n,j}$ and of $R_n(A)$ is given below. All expressions not listed should be taken to be zero. For $n = 0$, 
\begin{alignat}{1}
    R_0(A) = \frac{176}{15}P_0 - \frac{64}{3} P_3 + \frac{128}{5} P_5 \,.  
\end{alignat}
For $n = 1$, 
\begin{alignat}{1}
    Q_{1,1}(x) &= - \frac{32}{3} x \,, \qquad Q_{1,2} = \frac{8}{3} x \,, \nonumber \\
    R_1(A) &= -\frac{1168}{105} P_0 + \frac{128}{3}P_3 - \frac{2176}{15} P_5 + \frac{1280}{7} P_7 \,. 
\end{alignat}
For $n = 2$, 
\begin{alignat}{1}
    Q_{2,1}(x) &= -\frac{512}{15}x \,, \qquad Q_{2,2}(x) = \frac{8}{5}x \,, \nonumber \\
    Q_{2,3}(x) &= -\frac{8}{15} x(1+x)\,, \nonumber \\
    Q_{2,4}(x) &= \frac{2}{15}(x + 4x^2 + x^3) \,, \nonumber \\
    R_2(A) &= -\frac{64}{3}P_3 + \frac{640}{3}P_5 - 768 P_7 + \frac{14336}{15} P_{9}\,. 
\end{alignat}
And finally for $n = 3$, 
\begin{alignat}{2}
    Q_{3,1}(x) &=&& -\frac{416}{3}x \,, \qquad Q_{3,2}(x) = \frac{1184}{105}x \,, \nonumber \\
    Q_{3,3}(x) &=&& -\frac{256}{315}(x + x^2)\,, \nonumber \\
    Q_{3,4}(x) &=&& -\frac{2}{63}(x + 4x^2 + x^3) \,, \nonumber \\
    Q_{3,5}(x) &=&& - \frac{4}{315}(x + 11x^2 + 11x^3 + x^4) \,, \nonumber \\
    Q_{3,6}(x) &=&& \frac{1}{315}(x + 26x^2 + 66x^3 + 26x^4 + x^5) \,, \nonumber \\
    R_3(A) &=&& -\frac{512}{3465}P_0 - \frac{1408}{15}P_5 \nonumber \\
    & && \qquad+ \frac{6912}{7}P_7 - \frac{161792}{45}P_9 + \frac{49152}{11} P_{11} \,.
\end{alignat}
These are all the terms necessary to work at order $\mathcal{O}(\beta^6)$ and $\mathcal{O}(\xi^6)$. As before, if $A$ becomes small, we should expand Eq.~(\ref{eqn:electron_engloss_spec_expansion}) as
\begin{alignat}{1}
    \frac{dN_e}{d\Delta \, dt} = \frac{3 \sigma_T k_B^2 T^2}{32 \pi^2 \hbar^3 c^2} \sum_{n=0}^\infty \left[\sum_{j=0}^\infty D_{n,j} \beta^{2n} A^j + R_n(A)\right],
    \label{eqn:electron_engloss_spec_expansion2}
\end{alignat}
with the values of $D_{n,j}$ shown in Table~\ref{tab:D_n_j}.
\renewcommand{\arraystretch}{2}

\setlength{\tabcolsep}{7pt}

\begin{table*}[t!]
\begin{tabular}{c c c c c c c c}

\toprule
\hline
$D_{n,j}$ & $A$ & $A^2$ & $A^3$ & $A^5$ & $A^7$ & $A^9$ & $A^{11}$ \\
\hline
    $\beta^2$ & -8 & 16/3 & -10/9 & 7/270 & -1/1260 & 11/453600 & -13/17962560\\
    $\beta^4$ & -164/5 & 256/15 & -134/45 & 161/2700 & -19/9450 & 359/4536000 & -289/89812800 \\
    $\beta^6$ & -40676/315 & 208/3 & -1312/105 & 4651/18900 & -416/59535 & 989/4536000 & -173/22453200 \vspace{0.2cm}\\
\botrule
\end{tabular}
\caption{List of coefficients $D_{n,j}$ for use in Eq.~(\ref{eqn:electron_engloss_spec_expansion2}).}
\label{tab:D_n_j}
\end{table*}

These expressions are complicated, but can be checked in a similar fashion as the scattered photon spectrum by integrating over $\Delta \, d\Delta$ to obtain the mean energy loss rate of electrons scattering of a blackbody spectrum, given exactly in Eq.~(\ref{eqn:thomson_energy_loss_rate}). Using the fact that
\begin{alignat}{1}
    \int_0^\infty d\Delta \,\Delta \, P_k(A) = \frac{4 \pi^4 \beta^2 T^2}{15(n+2)}\,, 
\end{alignat}
one can verify that integrating the $\mathcal{O}(\beta^6)$ expansion gives
\begin{alignat}{1}
    \frac{dE_e}{dt} = \frac{4}{3} \sigma_T c U_\text{rad} \beta^2 (1 + \beta^2 + \beta^4 + \beta^6) \,,
\end{alignat}
which is precisely the Taylor expansion of Eq.~(\ref{eqn:thomson_energy_loss_rate}) in powers of $\beta$. 

\section{Positronium Annihilation Spectra}
\label{app:positronium_annihilation_spec}

The spin-triplet $^3S_1$ state of positronium annihilates to three photons, producing a photon spectrum per annihilation given by \cite{Ore:1949te}
\begin{multline}
    \left. \frac{dN_\gamma}{dE_\gamma} \right|_{^3S_1} = \frac{6}{(\pi^2 - 9) m_e} \bigg\{ \frac{2-x}{x} + \frac{x(1-x)}{(2-x)^2} \\
    + 2 \log(1 - x) \left[\frac{1-x}{x^2} - \frac{(1 - x)^2}{(2 - x)^3}\right] \bigg\} \,,
\end{multline}
where $x \equiv E_\gamma/m_e$. The kinematically allowed range is $0 \leq x \leq 1$. Assuming that the formation of positronium by low energy positrons populates all of the degenerate ground states equally, the averaged photon spectrum per annihilation is 
\begin{alignat}{1}
    \left. \frac{dN_\gamma}{dE_\gamma} \right|_\text{Ps} = \frac{1}{4} \delta(E_\gamma - m_e) + \frac{3}{4}  \left. \frac{dN_\gamma}{dE_\gamma} \right|_{^3S_1}.
\end{alignat}

\section{Cross Checks}
\label{app:cross_checks}

\subsection{Helium Deposition}
\label{app:helium_deposition}

\begin{figure}
  \centering
  \includegraphics[scale=0.53]{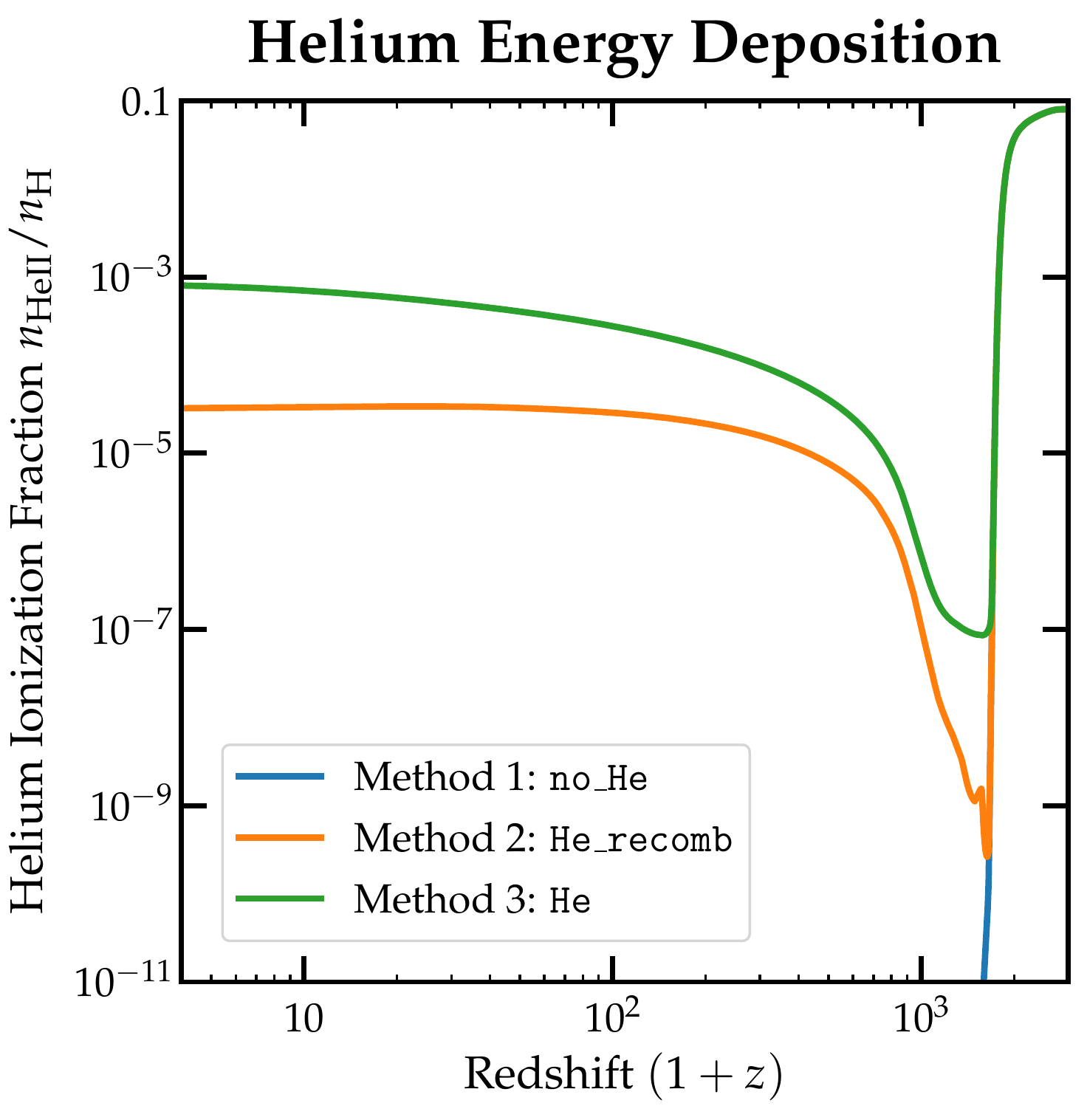}
  \caption{Helium ionization fraction with different helium energy deposition methods: (1) no tracking of the helium evolution (i.e.\ $x_\text{HeII}$ takes its baseline value) (blue) (2) all photoionized HeI atoms recombine, producing a photon that photoionizes hydrogen (orange), and (3) photoionized HeI atoms remain photoionized (green). The energy injection corresponds to \SI{100}{\mega \eV} DM decaying through $\chi \to \gamma \gamma$ with a lifetime of \SI{3e24}{s}.}
  \label{fig:He_f_method_xHeII}
\end{figure}

\begin{figure*}[t!]
  \begin{tabular}{cc}
      \includegraphics[scale=0.58]{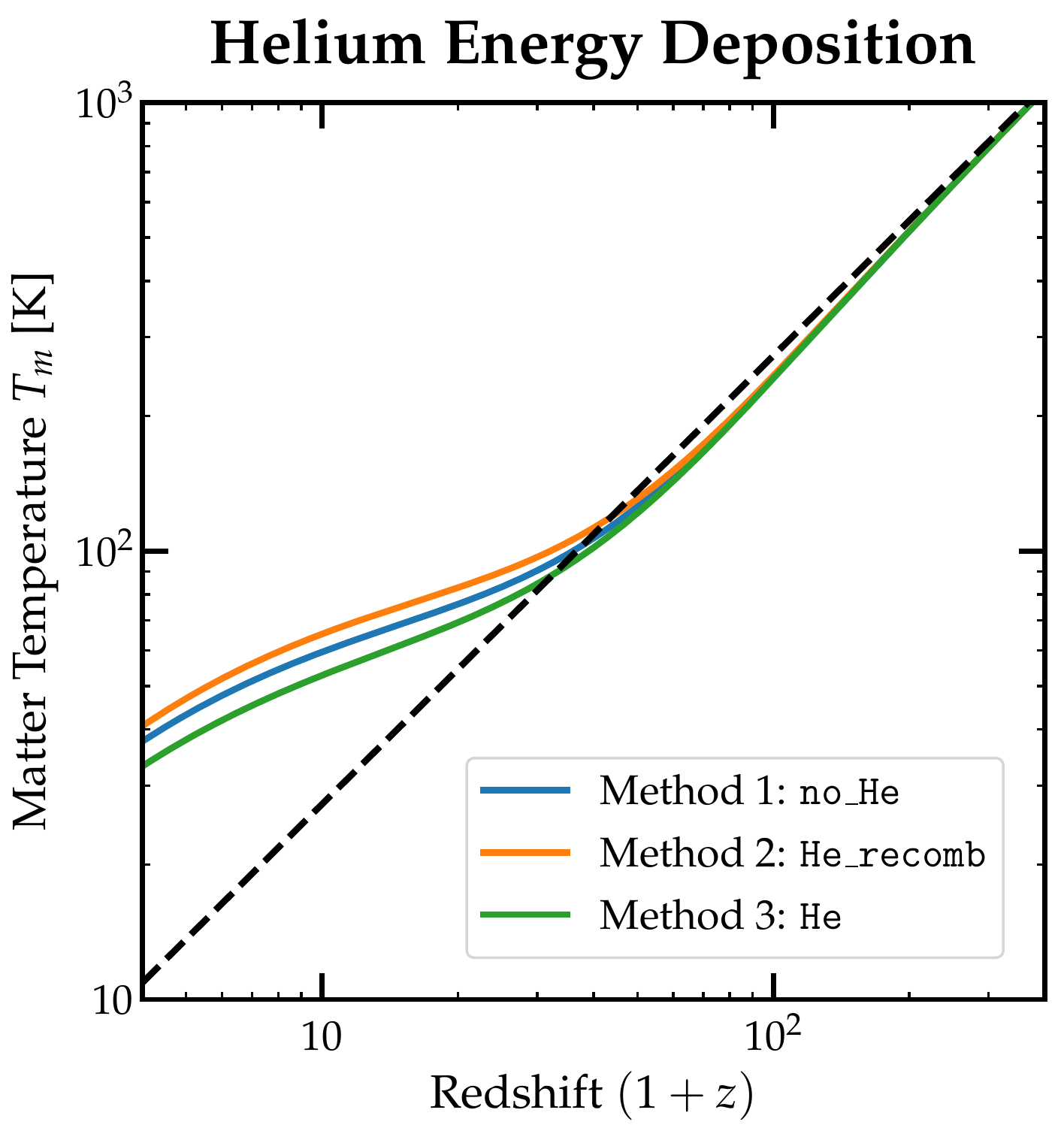} &
      \includegraphics[scale=0.58]{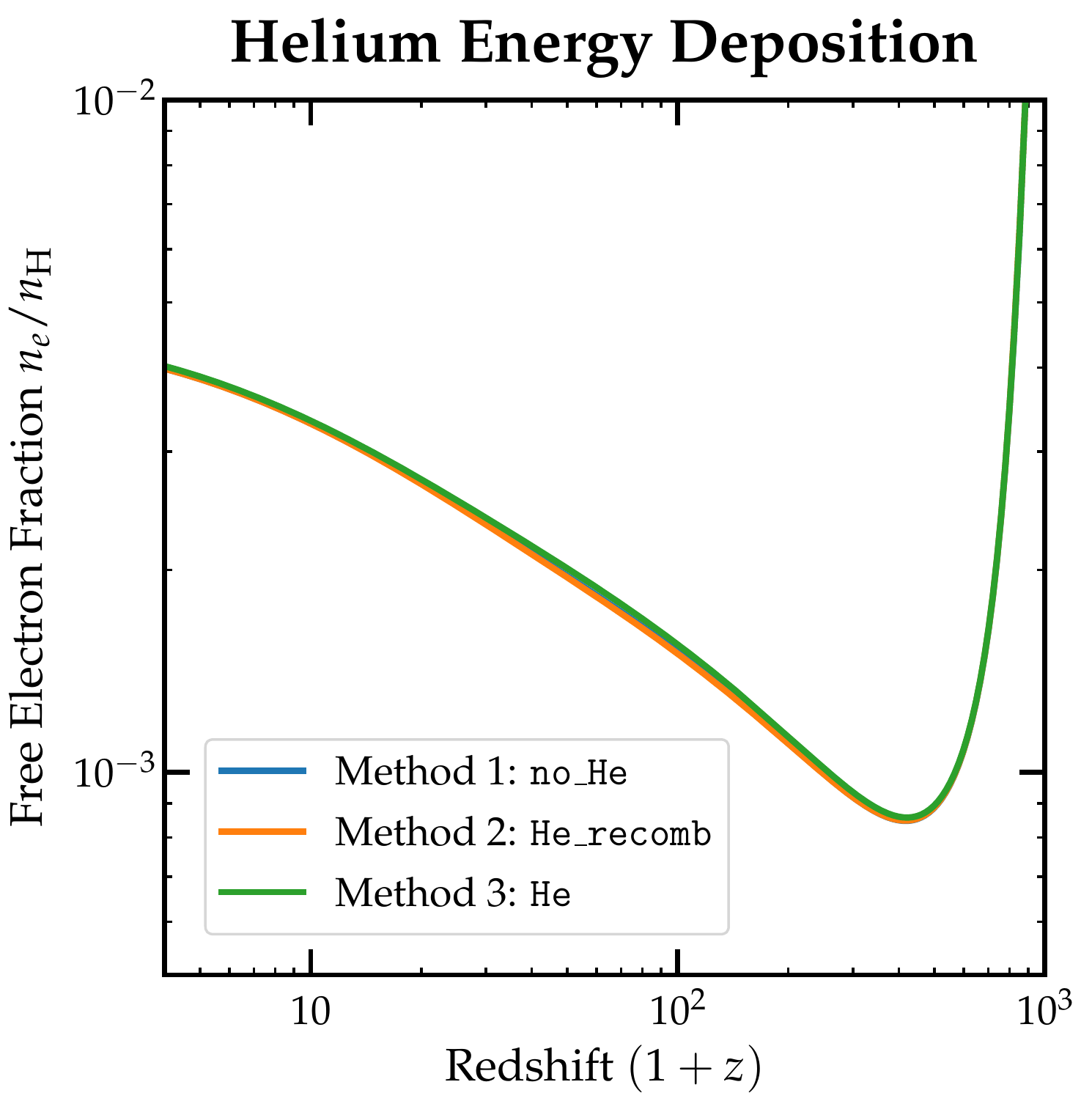} 
  \end{tabular}
  \caption{Matter temperature (left) and free electron fraction (right) evolution with different helium energy deposition methods: (1) no tracking of the helium evolution (i.e.\ $x_\text{HeII}$ takes its baseline value) (blue) (2) all photoionized HeI atoms recombine, producing a photon that photoionizes hydrogen (orange), and (3) photoionized HeI atoms remain photoionized (green). The CMB temperature is shown for reference (black, dashed). The energy injection corresponds to \SI{100}{\mega \eV} DM decaying through $\chi \to \gamma \gamma$ with a lifetime of \SI{3e24}{s}.}
\label{fig:He_f_method_histories}
\end{figure*}

In this section, we compare the various helium energy deposition methods discussed in Sec.~\ref{sec:photon_cooling}. We pick a dark matter candidate which decays to two photons with a lifetime of \SI{3e24}{\second} as an example, but the results are similar across different dark matter masses and energy injection rates. 

 The user can switch between methods by passing the keyword parameter \lstinline|compute_fs_method| to \texttt{evolve()} with the following strings for each method:  (1) \lstinline|'no_He'|, (2) \lstinline|'He_recomb'| and (3) \lstinline|'He'|, e.g.
\begin{lstlisting}
  helium_method_alt = main.evolve(
      DM_process='decay', mDM=1e8, 
      lifetime=3e24, primary='phot_delta',
      start_rs=3000., backreaction=True,
      helium_TLA=True, 
      compute_fs_method='He_recomb'
  )
\end{lstlisting}

Fig.~\ref{fig:He_f_method_xHeII} shows the helium ionization fraction $x_\text{HeII}$ as a function of redshift for each of the different methods. In method (1), $x_\text{HeII}$ is simply the baseline helium ionization fraction, which is almost entirely neutral once helium recombination is complete. No energy is assigned to helium iondization at all. Method (2) has no contribution to helium ionization from photons, since every ionized helium atom is assumed to recombine, producing a photon that photoionizes hydrogen instead (i.e.\ setting $q^\gamma_\text{He} = 0$ in Eq.~(\ref{eqn:He_ion_dep_phot})). The helium ionization level therefore deviates from the baseline only from energy injection in the $\text{He}_\text{ion}$ channel from low-energy electrons. On the other hand, method (3) assumes that all helium atoms that get photoionized stay ionized, maximizing the amount of energy into $\text{He}_\text{ion}$ from photons (i.e.\ setting $q^\gamma_\text{He} = 1-q$ in Eq.~(\ref{eqn:He_ion_dep_phot})). This explains the higher $x_\text{HeII}$ obtained. 

Despite these differences in $x_\text{HeII}$, the evolution of $x_e$ remains almost identical, due to the fact that the total number of ionization events between both hydrogen and helium remains the same regardless of method used. This in turn ensures only a small difference in $T_m$ between the methods. The ionization and temperature histories for all three methods for the particular channel we have chosen are shown in Fig.~\ref{fig:He_f_method_histories}. Users may bracket the uncertainty in the treatment of helium with methods (2) and (3). 

\subsection{Coarsening}

\begin{figure}
  \centering
  \includegraphics[scale=0.53]{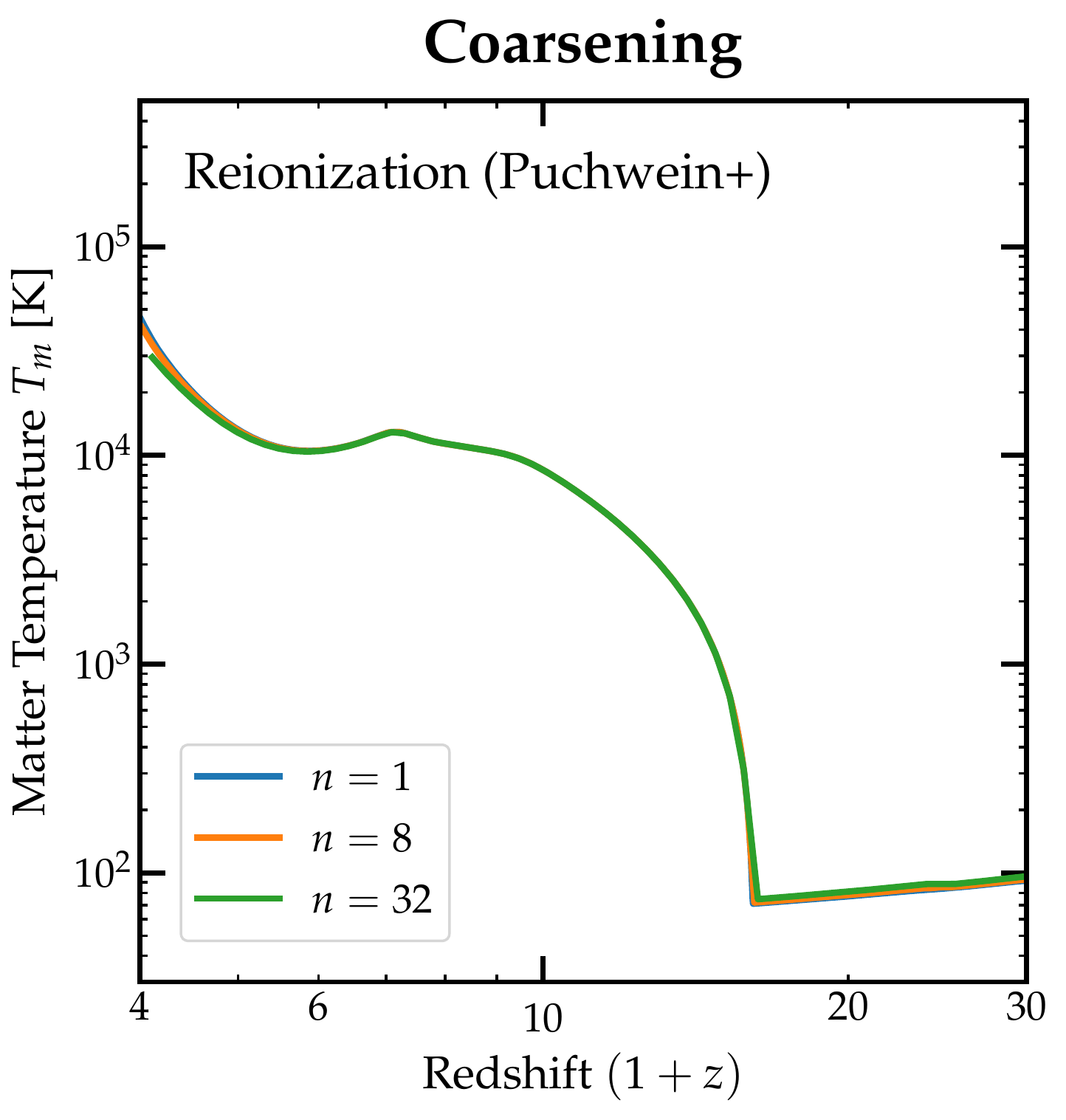}
  \caption{Matter temperature evolution with the default reionization model with no coarsening (blue), a coarsening factor of 8 (orange) and 32 (green).}
  \label{fig:coarsening}
\end{figure}

In the absence of reionization, a coarsening factor of up to 32 has been found to yield a small relative difference of between 5--10\% in the values of $f_c(z)$ across the full range of redshifts used in \texttt{DarkHistory}. With reionization, however, $T_m$ evolves more rapidly and attains larger values, and too much coarsening can lead to large absolute differences and somewhat larger relative differences in $T_m$, since we are averaging over the $T_m$ evolution over many redshift points. Fig.~\ref{fig:coarsening} shows the resultant temperature evolution as a function of redshift for the same model used in the previous section but with the default reionization model turned on, with coarsening factors of 1, 8 and 32. Once reionization starts, the difference in $T_m$ is $\sim 15\%$ for $n = 32$ compared to the uncoarsened result at $z \sim 4$, corresponding to an absolute error of $\sim \SI{5000}{\kelvin}$. Prior to reionization, the relative errors are slightly smaller at $\lesssim 10\%$. 

We therefore recommend using a coarsening factor of up to 32 if no reionization models are used, depending on the level of precision desired, and to use coarsening with care once reionization is included. We also emphasize that when using coarsening, it is best to check for convergence by comparing the result with less coarsening. 

\subsection{\texorpdfstring{$f_c(z)$ Contours}{f\_c(z) Contours}}

\begin{figure*}[t!]
  \centering
  \includegraphics[scale=0.22]{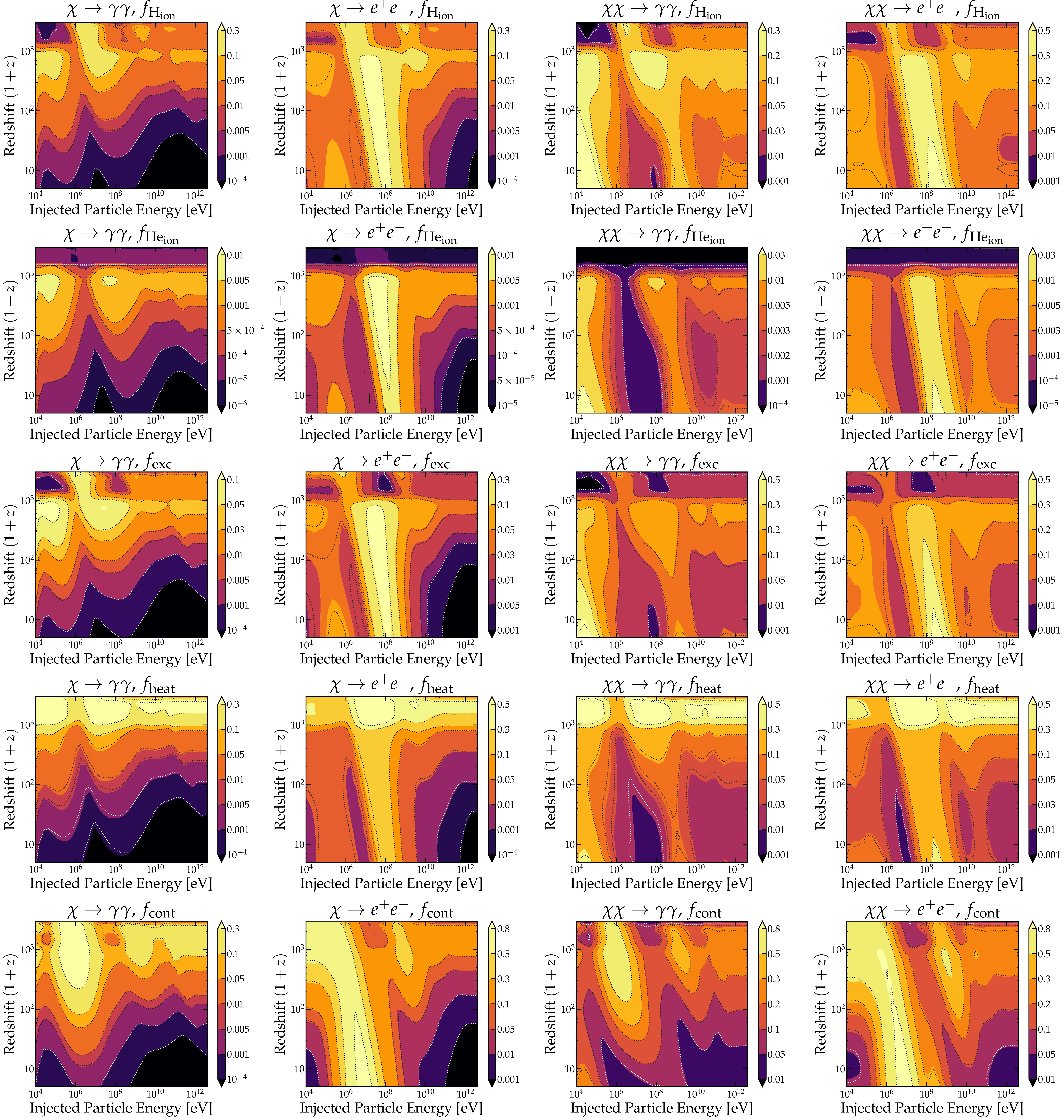}
  \caption{Computed $f_c(z)$ values without backreaction with \texttt{DarkHistory} for (from left to right) $\chi \to \gamma \gamma$ decays, $\chi \to e^+e^-$ decays, $\chi \chi \to \gamma \gamma$ annihilations and $\chi \chi \to e^+e^-$ annihilations (with no boost factor). The results from Refs.~\cite{Slatyer:2015kla,Liu:2016cnk} are shown for comparison (dashed lines). These contour plots agree with the previous results to within 10\% if all calculation methods are standardized between \texttt{DarkHistory} and Refs.~\cite{Slatyer:2015kla,Liu:2016cnk}, and represent an improved calculation of $f_c(z)$ neglecting backreaction.}
  \label{fig:f_contours_all}
\end{figure*}

Fig.~\ref{fig:f_contours_all} show the computed $f_c(z)$ contours within \texttt{DarkHistory} for all channels of interest without any backreaction.  The new $f_c(z)$ calculation by \texttt{DarkHistory} makes several small physics and numerical improvements over the previous calculation of these results~\cite{Slatyer:2015kla,Liu:2016cnk}, but still agree to within less than 10\% when methodologies (cosmological parameters, methods of interpolation etc.) are standardized between the code used in Ref.~\cite{Slatyer:2015kla} and \texttt{DarkHistory}. The new calculation also corrects a bug in earlier work in the treatment of prompt energy deposition from nonrelativistic and mildly relativistic injected electrons.  This accounts for the bulk of the visible differences in Fig.~\ref{fig:f_contours_all} between the current contours and those of Refs.~\cite{Slatyer:2015kla,Liu:2016cnk}, which are most pronounced for DM annihilation/decay to electrons and low injected particle energies.

\section{Table of Definitions}
\label{app:table}

Table~\ref{tab:variables} shows a list of variables and their definitions for reference. 

\clearpage

\begin{longtable*}{C{0.1 \textwidth}  C{0.2 \textwidth} L{0.6 \textwidth}}
  \toprule
  \hline
  \textbf{Category} & \textbf{Symbol} & \textbf{Definition}\\
  \hline
  \hline
  General 

  &$y$ & log-redshift, $y \equiv \log(1+z)$. \\
  
  & $\Delta y$, $\Delta t$ & log-redshift step size and associated time step size. \\
  &$\mathbf{x}$ & Ionization levels: $\mathbf{x} \equiv (x_\text{HII}, x_\text{HeII}, x_\text{HeIII}) \
\equiv (n_\text{HII}/n_\text{H}, n_\text{HeII}/n_\text{H}, n_\text{HeIII}/n_\text{H})$ i.e.\ the fractional abundance of ionized hydrogen atoms, singly-ionized helium atoms and doubly-ionized helium atoms with respect to the number of hydrogen atoms (both neutral and ionized). \\

& $\zeta_i$ & $\text{arctanh} \left[ (2/\chi_i) \left(n_i/n_\text{H} - \chi_i/2\right) \right]$ where $i \in \{\text{HII}, \text{HeII}, \text{HeIII}\}$, convenient reparametrization of $\mathbf{x}$ introduced for numerical purposes.\\

& $T_m$ & Temperature of the IGM. \\

& $T^{(0)}_m$, $x^{(0)}_\text{HII}(z)$ & Baseline temperature and ionization histories, obtained from Eq.~(\ref{eqn:TLA}). \\

& $m_\chi$, $\tau$, $\langle \sigma v \rangle$ & Dark matter mass, lifetime, and velocity-averaged annihilation cross section. \\

& $\left(\frac{dE}{dV \,dt}\right)_\text{inj}$ & Energy injection rate per volume for exotic forms of energy injection, given for dark matter annihilation/decay in Eq.~(\ref{eqn:energy_injection}). \\
\hline

Spectra 

& $G(z)$ & Conversion factor between the rate of injected events per volume to the number of injected events per baryon within a log-redshift step, as defined in Eq.~(\ref{eqn:per_baryon_to_dVdt}). \\

&$\overline{\mathbf{N}}_\text{inj}^\alpha[E_{\alpha,i}']$				
& Spectrum containing number of particles of type $\alpha \in \{\gamma, e\}$ injected into energy bin $E'_{\alpha, i}$ per annihilation event.\\ 

&$\mathbf{N}_\text{inj}^\alpha[E_{\alpha,i}', y']$						
& Spectrum containing the number of particles per baryon in a log-redshift step of type $\alpha$ injected into energy bin $E'_{\alpha, i}$ at log-redshift $y'$, as defined in Eq.~(\ref{eqn:injected_discretized_spec}).\\

&$\overline{\mathbf{N}}^\gamma_\text{pos}[E_{\gamma,i}']$		
& Spectrum of photons produced from a single positronium annihilation event.\\

&$\mathbf{N}^\gamma_\text{new}[E'_{\gamma,i}, y']$					
& Sum of the spectra of primary injected photons, and secondary photons produced by the cooling of electrons, as defined in Eq.~(\ref{eqn:new_inj_photons}).
\\
&$\mathbf{N}_\text{prop}^\gamma [E_{\gamma,i}', y']$				
& Spectrum of propagating photons with energies greater than $13.6$ eV that do not photoionize or get otherwise deposited into low-energy photons.
\\
&$\mathbf{N}^\gamma [E_{\gamma,i}', y']$						
& $\mathbf{N}_\text{prop}^\gamma [E_{\gamma,i}', y'] + \mathbf{N}^\gamma_\text{new}[E_{\gamma,i}', y']$, as defined in Eq.~(\ref{eqn:N_prop_plus_new}). 
\\
&$\mathbf{N}^\alpha_\text{low} [E_{\alpha,i}, y]$
& Low-energy photons ($\alpha = \gamma$) or electrons ($\alpha = e$) at log-redshift $y$.
\\
\hline
Photon Cooling

& $\overline{\mathsf{P}}^\gamma[E'_{\gamma,i}, E_{\gamma,j}, y', \Delta y, \mathbf{x}]$ & Transfer function for propagating photons, which multiplies $\mathbf{N}^\gamma[E_{\gamma,i}', y']$ and produces $\mathbf{N}^\gamma_\text{prop} [E_{\gamma,j}, y' - \Delta y]$, as defined in Eq.~(\ref{eqn:discretized_prop_tf}).  \\

&$\overline{\mathsf{D}}^e[E'_{\gamma,i}, E_{e,j}, y', \Delta y, \mathbf{x}]$ & Low-energy electron deposition transfer function, which multiplies $\mathbf{N}^\gamma[E_{\gamma,i}', y']$ and produces $\mathbf{N}^e_\text{low} [E_{e,j}, y'-\Delta y]$, as defined in Eq.~(\ref{eqn:lowengelec_tf}).\\

&$\overline{\mathsf{D}}^\gamma[E'_{\gamma,i}, E_{\gamma,j}, y', \Delta y, \mathbf{x}]$						
& Low-energy photon deposition transfer function, which multiplies $\mathbf{N}^\gamma[E_{\gamma,i}', y']$ and produces $\mathbf{N}^\gamma_\text{low} [E_{\gamma,j}, y'-\Delta y]$, as defined in Eq.~(\ref{eqn:lowengphot_tf}).\\

&$\overline{\mathsf{D}}_c^\text{high}[E'_{\gamma,i}, y', \Delta y, \mathbf{x}]$		
& High-energy deposition transfer matrix, which multiplies $\mathbf{N}^\gamma[E_{\gamma,i}', y']$ and returns the total energy that greater than \SI{3}{\kilo\eV} electrons produce during the cooling process deposit into channel $c \in \{$`ion', `exc', `heat'$\}$, as defined in Eq.~(\ref{eqn:highengdep_tf}) in the next log-redshift step at $y' - \Delta y$.\\

&$\left( \overline{\mathsf{P}}^\gamma_{1/2} \right)^n$		
& Coarsened propagating photon transfer function with a coarsening factor of $n$, as defined in Eq.~(\ref{eqn:prop_tf_coarsening}), which multiplies $\mathbf{N}^\gamma[E_{\gamma,i}', y']$ and produces $\mathbf{N}^\gamma_\text{prop} [E_{\gamma,j}, y' - n \Delta y]$. 
\\
\hline
Electron Cooling

& $\overline{\mathsf{N}}[E'_{e,i}, E_{e,j}]$ 
& Spectrum of secondary electrons produced due to the cooling of a single injected electron with initial energy $E'_{e,i}$. \\

&$\overline{\mathbf{R}}_c[E'_{e,i}]$ 		
& High-energy deposition vector containing the total energy deposited into channel $c \in $\{`ion', `exc', `heat'\} by a single injected electron with kinetic energy $E'_{e,i}$, as defined in Eq.~(\ref{eqn:elec_cooling_dep_tf}). \\

&$\overline{\mathbf{R}}_\text{CMB}[E_{e,i}']$				
& Total initial energy of CMB photons that are upscattered via ICS due to the cooling of a single electron of energy $E_{e,i}'$. \\

&$\overline{\mathsf{T}}_\text{ICS,0}[E_{e,i}', E_{\gamma,j}]$
& Spectrum of photons produced with energy $E_{\gamma,j}$ due to the cooling of a single electron of energy $E_{e,i}'$, as defined in Eq.~(\ref{eqn:ics_photons}). \\

&$\overline{\mathsf{T}}_\text{ICS}[E_{e,i}', E_{\gamma,j}]$ & The same as $\overline{T}_\text{ICS,0}$, but with the pre-scattering spectrum of upscattered CMB photons subtracted out, as defined in Eq.~(\ref{eqn:elec_cooling_ics}).\\

&$\overline{\mathsf{T}}_e [E'_{e,i}, E_{e,j}]$			
& Low-energy electron spectrum produced due to the cooling of a single electron of energy $E_{e,i}'$, as defined in Eq.~(\ref{eqn:elec_cooling_lowengelec}).\\
\hline
Low-energy deposition 
& $f_c(z, \mathbf{x})$ 
& Ratio of deposited to injected energy, as a function of redshift $z$ and the ionization level $\mathbf{x}$, into channels $c \in \{$`H ion', `He ion', `exc', `heat', `cont$\}$, as defined in Eq.~(\ref{eqn:fz}).\\

& $\left(\frac{dE^\alpha}{dV \,dt}\right)_c$				
& Energy deposited per volume and time by low-energy photons ($\alpha = \gamma$) or electrons ($\alpha = e$) into channel $c$.\\

&$E_c^\text{high}[y]$								
& Total amount of high-energy deposition into channels $c \in $\{`ion', `exc', `heat'\} at log-redshift $y$. \\

\botrule
\caption{A list of the important definitions used in \texttt{DarkHistory}.  In this table, all spectra are discretized spectra as described in Sec.~\ref{sec:discretization}.  Spectra without overlines are normalized so that their entries contain number (per baryon) of particles produced in a redshift step.  A primed energy denotes the energy of an injected particle, and by energy we mean kinetic energy.  In this table, when we refer to electrons we will always mean electrons plus positrons.}
\label{tab:variables}
\end{longtable*}

\bibliography{DarkHistory}

\end{document}